\documentclass{article}
\usepackage{tikz-cd}
\usepackage{amsfonts}
\usepackage{amsmath}
\usepackage{geometry}
\usepackage{comment}
\usepackage[utf8]{inputenc}

\newtheorem{theorem}{Theorem}

\newtheorem{remark}[theorem]{Remark}

\begin{document}

\begin{center}
{\large \textbf{\\[2mm]
Reductions of Topologically Massive Gravity II: }}

{\large \textbf{First Order Realizations of Second Order Lagrangians}}

\bigskip

\bigskip

Filiz \c{C}a\u{g}atay U\c{c}gun$^{1}$, O\u{g}ul Esen$^{2}$ and Hasan G\"{u}%
mral$^{3}$

\bigskip
\end{center}

$^{1}$Department of Software Engineering, Maltepe University, 34857 Maltepe, İstanbul, Turkey, filizcagatayucgun@maltepe.edu.tr 

$^{2}$Department of Mathematics, Gebze Technical University, 41400 Gebze,
Kocaeli, Turkey, oesen@gtu.edu.tr

$^{3}$Department of Mathematics, Yeditepe University, 34755 Ata{\c{s}}ehir, {\.{I}}stanbul, Turkey, hgumral@yeditepe.edu.tr

\bigskip

\textbf{Abstract:} Second order degenerate Cl\`{e}ment and Sar\i o\u{g}lu-Tekin  Lagrangians are casted into forms of various first order Lagrangians. Hamiltonian analysis of these equivalent formalisms are performed by means of Dirac-Bergmann constraint algorithm. 


\textbf{AMS2010:} 70H45, 70H50, 70H05, 83E05.

\tableofcontents

\section{Introduction}

A particularly interesting class of degenerate second order Lagrangians arises in the theory of topologically massive gravity \cite{DeJaTe82,DeJaTe82b}, 
namely Cl\'{e}ment, and Sar\i o\u{g}lu-Tekin Lagrangians. 
Cl\'{e}ment introduced
\begin{equation}
L^{C}=-\frac{m}{2}\zeta \mathbf{\dot{x}}^{2}-\frac{2m\Lambda }{\zeta }+\frac{\zeta
^{2}}{2\mu m}\mathbf{x}\cdot (\mathbf{\dot{x}}\times \mathbf{\ddot{x}})
\label{clemlag}
\end{equation}%in 
while searching particle like solutions of the topologically
massive gravity consisting of actions for cosmological gravity and Chern-Simons term \cite{cle92,
cle94a,cle94b}. The Cl\'{e}ment Lagrangian (\ref{clemlag}) is 
depending on positions $\mathbf{x}$, velocities $\mathbf{\dot{x}}$ and
accelerations $\mathbf{\ddot{x}.}$ Here, the inner product is defined by the
Lorentzian metric, $\zeta =\zeta (t)$ is a function which allows arbitrary
reparametrization of the variable $t$ whereas $\Lambda $ and $1/2m$ are the
cosmological and Einstein gravitational constants, respectively. Sar\i o\u{g}%
lu-Tekin Lagrangian \begin{equation}
L^{ST}=\frac{1}{2}\left[ a(\mathbf{\dot{x}}^{2}+\mathbf{\dot{y}}^{2})+\frac{2}{\mu }\dot{%
\mathbf{y}}\cdot \ddot{\mathbf{x}}-m^{2}(\mathbf{y}^{2}+\mathbf{x}^{2})%
\right],  \label{stlag}
\end{equation}
proposed in \cite{st06} while studying
actions consisting of Einstein-Hilbert, Chern-Simons and Pauli-Fierz terms.
Here, $a,\mu ,m$ are scalars, and $\mathbf{x},\mathbf{y}$ are the position
three-vectors. 

The
Legendre transformations of higher order Lagrangian formalisms can be achieved by
defining Ostrogradski momenta \cite%
{ost50}. If the Lagrangian is non-degenerate in the sense of Ostrogradski,
then the Legendre transformation is immediate. If it is degenerate, then one
additionally needs to employ the Dirac-Bergmann algorithm \cite%
{Be56,Dirac,Dirac1950,SuMu74} to arrive at a proper submanifold of momentum phase space in order to have a well-defined Hamilton's equations.

In our previous work \cite{CaEsGu18}, we obtained Hamiltonian formalisms of both the Cl\'{e}ment Lagrangian (\ref{clemlag}) and the Sar\i o\u{g}lu-Tekin
Lagrangian (\ref{stlag}) by writing
Ostrogradski momenta, and applying Dirac analysis on
associated momentum phase spaces.
In the present paper, we shall continue to the Hamiltonian analysis of Cl\'{e}ment (\ref{clemlag}) and Sar\i o\u{g}lu-Tekin (\ref{stlag}) Lagrangians by recasting the second order Lagrangians into first order forms on some iterated tangent bundles. Such reduction procedures are far from being unique and are mainly achieved by enlarging the configuration space with introductions of new coordinates and  Lagrange multipliers. In reduction procedures, the reduced first order Lagrangian would be
degenerate because of presence of the Lagrange multipliers  even if the second order one is not.  Hence,
in this work, while presenting the Hamiltonian formalizations of the Cl\'{e}ment and Sar\i o\u{g}lu-Tekin Lagrangians, we shall be facing with two kinds of
degeneracies. The first one is due to the functional structures of the
Lagrangians, that is degeneracies in the sense of Ostrogradski. The
second one is due to the degeneracies evolving in the reduction procedures. 

This paper is organized into four main sections. In the following section, some
preliminary aspects, such as the higher order Euler-Lagrange equations, the Legendre transformation, and the Dirac-Bergmann algorithm, will be exhibited for the sake of completeness. In the third section, we shall summarize three possible ways to recast a second order Lagrangian function into a first order form. In total reduction, this is achieved by considering both velocity and acceleration variables as independent coordinates. In this case, Lagrange multipliers will turn out to be Ostrogradski's momenta. In the second way, that we call partial reductions, only velocity variable is defined to be an independent variable. In the third way, only acceleration variable is considered as a new coordinate. This last kind of reduction considered in different contexts as the Schmidt method \cite{AnGoMaMa10,Sc94} or the Deriglazov's trick \cite{de2017}. We shall prove that these two  methods are the same for second order theories in the realm of acceleration bundle \cite{EsGu18}. Novelty of this work can be found in the the fourth, and the fifth sections which are reserved for applications of techniques presented in the third section to the Cl\'{e}ment (\ref%
{clemlag}) and the Sar\i o\u{g}lu-Tekin (\ref{stlag}) Lagrangians, respectively. For each case, reduced Lagrangian functions, and Legendre transformations are explicitly presented. Accordingly, total Hamiltonian functions, applications of the Dirac-Bergmann algorithm to presymplectic structures, and Dirac brackets are exhibited.

\section{Fundamentals}

\subsection{Geometry of iterated tangent bundles}

Let $Q$ be a manifold. Consider the set $%
C_{q}(Q)$ of smooth curves passing through a point $q$ in $Q$. Two curves $\gamma$ and $\gamma'$ are called $k$-equivalent and denoted by $\gamma\sim^{k}_q \gamma'$, if they agree up to their $k-$%
th order derivatives at $q$ that is if
\begin{equation}
D^r (f\circ \gamma) (0)=D^r (f\circ\gamma') (0), \qquad r=0,1,2,...,k,
\label{ERk}
\end{equation}
for all real valued functions $f$ defined on $Q$ \cite{LeRo11}. Here, we assumed that $\gamma(0)=\gamma'(0)=q$.
Under the equivalence relationship $\sim^{k}_q$, an
equivalence class, denoted by $\mathfrak{t}^{k}\gamma(0)$, is called a $k$-th order tangent vector at $q$. The set of all equivalence classes of curves,
that is the set of all $k$-th order tangent vectors at $q$ defines $k$-th
order tangent space $T_{q}^{k}Q$. The union
\begin{equation*}
T^{k}Q=\bigsqcup_{q\in Q} T_{q}^{k}Q
\end{equation*}
 of all $k$-th order tangent spaces is $k$-th order tangent bundle of $Q$.

There exists a projection from the $k$-th order tangent bundle $T^{k}Q$ to the manifold $Q$ defined as
\begin{equation}
\tau^k_{Q}:T^{k}Q\longrightarrow Q:\mathfrak{t}^{k}\gamma(0) \longrightarrow \gamma(0)\label{tbp}.
\end{equation}
All possible triples
\begin{equation*}
(T^{k}Q,\tau_{Q}^{k},Q)
\end{equation*}
are fiber bundles with total space $T^{k}Q$, the projection $\tau_{Q}^{k}$, and the base space $Q$.
If $k=1$ then we arrive at the tangent bundle $(TQ,\tau_{Q},Q)$. $TQ$ admits a vector bundle structure on $Q$ but in general $T^kQ$ does not necessarily admit a vector bundle structure on $Q$ for $k\geq2$ \cite{Su17}.

Assume that $Q$ is an $n$-dimensional manifold with local coordinates $( \mathbf{q} )=(q^1,...,q^n)$, then the first order tangent bundle $TQ$ is a $2n$-dimensional manifold with induced coordinates
\begin{equation}
( \mathbf{q} ,\mathbf{\dot q}):TQ\longrightarrow \mathbb{R}^{2n}:\mathfrak{t}\gamma(0)\longrightarrow ( \mathbf{q} \circ \gamma(0),D( \mathbf{q} \circ \gamma) (0)).
\end{equation}
The $k-$th order tangent bundle $T^{k}Q$ is a $[( k+1)n]$-dimensional
manifold with induced coordinates
\begin{equation} \label{coorT}
( \mathbf{q} ,\mathbf{\dot q},..., \mathbf{q} ^{(k)}):T^kQ\longrightarrow \mathbb{R}^{( k+1)n} \notag :\mathfrak{t}^k\gamma(0)\longrightarrow ( \mathbf{q} \circ \gamma(0),D( \mathbf{q} \circ \gamma) (0),...,D^k( \mathbf{q} \circ \gamma)).
\end{equation}

There are two possible ways to write the iterated bundle $TTQ$ as a vector bundle over the tangent bundle $TQ$. The geometry of this double vector bundle structure \cite{GrRo09} can be summarized in the following commutative diagram, called as tangent rhombic,
\begin{equation} \label{tr}
 \begin{tikzcd}[column sep=tiny,row sep=huge]
 &&TTQ \arrow[lld, "\tau_{TQ}"] \arrow[rrd, "T\tau_{Q}"] & 
 \\
  TQ \arrow[rrd,"\tau_{Q}"]&& & &TQ \arrow[dll,"\tau_{Q}"] \\
& &Q
& &
\end{tikzcd}
\end{equation}
%\begin{equation} \label{tr}
 %\begin{tikzcd}[column sep=tiny,row sep=huge]
 %&&TTQ \arrow[lld, "\tau_{TQ}"] \arrow[rrd, "T\tau_{Q}"] & 
 %\\
  %TQ \arrow[rrd,"\tau_{Q}"]&& & &TQ \arrow[dll,"\tau_{Q}"] %\\
%& &Q
%& &
%\end{tikzcd}
%\end{equation}
where $\tau_{TQ}$ is the tangent bundle projection whereas $T\tau_Q$ is the tangent lift of $\tau_Q$. 
In terms of a local coordinate system $\mathbf{(q,v,\dot{q},\dot{v})}$ on $TTQ$, these projections are computed to be
\begin{eqnarray}
\tau_{TQ}:TTQ \longrightarrow TQ &:& \mathbf{(q,v,\dot{q},\dot{v})} \mapsto \mathbf{(q,v)},
\\
T\tau_{Q}:TTQ \longrightarrow TQ &:& \mathbf{(q,v,\dot{q},\dot{v})} \mapsto \mathbf{(q,\dot{q})}.
\end{eqnarray}
In terms of these projections, we present the embedding of the second order tangent bundle $T^2Q$ into $TTQ$ as follows
\begin{equation} \label{T2Q-}
T^2Q=\{\mathbf{z}\in TTQ: \tau_{TQ}(\mathbf{z})=T\tau_{Q}(\mathbf{z})\}
\end{equation}
which, in a local frame $\mathbf{(q,\dot{q},v,\dot{v})}$ on $TTQ$, reads
\begin{eqnarray} \label{sot}
T^2Q=\{\mathbf{(q,v,\dot{q},\dot{v})}\in TTQ : \mathbf{\dot{q}-v=0}\}.
\end{eqnarray}
Existence of this embedding leads us to write a second order Lagrangian defined on $T^2Q$ as a first order Lagrangian function defined on $TTQ$ along with the constraint $\mathbf{\dot{q}-v=0}$. Such a first order realization of a second order Lagrangian function fails to be unique due to the existence of the double vector bundle structure of $TTQ$ pictured in (\ref{tr}). We discuss this duality in the framework of Lagrangian dynamics by labeling them as partial reduction I and partial reduction II in Subsections (\ref{parred1}) and (\ref{parred2}), respectively. They will provide two possible ways to define velocity as a new coordinate in a second order framework.

\subsection{Acceleration bundle}

It is interesting to note a submanifold of $TTQ$ defined by 
\begin{equation} \label{AQ}
AQ=\{\mathbf{z}\in TTQ:\tau _{TQ}(\mathbf{z})=T\tau _{Q}\left( \mathbf{z}\right)=\mathbf{0} \}.
\end{equation}
We call $AQ$ as acceleration bundle \cite{EsGu18}.
Comparing the definitions of $T^2Q$ and $AQ$ given in (\ref{T2Q-}) and (\ref{AQ}), respectively,  one immediately observes that $AQ$ is also a submanifold of $T^2Q$. Accordingly, $AQ$ can be identified with the intersection
\begin{equation}
AQ=VTQ\cap T^2Q,
\end{equation}
where $VTQ$ is the vertical bundle consisting of vectors on $TQ$
 projecting to the zero vectors on $Q$ via the mapping $T\tau_{Q}$.  

Here is a direct way to define $AQ$ without referring to $TTQ$. Consider the set $C_{q}(Q)$ of smooth curves passing through the point $q$ in $Q$. We
define a subset $K_q( Q)$ of $C_{q}\left( Q\right)$ by
considering only the curves whose first derivatives are vanishing at $q$.
More formally,
\begin{equation}
K_q(Q)=\left\{\gamma\in C_{q}(Q):D(f\circ \gamma) (0)=0, \quad \forall f:Q\mapsto\mathbb{R} \right\}.
\end{equation}
It is worthless to say that since
vanishing of the first derivative is asked only at a single point, the
curve $\gamma$ in $K_q(Q)$ does not have to be a constant curve. We are now defining an equivalence relation on  $K_q(Q)$. We call two curves $\gamma$ and $\gamma'$ in $K_q(Q)$ equivalent if the second derivatives of $\gamma$ and $\gamma'$ are equal at $q$, that is if
\begin{equation*}
D^2(f\circ \gamma) (0)=D^2 (f\circ\gamma') (0),
\end{equation*}
for all real valued functions $f$ on $Q$. Here, it is assumed that $\gamma(0)=\gamma'(0)=q$. An equivalence class is denoted by $\mathfrak{a}\gamma(0)$. The set of all equivalence classes in $K_q(Q)$ is called  acceleration space $A_{q}Q$ at $q\in Q$.
If $Q$ is an $n$-dimensional manifold with local coordinates $(\mathbf q)$, then union of all acceleration spaces
\begin{equation*}
AQ=\bigsqcup_{q\in Q}
A_{q}Q.
\end{equation*}
is a $2n$-dimensional manifold with induced local coordinates
\begin{align}
( \mathbf{q} ,     \mathbf{a}     ):AQ\longrightarrow \mathbb{R}^{2n}:&\mathfrak {a}\gamma(0)\longrightarrow ( \mathbf{q} \circ \gamma(0),D^2( \mathbf{q} \circ \gamma)(0)).
\end{align}
Assume that $( \mathbf{q} )$ and $(\mathbf{x})$ be two compatible charts around a point $q$ in $Q$. Then, the induced local charts on $AQ$, given by $(\mathbf{q,a})$ and $(\mathbf{x,b})$, are also compatible. Transformations relating these two local pictures are computed to be
\begin{align} \label{locT2Q}
\mathbf{x}=\mathbf{x}( \mathbf{q} ),
\qquad
\mathbf{b}=\left (\mathbf{a} \cdot \frac{\partial }{\partial  \mathbf{q} }\right ) \mathbf{x}.
\end{align}
These coordinate transformations suggest a vector bundle structure of $ AQ$ over $Q$ with projection
\begin{equation} \label{accpro}
\alpha_{Q}: AQ\rightarrow Q:\mathfrak{a}\gamma(0)\longrightarrow \gamma(0).
\end{equation}

Acceleration bundle geometry will lead us to introduce an alternative reduction procedure to write a second order Lagrangian function as a first order Lagrangian function by labeling the acceleration as a new coordinate instead of the velocity. Accordingly, in Subsection (\ref{acce}), we shall elaborate the geometries of both Schmidt's method and Deriglazov's trick in the realm of $AQ$.

\subsection{Higher order Euler-Lagrange equations}

A Lagrangian function $L$ is a real valued function on $TQ$, and generates the Euler-Lagrange equations
\begin{equation}
\frac{\partial L}{\partial \mathbf{q}}-\frac{d}{dt}\frac{\partial L}{%
\partial \mathbf{\dot{q}}}=\mathbf{0}. \label{EL}
\end{equation}%
Second order tangent bundle 
$T^{2}Q$ of $Q$ is $3n$-dimensional manifold with local
coordinates $(\mathbf{q;\dot{q};\ddot{q}})$. A second order Lagrangian function $L=L(\mathbf{q;\dot{q};\ddot{q}})$ is a
real-valued function on $T^{2}Q$. The second order Euler-Lagrange equations are
\begin{equation}
\frac{\partial L}{\partial \mathbf{q}}-\frac{d}{dt}\frac{\partial L}{%
\partial \mathbf{\dot{q}}}+\frac{d^{2}}{dt^{2}}\frac{\partial L}{\partial 
\mathbf{\ddot{q}}}=\mathbf{0}.  \label{EL2}
\end{equation}%
Notice that, the set (\ref{EL2}) of second order Euler-Lagrange equations
consists of the fourth order differential equations only if one of the components of $\partial
L/\partial \mathbf{\ddot{q}}$ depends on the acceleration $\mathbf{\ddot{q}}$. While presenting Schmidt's method in Subsection (\ref{schmidt}), we shall need the Euler-Lagrange equations for the third order Lagrangians. Accordingly, we record here the third order Euler-Lagrange equations
\begin{equation}
\frac{\partial L}{\partial \mathbf{q}}-\frac{d}{dt}\frac{\partial L}{%
\partial \mathbf{\dot{q}}}+\frac{d^{2}}{dt^{2}}\frac{\partial L}{\partial 
\mathbf{\ddot{q}}}-\frac{d^{3}}{dt^{3}}\frac{\partial L}{\partial 
\mathbf{q}^{(3)}}=\mathbf{0}.  \label{EL3}
\end{equation}%
generated by a third order Lagrangian function $L=L(\mathbf{q;\dot{q};\ddot{q};q}^{(3)})$.

\subsection{The Legendre transformation and the Dirac-Bergmann algorithm}

Hamiltonian representation of a physical system is defined on the cotangent bundle $T^*Q$ of the configuration manifold $Q$ \cite{AbMa78}. Physically, $T^*Q$ corresponds to the momentum-phase space of a physical system. $T^*Q$ carries a canonical symplectic two-form $\Omega_Q$, hence a non-degenerate Poisson structure $\{\bullet,\bullet\}$. Dynamics of an observable $f$ governed by a Hamiltonian function $H$ is determined by the Hamilton's equations
\begin{equation}
\dot{f}=\{f,H\}.
\end{equation}
If we equip $T^*Q$ with the Darboux' coordinates $(\mathbf{q},\mathbf{p})$ then the canonical Poisson bracket relations can be determined by
$$\{\mathbf{p},\mathbf{q}\}=\mathbb{I},$$ and the rest of all possible brackets are identically zero. Here, $\mathbb{I}$ is the identity matrix. In this coordinate frame, the Hamilton's equations turn out to be 
\begin{equation}
\dot{\mathbf{q}}=\frac{\partial H}{\partial \mathbf{p}},\qquad \dot{\mathbf{p}} =-\frac{%
\partial H}{\partial \mathbf{q}}.  \label{HamEq2}
\end{equation}

For Hamiltonian formulation of the first order Euler-Lagrange equations (\ref{EL}), one needs to define a passage from the tangent bundle $TQ$ to the cotangent bundle $T^*Q$. This is achieved by means of the fiber derivative of a Lagrangian function $L$ that is
\begin{equation} \label{FdL}
\mathbb{F}L:TQ\rightarrow T^{\ast }Q:\left( \mathbf{q},\dot{\mathbf{q}}\right) \rightarrow
\left( \mathbf{q},\frac{\partial L}{\partial \dot{\mathbf{q}}}\right) . 
\end{equation}%
It is evident that, in order to make the transformation (\ref{FdL})
invertible, one needs to employ a non-degeneracy condition, called the Hessian
condition,
\begin{equation}
\det \frac{\partial ^{2}L}{\partial \dot{\mathbf{q}}^{2}}\neq 0.
\label{Hessian}
\end{equation}%
In this case, the
velocities $\dot{\mathbf{q}}$ can uniquely be written as functions of position and momenta, and the Hamiltonian function on $T^{\ast }Q$ is defined to be
\begin{equation} \label{cHF}
H\left(\mathbf{q},\mathbf{p}\right) ={\mathbf{p}}\cdot \dot{\mathbf{q}} \left(
\mathbf{q},\mathbf{p}\right) -L\left( \mathbf{q},\dot{\mathbf{q}}\left(
\mathbf{q,p}\right) \right).
\end{equation}
 Notice that, the Hamilton's equations (\ref{HamEq2}) generated by the Hamiltonian function presented in  (\ref{cHF}) equal to the first order Euler-Lagrange
equations (\ref{EL}).

If the Hessian condition (\ref{Hessian}) is not satisfied then one cannot solve the velocities $\dot{\mathbf{q}}$ in terms of momenta $\mathbf{p}$. Instead, one arrives at an immersed submanifold, called primary constraint submanifold, $C$ of $T^{\ast
}Q $. We are assuming that $C$ can be defined as a set of constraint functions, called primary constraints, $\Phi _{a}\approx 0$ on $T^*Q$. Here $a$ is ranging from $1$ to the codimension $r$ of $C$. The equalities in the primary constraints are weak in the sense
that they will be ignored during set up of Dirac formalism, and will actually
vanish in any solutions to equations of motion.  The dynamics on primary constraint submanifold
is not well-defined by the canonical Hamiltonian function (\ref{cHF}),
it is rather governed by the total Hamiltonian 
\begin{equation*}
H_{T}=H+u^{a}\Phi _{a}
\end{equation*}%
which contains linear combinations of the primary constraints with Lagrange
multipliers $u^{a }$.  The requirement that the solutions of
Euler-Lagrange equations remain on the constraint submanifold is described by
the weak equality%
\begin{equation}
\dot{\Phi}_{b}=\{ \Phi_b,H_T \} =\{\Phi _{b},H\}+u^{a }\{\Phi _{b},\Phi
_{a}\}\approx 0.
\end{equation}%
These consistency conditions may lead
to determination of Lagrange multipliers if the left hand sides contain $%
u^{a }$. In this case, one solves for $u^{a}$ through the set of
linear equations%
\begin{equation*}
\{\Phi _{b },\Phi _{a }\}u^{a}=-\{\Phi _{b},H\}
\end{equation*}%
for which the solution set, namely, number of multipliers that can be solved
is characterized by the rank of the skew-symmetric matrix $\{\Phi _{b
},\Phi _{a}\}$ of Poisson brackets. Obviously, if the number of
primary constraints is odd then Lagrange multipliers cannot be solved completely and
one aspects more constraints to determine $H_{T}$ in terms of phase space
variables. This secondary constraints follow if the left hand sides does not
contain $u^{a}$ or, $n-r$ is odd. Repeating this process, one enlarges
the primary constraint set with the new (secondary, tertiary, etc.)
constraints, redefines $H_{T}$ by introducing new Lagrange multipliers for
new constraints and, repeats the consistency computations.
Iterated applications of consistency computations lead to a complete set of
constraints $\Phi _{\alpha }:\alpha =1,...,k$. Let 
\begin{equation*}
\mathcal{M}_{\alpha \beta}=\left\{ \Phi _{\alpha },\Phi _{\beta }\right\}
\end{equation*}%
be the matrix of Poisson brackets of constraints modulo all constraints. If $%
rank(M_{\alpha \beta })=s$, then $ker(M_{\alpha \beta })$ is $\left(
k-s\right) -$dimensional. A basis for the kernel can be constructed from
linear combinations $\psi _{\alpha }$ of $\Phi _{\alpha }$ satisfying%
\begin{equation*}
\left\{ \psi _{\alpha },\psi _{\beta }\right\} \approx 0,\text{ \ }\alpha
,\beta =1,...,k-s
\end{equation*}%
and are called as the first class constraints. Note that the number of Lagrange
multipliers which can be solved is also determined by the matrix of all
constraints. Let $\chi _{\alpha }:\alpha =1,...,s$ be the second class
constraints whose Poisson brackets does not vanish (modulo constraints).
Define the $s\times s-$matrix%
\begin{equation}
C_{\alpha \beta }=\left\{ \chi _{\alpha },\chi _{\beta }\right\} ,\text{ \ }%
\alpha ,\beta =1,...,s\label{cmatrix}
\end{equation}%
which is invertible by construction. Define the Dirac bracket%
\begin{equation}
\{f,g\}_{DB}=\{f,g\}-\{f,\chi _{\alpha }\}(C^{-1})^{\alpha \beta }\{\chi
_{\beta },g\}  \label{diracbrac}
\end{equation}%
\cite{SuMu74}. Note that, since $\{f,\chi _{\alpha }\}_{DB}=0$ for arbitrary
function $f$, second class constraints can be set to zero either before or
after evaluation of Dirac bracket. The initial $2n$ dimensional Hamiltonian
system with $(k-s)$ number of first class and $s$ number of second class constraints reduces to  $%
2n-2(k-s)-s=2n-2k+s$ dimensional Hamiltonian system equipped with the Dirac
bracket and with the total Hamiltonian function. The final bracket
eliminates the second class constraints from the set of all constraints
leaving a complete set of first class constraints. First class constraints
form a closed local symmetry algebra for the system. Computing%
\begin{equation*}
\left\{ \psi _{\alpha },H\right\} =c_{\alpha }^{\beta }\psi _{\beta },\text{
\ \ }\left\{ \psi _{\alpha },\psi _{\beta }\right\} =c_{\alpha \beta
}^{\gamma }\psi _{\gamma }
\end{equation*}%
one finds the structure constants of this algebra \cite{Dirac,Dirac1950}.

\section{Reduction of the second order Lagrangian theories to the first order ones} \label{red}

\subsection{Total reduction} \label{totred}

Given a second order Lagrangian function
\begin{equation} \label{sol}
L=L\left( \mathbf{q;\dot{q};%
\ddot{q}}\right)
\end{equation}
on the second order tangent bundle $T^{2}Q$, define the $3n$-dimensional configuration space $M=T^{2}Q$
with coordinates 
\begin{equation} \label{local}
\mathbf{q}_{\left( 1\right) }=\mathbf{q}\text{, \ \ }\mathbf{q}_{\left(
2\right) }=\mathbf{\dot{q}}\text{, \ \ }\mathbf{q}_{\left( 3\right) }=%
\mathbf{\ddot{q},}
\end{equation}%
\cite{GoRa94, NaHa96}. In order not to forget the constitutional relationships in the
coordinates of $M$, we impose the sets of constraints 
\begin{equation} \label{consM}
\mathbf{\dot{q}}%
_{\left( 1\right) }-\mathbf{q}_{\left( 2\right) }=0, \qquad \mathbf{\dot{q}}%
_{\left( 2\right) }-\mathbf{q}_{\left( 3\right) }=0
\end{equation}
on the tangent bundle $%
TM$ equipped with coordinates 
\begin{equation*}
\left( \mathbf{q}_{\left( 1\right) }\mathbf{,\mathbf{q}}_{\left( 2\right) }%
\mathbf{,q}_{\left( 3\right) };\mathbf{\dot{q}}_{\left( 1\right) }\mathbf{,%
\dot{q}}_{\left( 2\right) }\mathbf{,\dot{q}}_{\left( 3\right) }\right) \in
TM.
\end{equation*}%
We define the first order Lagrangian function%
\begin{equation}
L_{C}={L}\left( \mathbf{q}_{\left( 1\right) },\mathbf{q}_{(
2) },\mathbf{q}_{(3)}\right)+\mathbf{\lambda }\cdot \left( 
\mathbf{\dot{q}}_{\left( 1\right) }-\mathbf{q}_{\left( 2\right) }\right) +%
\mathbf{\beta \cdot }\left( \mathbf{\dot{q}}_{\left( 2\right) }-\mathbf{q}%
_{\left( 3\right) }\right) ,  \label{L-1}
\end{equation}%
where $\mathbf{\lambda }\in 
%TCIMACRO{\U{211d} }%
%BeginExpansion
\mathbb{R}
%EndExpansion
^{n}$ and $\mathbf{\beta }\in 
%TCIMACRO{\U{211d} }%
%BeginExpansion
\mathbb{R}
%EndExpansion
^{n}$ are Lagrange multipliers. We consider $L_{C}$ as a Lagrangian function
on the extented space $T\left( M\times 
%TCIMACRO{\U{211d} }%
%BeginExpansion
\mathbb{R}
%EndExpansion
^{n}\times 
%TCIMACRO{\U{211d} }%
%BeginExpansion
\mathbb{R}
%EndExpansion
^{n}\right) $ by letting $\mathbf{\lambda }$ and $\mathbf{\beta }$ be
variables as well. For the first order Lagrangian function $L_{C}$ in (\ref%
{L-1}), the first order Euler-Lagrange equations (\ref{EL}) obtained by variations of $\mathbf{q}_{\left( 1\right) },\mathbf{q}_{\left( 2\right) },\mathbf{q}_{\left( 3\right) },\mathbf{\lambda }$ and $\mathbf{\beta }$ are
\begin{equation} \label{EL5}
\frac{\partial {L}}{\partial \mathbf{q}_{\left(
1\right) }}-\mathbf{\dot{\lambda}}=\mathbf{0},  \qquad 
 \frac{\partial {L}}{\partial \mathbf{q}_{\left(
2\right) }}-\mathbf{\lambda }-\mathbf{\dot{\beta}}=\mathbf{0},  \qquad 
 \frac{\partial {L}}{\partial \mathbf{q}_{\left(
3\right) }}-\mathbf{\beta }=\mathbf{0},  \qquad 
\mathbf{\dot{q}}_{\left( 1\right) }-\mathbf{q}_{\left( 2\right) }=\mathbf{0,}  \qquad 
 \mathbf{\dot{q}%
}_{\left( 2\right) }-\mathbf{q}_{\left( 3\right) }=\mathbf{0.}  \notag
\end{equation}%
Here, the last two equations are the constraints presented in (\ref{consM}) whereas the second and third
equations define the Lagrange multipliers 
\begin{equation}
\mathbf{\lambda }= \frac{\partial {L}}{\partial \mathbf{q}_{\left( 2\right) }}%
-\frac{d}{dt}\frac{\partial {L}}{\partial \mathbf{q}_{\left( 3\right) }}%
,\qquad \mathbf{\beta }=\frac{\partial {L}}{\partial \mathbf{q}%
_{\left( 3\right) }}.  \label{constraints}
\end{equation}%
This shows that, the Lagrange multipliers $\mathbf{\lambda }$ and $\mathbf{%
\beta }$ cannot be indepedent variables, instead they are determined by the
Lagrangian function $L$. By substituting the definitions of $\mathbf{%
\lambda }$ and $\mathbf{\beta }$ into the constraint Lagrangian $L_{C}$ in (%
\ref{L-1}) we arrive at a Lagrangian function 
$$L_{U}={L}\left( \mathbf{q}_{\left( 1\right) },\mathbf{q}_{(
2) },\mathbf{q}_{(3)}\right)+\left( \frac{\partial {L}}{\partial \mathbf{q}_{\left( 2\right) }}%
-\frac{d}{dt}\frac{\partial {L}}{\partial \mathbf{q}_{\left( 3\right) }}\right)\cdot \left( 
\mathbf{\dot{q}}_{\left( 1\right) }-\mathbf{q}_{\left( 2\right) }\right) +%
\left(\frac{\partial {L}}{\partial \mathbf{q}%
_{\left( 3\right) }}\right)\cdot \left( \mathbf{\dot{q}}_{\left( 2\right) }-\mathbf{q}%
_{\left( 3\right) }\right) $$ 
defined precisely on
the tangent bundle $TM$. We shall call $L_{U}$ as unconstraint Lagrangian
function. Note that, by pulling the constraints in (\ref{constraints}) back
to $T^{2}Q$ we arrive at that 
\begin{equation}
\mathbf{\lambda }=\frac{\partial {L}}{\partial \mathbf{\dot{q}}}-\frac{d}{dt}%
\frac{\partial {L}}{\partial \mathbf{\ddot{q}}}\text{ \ \ and \ \ }\mathbf{%
\beta }=\frac{\partial {L}}{\partial \mathbf{\ddot{q}}}.  \label{cons2}
\end{equation}%
It is a matter of direct
calculation to show that the first order Euler-Lagrange dynamics generated by
both the constraint $L_{C}$ and the unconstraint $L_{U}$ Lagrangian functions
coincide with the second order Euler-Lagrange equation (\ref{EL2}).

For the Hamilton formalism, from the definitions of conjugate momenta for the first order Lagrangian (\ref{L-1}), the primary constraints  are defined as $$\boldsymbol{\Phi}^{(1)}=\mathbf{p}^{(1)}-\boldsymbol{\lambda},\quad \boldsymbol{\Phi}^{(2)}=\mathbf{p}^{(2)}-\boldsymbol{\beta},\quad \boldsymbol{\Phi}^{(3)}=\mathbf{p}^{(3)},\quad \boldsymbol{\Phi}^{(\lambda)}=\mathbf{p}^{(\lambda)},\quad \boldsymbol{\Phi}^{(\beta)}=\mathbf{p}^{(\beta)}.$$
In this case, the total Hamiltonian function becomes
\begin{align}
H_T&=H+\boldsymbol{\Phi}^{(1)}\cdot\mathbf{u}_{(1)}+\boldsymbol{\Phi}^{(2)}\cdot\mathbf{u}_{(2)}+\boldsymbol{\Phi}^{(3)}\cdot\mathbf{u}_{(3)}+\boldsymbol{\Phi}^{(\lambda)}\cdot\mathbf{u}_{\lambda}+\boldsymbol{\Phi}^{(\beta)}\cdot\mathbf{u}_{\beta}
\end{align}
where the canonical Hamiltonian is
$$H=-{L}\left( \mathbf{q}_{\left( 1\right) },\mathbf{q}_{(
2) },\mathbf{q}_{(3)}\right)+\boldsymbol{\lambda}\cdot\mathbf{q}_{(2)}+\boldsymbol{\beta}\cdot\mathbf{q}_{(3)}.$$
Note that, the Hamilton function for the present case is only depending on $ \mathbf{q}_{\left( 1\right) },\mathbf{q}_{(
2) },\mathbf{q}_{(3)},\boldsymbol{\lambda}$ and $\boldsymbol{\beta}$.

\subsection{Velocity as a new coordinate}
\subsubsection{Partial reduction I} \label{parred1}

Define a $2n$-dimensional configuration space $N=TQ$ with
coordinates 
\begin{equation}
\mathbf{q}_{\left( 1\right) }=\mathbf{q}\text{, \ \ }\mathbf{q}_{\left(
2\right) }=\mathbf{\dot{q}}\text{.}  \label{TN-iden}
\end{equation}%
We are imposing the set
of constraints $\mathbf{\dot{q}}_{\left( 1\right) }-\mathbf{q}_{\left(
2\right) }=0$ in the local frame
\begin{equation*}
\left( \mathbf{q}_{\left( 1\right) }\mathbf{,\mathbf{q}}_{\left( 2\right) };%
\mathbf{\dot{q}}_{\left( 1\right) }\mathbf{,\dot{q}}_{\left( 2\right)
}\right) \in TN.
\end{equation*}%
$TN$ can be identified with $4n$-dimensional iterated tangent bundle $TTQ$.
An element of $TN=TTQ$ satisfying the constraint $%
\mathbf{\dot{q}}_{\left( 1\right) }-\mathbf{q}_{\left( 2\right) }=0$ is
called a second order vector field \cite{AbMa78}. Compare this with the embedding in (\ref{T2Q-}). Define a Lagrangian function%
\begin{equation} \label{LC1}
L_{C_{1}}={L}\left( \mathbf{q}_{\left( 1\right) }\mathbf{,\mathbf{q}}%
_{\left( 2\right) }\mathbf{,\mathbf{\dot{q}}}_{\left( 2\right) }\right) +%
\boldsymbol{\lambda }_1\cdot \left( \mathbf{\dot{q}}_{\left( 1\right) }-\mathbf{q}%
_{\left( 2\right) }\right)  
\end{equation}%
on the space $T(TQ\times 
\mathbb{R}
^{n})$ with $\boldsymbol{\lambda }_1$ being Lagrange multipliers. The first order
Euler-Lagrange equations (\ref{EL}) for $L_{C_1}$ turn out to be 
\begin{equation}
\frac{\partial {L}}{\partial \mathbf{q}_{\left( 1\right) }}-\boldsymbol{\dot{%
\lambda}}_1=\mathbf{0},\qquad \frac{\partial {L}}{\partial \mathbf{q}
_{\left( 2\right) }}-\frac{d}{dt}\frac{\partial {L}}{\partial \mathbf{\dot{q}%
}_{\left( 2\right) }}=\boldsymbol{\lambda }_1, \qquad \mathbf{\dot{q}}_{\left(
1\right) }-\mathbf{q}_{\left( 2\right) }=\mathbf{0}.  \label{Red1}
\end{equation}%
A direct calculation proves that the
system (\ref{Red1}) is equivalent to the second order Euler-Lagrange equations (\ref
{EL2}).

Let $T^*(TQ\times \mathbb{R}^n)$ be endowed with coordinates $(\mathbf{q}_{(1)},\mathbf{q}_{(2)},\boldsymbol{\lambda}_1;\mathbf{p}^{(1)},\mathbf{p}^{(2)},\mathbf{p}^{\lambda_1})$ and define the conjugate momenta as fiber derivatives of $L_{C_1}$. This results in a set of primary  constraints
\begin{equation} \label{cons1} 
\boldsymbol{\Phi}^{(1)}=\mathbf{p}^{(1)}-\boldsymbol{\lambda}_{1}\approx \mathbf{0}, \qquad \boldsymbol{\Phi}^{(2)}=\mathbf{p}^{(2)}-\frac{\partial L}{\partial \mathbf{\dot{q}}_{2}}\approx \mathbf{0}, \qquad \boldsymbol{\Phi}^{(\lambda_1)}=\mathbf{p}^{\lambda_1}\approx \mathbf{0}.
\end{equation}
In this case, the canonical Hamiltonian is computed to be
\begin{equation}
H_{C_1}=\mathbf{p}^{(i)}\cdot \mathbf{\dot{q}}_{(i)}+\boldsymbol{\dot{\lambda}}_1\cdot \mathbf{p^{\lambda_1}}- L_{C_1}=\mathbf{p}^{(i)}\cdot \mathbf{\dot{q}}_{(i)}-L,\label{CanHam1}
\end{equation}
where $i$ runs from $1$ to $2$. We further define the total Hamiltonian by adding the constraints in (\ref{cons1}) as follows
\begin{equation}
H_{T_1}=H_{C_1}+\mathbf{u}_{(\alpha)}\cdot \Phi^{(\alpha)}\label{TotHam1}
\end{equation}
where $\alpha$ takes values $\{1,2,{\boldsymbol{\lambda}_1}\}$ and $\mathbf{u}_{(\alpha)}$ s are the Lagrange multipliers to be defined later. The consistency checks for two sets of  constraints under the Hamiltonian dynamics generated by $H_{T_1}$ result in
\begin{equation}
\{H_{T_1},\boldsymbol{\Phi}^{(1)}\}=-\frac{\partial H_{C_1}}{\partial \mathbf{q}_{(1)}}-\mathbf{u}_{(\lambda_1)}\approx 0, \qquad \{H_{T_1},\boldsymbol{\Phi}^{(\lambda_1)}\}=\mathbf{u}_{(1)}\approx 0.
\end{equation}
These equations determine $\mathbf{u}_{(\lambda_1)}$ and $\mathbf{u}_{(1)}$ explicitly. To determine the Lagrange multiplier $\mathbf{u}_{(2)}$, one addresses the functional structure of the Lagrangian $L$ because to single out $\mathbf{u}_{(2)}$ from consistency check of the constraint $\boldsymbol{\Phi}^{(2)}$ requires some non-degeneracy conditions on $L$. 

From the Euler-Lagrange equations (\ref{Red1}), Lagrange multiplier $\boldsymbol{\lambda }_1$ can be solved in terms of the Lagrangian function $L$. Substituting this definition of $\boldsymbol{\lambda }_1$ into $L_{C_1}$, we obtain an unconstraint Lagrangian function
\begin{equation} \label{LU1}
L_{U_1}= {L}\left( \mathbf{q}_{\left( 1\right) },\mathbf{q}%
_{\left( 2\right) },\mathbf{\dot{q}}_{\left( 2\right) }\right) + \left (
\frac{\partial {L}}{\partial \mathbf{q}%
_{\left( 2\right) }}-\frac{d}{dt}\frac{\partial {L}}{\partial \mathbf{\dot{q}%
}_{\left( 2\right) }} \right ) \cdot \left( \mathbf{\dot{q}}_{\left( 1\right) }-\mathbf{q}%
_{\left( 2\right) }\right).
\end{equation}
In general, $L_{U_1}$ is of the second order,  since it involves the total time derivative of $ \partial {L} / \partial \mathbf{\dot{q}%
}_{\left( 2\right) }$ which may result with an appearance of the acceleration terms. For a non-degenerate Lagrangian function $L$, $L_{U_1}$ is certainly a second order Lagrangian involving the term $\mathbf{\ddot{q}}_{(2)}$. On the other hand, if Lagrangian function is totally degenerate, that is, if $ \partial {L} / \partial \mathbf{\dot{q}%
}_{\left( 2\right) }$ does not depend on $\mathbf{\dot{q}%
}_{\left( 2\right) }$, then the term $ \partial {L} / \partial \mathbf{\dot{q}%
}_{\left( 2\right) }$ may only depend on $(\mathbf{q}_{\left( 1\right) },\mathbf{q}%
_{\left( 2\right) })$. In this case, the total time derivative will only give the first order terms and $L_{U_1}$ turns out to be in the first order form. There still remains some other options on the degree of the unconstraint Lagrangian function from these two extreme cases. All these are related with the degeneracy level of $L_{U_1}$. In the forthcoming sections, we shall explicitly present some of these cases on concrete examples. It is needless to say that the dynamics generated by the unconstraint Lagrangian $L_{U_1}$ is equivalent to the second order Euler-Lagrange equations (\ref
{EL2}) independent of its degree.

\subsubsection{Partial reduction II} \label{parred2}

It is interesting to note a tricky point in the definition of $L_{C_1}$. In the previous subsection, we have preferred to substitute the velocity component $%
\mathbf{\dot{q}}$ in $L=L\left( \mathbf{q;\dot{q};%
\ddot{q}}\right)$ by $\mathbf{q}_{\left( 2\right) }$  in (\ref{LC1}). An alternative way is to take $\mathbf{\dot{q}}$ as $%
\mathbf{\mathbf{\dot{q}}}_{\left( 1\right) }$. This leads us to another Lagrangian function
\begin{equation}
L_{C_{2}}={L}\left( \mathbf{q}_{\left( 1\right) }\mathbf{,\mathbf{\dot{q}}}%
_{\left( 1\right) }\mathbf{,\mathbf{\dot{q}}}_{\left( 2\right) }\right) +%
\boldsymbol{\lambda}_2 \cdot \left( \mathbf{\dot{q}}_{\left( 1\right) }-\mathbf{q}%
_{\left( 2\right) }\right),\label{LC2}
\end{equation}
where $\boldsymbol{\lambda}_2$ being a new set of Lagrange multiplier. The Euler-Lagrange equations generated by $%
L_{C_{2}}$ are computed to be
\begin{equation}
\frac{\partial {L}}{\partial \mathbf{q}_{\left( 1\right) }}-\frac{d}{dt}%
\frac{\partial {L}}{\partial \mathbf{\dot{q}}_{\left( 1\right) }}-\mathbf{%
\dot{\boldsymbol{\lambda}}}_2=\mathbf{0}\text{, \ \ }-\frac{d}{dt}\frac{\partial {L}}{%
\partial \mathbf{\dot{q}}_{\left( 2\right) }}=\boldsymbol{\lambda}_2\text{, \ \ }%
\mathbf{\dot{q}}_{\left( 1\right) }-\mathbf{q}_{\left( 2\right) }=\mathbf{0}%
\text{.}  \label{Red2}
\end{equation}%
It is immediate to observe that this system is equivalent to the second order Euler-Lagrange equations (\ref{EL2}). 
Even though the Euler-Lagrange equations generated by both $L_{C_{1}}$
and $L_{C_{2}}$ will eventually be the same, the functional structures of the Lagrange multipliers and the unconstraint formalisms will not be the same. The unconstraint Lagrangian for this case is 
\begin{equation}
L_{U_{2}}={L}\left( \mathbf{q}_{\left( 1\right) }\mathbf{,\mathbf{\dot{q}}}%
_{\left( 1\right) }\mathbf{,\mathbf{\dot{q}}}_{\left( 2\right) }\right) -\frac{d}{dt}\frac{\partial {L}}{%
\partial \mathbf{\dot{q}}_{\left( 2\right) }} \cdot \left( \mathbf{\dot{q}}_{\left( 1\right) }-\mathbf{q}%
_{\left( 2\right) }\right).\label{LU2}
\end{equation}

Two different reductions, namely $L_{C_{1}}$ and $L_{C_{2}}$, of the second order Lagrangian $L$ is a manifestation of the double vector bundle structure (\ref{tr}) of the
iterated tangent bundle $TTQ$ over the base manifold $TQ$. The first
constraint function $L_{C_{1}}$ is the result of the fibration $\tau _{TQ}$
whereas the second one is the result of the fibration $T\tau _{Q}$.  Locally, if we choose a local coordinate chart $( \mathbf{q}_{\left( 1\right) },\mathbf{q}_{\left( 2\right) };\mathbf{\dot{q}}_{\left( 1\right) }\mathbf{,%
\dot{q}}_{\left( 2\right) })$ on $TTQ$,
 the projections are defined by the two-tuples $( \mathbf{q}_{\left( 1\right) },\mathbf{q}_{\left( 2\right) })$ and $( \mathbf{q}_{\left( 1\right) },\mathbf{\dot{q}}_{\left( 1\right) })$, respectively. 
As a result, we may say that recasting a second order Lagrangian as a first order function geometrically corresponds to deciding the base manifold of the iterated bundle $TTQ$. In the following subsection, we shall present another way to define a submanifold of $TTQ$ and the related first order Lagrangian dynamics.

\subsection{Acceleration as a new coordinate} \label{acce}

The main understanding of the previous section is to call the velocity as a new variable. This is not the only way to reduce a second order Lagrangian function to a first order one. In this section, we shall call the acceleration as a new variable. 

\subsubsection{Schmidt's method}
\label{schmidt}

Let us start this by the following construction, called as Schmidt's method \cite{Sc94,Sc95}. We refer \cite{AnGoMaMa10,AnGoMa07,EsGu18} for discussions on the relationship between the methods of Ostrogradski and Schmidt in pure geometrical terms. Recall acceleration bundle $AQ$ defined in \eqref{AQ} and, in the light of the local coordinates in (\ref{local}), define a local chart $(\mathbf{q}_{\left( 1\right) },\mathbf{q}_{(3)})$ for $AQ$ consisting of position $\mathbf{q}_{\left( 1\right) }$ and acceleration $\mathbf{q}_{\left( 3\right) }$.
The induced coordinates on the tangent bundle $TAQ$ of $AQ$ are  
\begin{equation*}
(\mathbf{q}_{\left( 1\right) },\mathbf{q}_{\left( 3\right) },\mathbf{\dot{q}}_{\left( 1\right) },\mathbf{\dot{q}}_{\left( 3\right) } )\in TAQ.
\end{equation*}

Start with a second order Lagrangian function $L$. In order to link the acceleration $\mathbf{q}_{(3)}$ with the
derivative of the velocity $\mathbf{\dot{q}}_{(1)}$, introduce a
trivial bundle structure $T (AQ\times R)$ over the tangent bundle $T AQ$. Here, $R$ is
an $n-$dimensional manifold with local coordinates $(\mathbf{r})$. Define a first order Lagrangian function  
\begin{equation}
L_{3}\left( \mathbf{q}_{( 1)},\mathbf{q}_{(3)},\mathbf{\dot{q}}_{( 1)},\mathbf{\dot{q}}_{(3)}, \mathbf{r},\dot{\mathbf{r}}\right) ={L}\left( \mathbf{q}_{( 1)},\mathbf{\dot{q}}_{( 1)},\mathbf{q}_{(3)}\right) +\frac{\partial F}{\partial\mathbf{q}_{(1)}}\cdot\mathbf{%
\dot{q}}_{(1)}+\frac{\partial F}{\partial\mathbf{%
\dot{q}}_{(1)}}\cdot\mathbf{%
\mathbf{q}}_{(3)}+\frac{\partial F}{\partial \mathbf{q}_{(3)}}\cdot\mathbf{
\dot{q}}_{(3)}+\frac{\partial F}{\partial\mathbf{\mathbf{r}}}\cdot\mathbf{%
\dot{r}}  \label{L3}
\end{equation}
on the total space $T (AQ\times R)$. Here, $F$
is an arbitrary function depending on $(\mathbf{q}_{( 1)},\mathbf{q}_{(3)},\mathbf{\dot{q}}_{( 1)}, \mathbf{r})$. A direct calculation proves that the first order Euler-Lagrange equations generated by $L_{3}$ on $T(AQ\times M)$ is equivalent to the second order Euler-Lagrange equations (\ref{EL3}) 
only if the matrix 
$
[\partial^{2}F/\partial\mathbf{\mathbf{\mathbf{\dot{q}}}_{(1)}\partial r}]$
is non-degenerate. In order to satisfy this condition, one may simply choose the auxiliary function as $F=\mathbf{\dot{q}}_{(1)} \cdot \mathbf{r}$. Let us proceed with this choice. In this particular case, the Lagrangian function $L_{3}$ in (\ref{L3}) becomes 
\begin{equation}
L_{3}\left( \mathbf{q}_{( 1)},\mathbf{q}_{(3)},\mathbf{\dot{q}}_{( 1)},\mathbf{\dot{q}}_{(3)}, \mathbf{r},\dot{\mathbf{r}}\right) ={L}\left( \mathbf{q}_{( 1)},\mathbf{\dot{q}}_{( 1)},\mathbf{q}_{(3)}\right) +\mathbf{r}\cdot\mathbf{%
\mathbf{q}}_{(3)}+\mathbf{\dot{q}}_{(1)}\cdot\mathbf{%
\dot{r}}.  \label{L3-2}
\end{equation}
It is immediate to see that the first order Euler-Lagrange equations generated by $L_3$ in (\ref{L3-2}) is equivalent to the second order Euler-Lagrange equations generated by $L$. 

Let us define the conjugate momentum coordinates on the cotangent bundle $T^*(AQ\times R)$ by three-tuple $(\mathbf{p}_{(1)}, \mathbf{p}_{(2)},\mathbf{p}_{(r)})$  which can be computed as
\begin{align} \label{mom-Sch} 
\mathbf{p}^{(1)} =\frac{\partial{L}}{\partial \mathbf{\dot{q}}_{(1)}}\left( \mathbf{q}_{( 1)},\mathbf{\dot{q}}_{( 1)},\mathbf{q}_{(3)}\right)+\mathbf{\dot{r}}, \qquad \mathbf{p}^{(3)}  = \mathbf{0} 
, \qquad 
\mathbf{p}^{(r)}  =\mathbf{\dot{q}}^{(1)}. 
\end{align}
It is evident that, we can solve $\mathbf{\dot{r}}$ in terms of $(\mathbf{\dot{q}}_{(1)},\mathbf{\dot{q}}_{(3)},\mathbf{p}^{(1)},\mathbf{p}^{(r)})$. See also that $ \varphi_1=\mathbf{p}^{(3)}\approx 0 $ is a primary constraint. Accordingly, we define the following total Hamiltonian function 
\begin{equation} \label{ham-Sch}
H_{T}=\mathbf{p}^{(1)} \cdot \mathbf{p}^{(r)} - L\left( \mathbf{q}_{( 1)},\mathbf{p}^{(r)},\mathbf{q}_{(3)}\right)-\mathbf{r}\cdot\mathbf{q}_{(3)}+\mathbf{u}\cdot\mathbf{p}^{(3)}, 
\end{equation}
where $\mathbf{u}$ being Lagrange multipliers. Check the consistency of the primary constraint as
\begin{equation}
\varphi_2=\left\{ \mathbf{p}^{(3)},H_{T}\right\} =\frac {\partial%
{L}}{\partial \mathbf{q}_{(3)}}+\mathbf{r}=\mathbf{0}.
\end{equation}
Consistency of the secondary constraint $\varphi_{2}$ results with a tertiary constraint
\begin{equation}
\varphi_3=\left\{ \varphi_{2},H_{T}\right\}=\left ( \mathbf{u}\cdot \frac{\partial}{\partial\mathbf{q}_{(3)}} +\mathbf{q}_{(3)}\cdot \frac{\partial}{\partial \mathbf{p}^{(r)}}+\mathbf{p}^{(r)}\cdot\frac{\partial }{\partial \mathbf{q}_{(1)}} \right) \left( \frac{\partial{L}}{\partial\mathbf{q}_{(3)}}  \right) + \mathbf{p}^{(r)}=0.
\end{equation}
If the Lagrangian is non-degenerate in the sense of Ostrogradski then this step determines
the Lagrange multipliers $\mathbf{u}$, and the constraint
algorithm is finished up. If the Lagrangian is degenerate further steps may
be needed to determine the Lagrange multipliers as well as to close up the
Poisson algebra. 

\subsubsection{Deriglazov's trick} 
\label{alexei}

Now, we apply a trick due to Deriglazov  to reduce the second order Lagrangian function $L=L(\mathbf{q,\dot{q},\ddot{q}})$ to a first order one by following \cite{de2017}. We introduce the action integral
\begin{equation}
\int  \left( L\left( \mathbf{q}_{( 1)},\mathbf{\dot{q}}_{( 1)},\mathbf{q}_{(3)}\right) +\boldsymbol{\gamma}\cdot (\mathbf{\ddot{q}}_{(1)}- \mathbf{%
\mathbf{q}}_{(3)})\right)  dt, 
\end{equation} 
where $\boldsymbol{\gamma}$ is a set of Lagrange multipliers. 
Applying the by-parts technique to the second term inside the integral, we arrive at the following reduced Lagrangian function
\begin{equation}
{L_4}=L\left( \mathbf{q}_{( 1)},\mathbf{\dot{q}}_{( 1)},\mathbf{q}_{(3)}\right) -\boldsymbol{\dot{\gamma}}\cdot \mathbf{\dot{q}}_{(1)} -
\boldsymbol{\gamma}\cdot \mathbf{q}_{(3)}  \label{AlexeiLag}
\end{equation} 
on the extended velocity phase space $T(AQ\times R)$ with coordinates $(\mathbf{q}_{(1)},\mathbf{q}_{(3)},\boldsymbol{\gamma};\mathbf{\dot{q}}_{(1)},\mathbf{\dot{q}}_{(3)},\dot{\boldsymbol{\gamma}})$. It is immediate to observe that the first order Euler-Lagrange equations generated by $L_4$ is equivalent to the second order Euler-Lagrange equations (\ref{EL2}) generated by $L$. 

The Schmidt's method and Deriglazov's trick are the same in the particular cases we are interested in, that is degenerate second order Lagrangian functions. To see this, compare the Lagrangian functions $L_3$ and $L_4$ presented in (\ref{L3-2}) and (\ref{AlexeiLag}). These two Lagrangian functions are the same if a simple identification $\boldsymbol{\gamma}=-\mathbf{r}$ is performed. This results with the identification $\dot{\boldsymbol{\gamma}}=-\mathbf{\dot{r}}$ as well.  

\section{Cl\`{e}ment Lagrangian \label{Clement-Lag}}

Let us recall here degenerate second order Cl\`{e}ment's Lagrangian function 
\begin{equation}
L^{C}[\mathbf{x}]=-\frac{m}{2}\zeta ||\mathbf{\dot{x}}||^{2}-\frac{2m\Lambda }{%
\zeta }+\frac{\zeta ^{2}}{2\mu m}\mathbf{x}\cdot (\mathbf{\dot{x}}\times 
\mathbf{\ddot{x}})  \label{LC}
\end{equation}%
on the second order tangent bundle $T^{2}Q$ with local coordinates $[\mathbf{%
x}]=(\mathbf{x,\dot{x},\ddot{x})}$. Here, the inner product $||\mathbf{x}%
||^{2}=T^{2}-x^{2}-y^{2}$ is defined by the Lorentzian metric and the triple
product is $\mathbf{x}\cdot (\mathbf{\dot{x}}\times \mathbf{\ddot{x}}%
)=\epsilon _{ijk}x^{i}\dot{x}^{j}\ddot{x}^{k}$ where $\epsilon _{ijk}$ is
the completely antisymmetric tensor of rank three. Dot denotes the
derivative with respect to the variable $t$ and $\zeta =\zeta (t)$ is a
function which allows arbitrary reparametrization of the variable $t$. $%
\Lambda $ and $1/2m$ are cosmological and Einstein gravitational constants,
respectively.

The variation of Cl\`{e}ment Lagrangian (\ref{LC}) with respect to $\zeta $
gives the energy constraint%
\begin{equation}
E^{C}[\mathbf{x}]=-\frac{m}{2}||\mathbf{\dot{x}}||^{2}+2\frac{m\Lambda }{\zeta
^{2}}+\frac{\zeta }{m\mu }\mathbf{x}\cdot (\mathbf{\dot{x}}\times \mathbf{%
\ddot{x}})=0.  \label{lcon}
\end{equation}%
whereas the variation of the Lagrangian (\ref{LC}) with respect to $\mathbf{x%
}$ results with the Euler-Lagrange equations 
\begin{equation}
2m^{2}\mu \mathbf{\ddot{x}}+3\mathbf{\dot{x}}\times \mathbf{\ddot{x}}+2%
\mathbf{x}\times \mathbf{x}^{(3)}=\mathbf{0}  \label{clee}
\end{equation}
which is a third order differential equation. 
In the Euler-Lagrange equations (\ref{clee}), we set the reparametrization
function $\zeta $ equal to one. The Cl\`{e}ment Lagrangian (\ref{LC}) is
invariant under translations in $t$ and pseudo-rotations in space. Time
translation symmetry gives the conservation of energy.

In the following subsections, we shall apply four reductions methods, namely, the total reduction, partial reduction I, partial reduction II, and Deriglazov's trick / Schmidt's method, presented in the previous section to the case of the Cl\`{e}ment Lagrangian (\ref{LC}). Then we shall obtain their Hamiltonian realizations by employing the Dirac-Bergmann constraint algorithm. For each of the methods, we shall present the Dirac brackets. In addition, we shall exhibit the unconstraint Lagrangian realizations for the case of partial reductions and write the associated Hamiltonian formalisms.

\subsection{Total Reduction}
Recall that, in (\ref{totred}), we have presented the total reduction of a second order Lagragian function. For the case of the Cl\`{e}ment Lagrangian in (\ref{LC}), this reads the following first order Lagrangian function 
\begin{eqnarray} \label{totclemlag}
L_C=-\frac{m\zeta}{2} ||{\bf{q}}_{(2)} ||^2+\frac{\zeta^2}{2\mu m}{\bf{q}}_{(1)}\cdot{\bf{q}_{(2)}}\times {\bf{q}}_{(3)}+\boldsymbol{\lambda}_1\cdot(\dot{\bf{q}}_{(1)}-{\bf{q}}_{(2)}
)+\boldsymbol{\lambda}_2\cdot(\dot{\bf{q}}_{(2)}-{\bf{q}}_{(3)})
\end{eqnarray}
with coordinates $\mathbf{q}_{\left( 1\right) }=\mathbf{x}$, $\mathbf{q}_{\left(
2\right) }=\mathbf{\dot{x}}$, and $\mathbf{q}_{\left( 3\right) }=%
\mathbf{\ddot{x}}$. Here, $\boldsymbol{\lambda}_1$ and $\boldsymbol{\lambda}_2$ are Lagrange multipliers. Variation of $L_C$ with respect to $\mathbf{q}_{\left( 1\right) },\mathbf{q}_{\left( 2\right) }$ and $\mathbf{q}_{\left( 3\right) }$  results in the Euler-Lagrange equations 
\begin{align}
\frac{\zeta^2}{2\mu m}\mathbf{q}_{\left( 2\right) }\times\mathbf{q}_{\left( 3\right) }-\boldsymbol{\dot \lambda}_1=\mathbf{0},\quad -m\zeta \mathbf{q}_{\left( 2\right) }+ \frac{\zeta^2}{2\mu m}\mathbf{q}_{\left( 3\right) }\times\mathbf{q}_{\left( 1\right) }-\boldsymbol{ \lambda}_1-\boldsymbol{\dot \lambda}_2=\mathbf{0} ,\quad \frac{\zeta^2}{2\mu m}\mathbf{q}_{\left( 1\right) }\times\mathbf{q}_{\left( 2\right) }-\boldsymbol{ \lambda}_2=\mathbf{0}.
\end{align} 
We introduce the conjugate momenta $({\bf{p}}^{(1)},{\bf{p}}^{(2)},{\bf{p}}^{(3)},{\bf{p}}^{(\lambda_1)},{\bf{p}}^{(\lambda_2)})$. The Legendre transformation results with the following identities 
\begin{eqnarray}
{\bf{p}}^{(1)}=\frac{\partial L_C}{\partial\dot{\mathbf{q}}_{(1)}}=\boldsymbol{\lambda}_1,~~{\bf{p}}^{(2)}=\frac{\partial L_C}{\partial\dot{\mathbf{q}}_{(2)}}=\boldsymbol{\lambda}_2,~~{\bf{p}}^{(3)}=\frac{\partial L_C}{\partial\dot{\mathbf{q}}_{(3)}}=0,~~{\bf{p}}^{(\lambda_1)}=\frac{\partial L_C}{\partial\dot{\boldsymbol{\lambda}}_{(1)}}=0,~~{\bf{p}}^{(\lambda_2)}=\frac{\partial L_C}{\partial\dot{\boldsymbol{\lambda}}_{(2)}}=\textbf{0}.\label{PC}
\end{eqnarray}
Definition of momenta in (\ref{PC}) imply the set of primary constraints  
\begin{equation}
{\bf\Phi}^{(1)}={\bf{p}}^{(1)}-\boldsymbol{\lambda}_1,~~{\bf\Phi}^{(2)}={\bf{p}}^{(2)}-\boldsymbol{\lambda}_2,~~{\bf\Phi}^{(3)}={\bf{p}}^{(3)},~~{\bf\Phi}^{(\lambda_1)}={\bf{p}}^{(\lambda_1)},~~{\bf\Phi}^{(\lambda_2)}={\bf{p}}^{(\lambda_2)}.
\end{equation}
Let us now introduce the canonical Hamiltonian function
\begin{eqnarray}
H&=&{\bf{p}}^{(1)}\cdot{\dot{\mathbf{q}}_{(1)}}+{\bf{p}}^{(2)}\cdot{\dot{\mathbf{q}}_{(2)}}+{\bf{p}}^{(3)}\cdot{\dot{\mathbf{q}}_{(3)}}+{\bf{p}}^{(\lambda_1)}\cdot{\dot{\boldsymbol{\lambda}}_{(1)}}+{\bf{p}}^{(\lambda_2)}\cdot{\dot{\boldsymbol{\lambda}}_{(2)}}-L_C\notag\\
&=&{\bf{p}}^{(1)}\cdot{\dot{\mathbf{q}}_{(1)}}+{\bf{p}}^{(2)}\cdot{\dot{\mathbf{q}}_{(2)}}+{\bf{p}}^{(3)}\cdot{\dot{\mathbf{q}}_{(3)}}+{\bf{p}}^{(\lambda_1)}\cdot{\dot{\boldsymbol{\lambda}}_{(1)}}+{\bf{p}}^{(\lambda_2)}\cdot{\dot{\boldsymbol{\lambda}}_{(2)}}+\frac{m\zeta}{2}||{\bf{q}}_{(2)}||^2-\frac{\zeta^2}{2\mu m}{\bf{q}}_{(1)}\cdot{\bf{q}_{(2)}}\times {\bf{q}}_{(3)}\notag\\
&-&\boldsymbol{\lambda}_1\cdot(\dot{\bf{q}}_{(1)}-{\bf{q}}_{(2)}
)-\boldsymbol{\lambda}_2\cdot(\dot{\bf{q}}_{(2)}-{\bf{q}}_{(3)}).
\end{eqnarray}
After some algebraic manipulations we arrive at 
\begin{eqnarray}
H&=&({\bf{p}}^{(1)}-\boldsymbol{\lambda}_1)\cdot{\dot{\mathbf{q}}_{(1)}}+({\bf{p}}^{(2)}-\boldsymbol{\lambda}_2)\cdot{\dot{\mathbf{q}}_{(2)}}+{\bf{p}}^{(3)}\cdot{\dot{\mathbf{q}}_{(3)}}+{\bf{p}}^{(\lambda_1)}\cdot{\dot{\boldsymbol{\lambda}}_{(1)}}+{\bf{p}}^{(\lambda_2)}\cdot{\dot{\boldsymbol{\lambda}}_{(2)}}\notag\\
&+&\frac{m\zeta}{2}||{\bf{q}}_{(2)}||^2-\frac{\zeta^2}{2\mu m}{\bf{q}}_{(1)}\cdot{\bf{q}_{(2)}}\times {\bf{q}}_{(3)}+\boldsymbol{\lambda}_1\cdot{\bf{q}}_{(2)}
+\boldsymbol{\lambda}_2\cdot{\bf{q}}_{(3)}.\label{HamforTotClem}
\end{eqnarray}
By substituting the primary constraints, we define the total Hamiltonian function as
\begin{equation}
H_T=H+\boldsymbol{\Phi}^{(1)}\cdot{\bf U}_1+\boldsymbol{\Phi^{(2)}}\cdot{\bf U}_2+{\bf\Phi}^{(3)}\cdot{\bf U}_3+{{\bf\Phi}^{(\lambda_1)}\cdot{\bf U}_{\lambda_1}}+{\bf\Phi}^{(\lambda_2)}\cdot{\bf U}_{\lambda_2}\label{TotCle1}
\end{equation}
where \begin{align}
H=\frac{m\zeta}{2}||{\bf{q}}_{(2)}||^2-\frac{\zeta^2}{2\mu m}{\bf{q}}_{(1)}\cdot{\bf{q}_{(2)}}\times {\bf{q}}_{(3)}+\boldsymbol{\lambda}_1\cdot{\bf{q}}_{(2)}
+\boldsymbol{\lambda}_2\cdot{\bf{q}}_{(3)}\label{HamforTotClem}
\end{align} 
is the canonical Hamiltonian function, and ${\bf U}_1,{\bf U}_2,{\bf U}_3,{\bf U}_{\lambda_1}, {\bf U}_{\lambda_2}$ are the Lagrange multipliers. Here are the steps of the Dirac-Bergmann Constraint algorithm and the computation of the total Hamiltonian function.

\bigskip

\noindent \textbf{Dirac-Bergmann constraint algorithm step 1:}  Consistency checks of the primary constraints ${\bf\Phi}^{(1)}$, ${\bf\Phi}^{(2)}$, ${\bf\Phi}^{(\lambda_1)}$ and ${\bf\Phi}^{(\lambda_2)}$ result with
\begin{equation}
\begin{split}
\dot{{\bf\Phi}}^{(1)}&=\{\boldsymbol{\Phi}^{(1)},H_{T}\} \approx \frac{\zeta^2}{2\mu m}{\bf{q}}_{(2)}\times {\bf{q}}_{(3)}-{\bf U}_{\lambda_1},
\\
 \dot{{\bf\Phi}}^{(2)}&=\{\boldsymbol{\Phi}^{(2)},H_{T}\}\approx -m\zeta {\bf{q}}_{(2)}+\frac{\zeta^2}{2\mu m}{\bf{q}}_{(3)}\times {\bf{q}}_{(1)}-\boldsymbol{\lambda}_1-{\bf U}_{\lambda_2}
\\
\dot{{\bf\Phi}}^{(\lambda_1)}&=\{\boldsymbol{\Phi}^{(\lambda_1)},H_{T}\} \approx {\bf U}_1-{\bf{q}}_{(2)},
\\
 \dot{{\bf\Phi}}^{(\lambda_2)}&=\{\boldsymbol{\Phi}^{(\lambda_2)},H_{T}\}\approx {\bf U}_2-{\bf{q}}_{(3)}
 \end{split}
\end{equation}
which lead to determine ${\bf U}_1,{\bf U}_2,{\bf U}_{\lambda_1},{\bf U}_{\lambda_2}$. Consistency checks of the primary constraints ${\bf\Phi}^{(3)}$ determine a set of secondary constraints 
\begin{equation} \label{Cl-1-2}
\boldsymbol{\Phi}=\dot{\boldsymbol{\Phi}}^{(3)}=\{\boldsymbol{\Phi}^{(3)},H_{T}\}\approx \frac{\zeta^2}{2\mu m}{\bf{q}}_{(1)}\times {\bf{q}}_{(2)}-\boldsymbol{\lambda}_2.
\end{equation}
Accordingly, we revise the total Hamiltonian $H_T$ by adding the secondary constraint $\boldsymbol{\Phi}$ with a set $\mathbf{U}$ of Lagrange multipliers. This reads $H_T^1=H_T+\mathbf{U}\cdot \boldsymbol{\Phi}$.
\bigskip

\noindent \textbf{Dirac-Bergmann constraint algorithm step 2:}   Consistency checks of the secondary constraints ${\bf\Phi}$ determine another set of constraints 
\begin{equation}
\boldsymbol{\varphi}=\dot{{\boldsymbol \Phi}}=\{\boldsymbol{\Phi},H_{T}^1\} \approx m\zeta{\bf{q}}_{(2)}+\frac{\zeta^2}{\mu m}{\bf{q}}_{(1)}\times {\bf{q}}_{(3)}+\boldsymbol{\lambda}_1.
\end{equation} 
We revise the total Hamiltonian function as $H_T^2=H^1_T+\mathbf{V}\cdot \boldsymbol{\varphi}$ where $\mathbf{V}$ being a set of Lagrange multipliers.

\bigskip

\noindent \textbf{Dirac-Bergmann constraint algorithm step 3:}
The conservation of $\boldsymbol{\varphi}$ gives
\begin{eqnarray}
\dot{{\boldsymbol{\varphi} }}=\{\boldsymbol{\varphi},H_{T}^2\}\approx m\zeta{\bf{q}}_{(3)}+\frac{3\zeta^2}{2\mu m}{\bf{q}}_{(2)}\times {\bf{q}}_{(3)}+\frac{\zeta^2}{\mu m}{\bf{q}}_{(1)}\times {\bf{U}}_{(3)}.
\end{eqnarray}
From this, we can determine two components of ${\bf{U}}_{(3)}$ while we arrive at a scalar constraint by simply taking the dot product of $\dot{{\boldsymbol{\varphi} }}$ with ${\bf{q}}_{(1)}$, that is 
\begin{equation}
\chi={\bf{q}}_{(1)}\cdot(m\zeta{\bf{q}}_{(3)}+\frac{3\zeta^2}{2\mu m}{\bf{q}}_{(2)}\times {\bf{q}}_{(3)}).
\end{equation} 
We arrive at the total Hamiltonian function $H^3_T=H^2_T+W\chi$. Here,  $W$ is a Lagrange multiplier.

\bigskip

\noindent \textbf{Dirac-Bergmann constraint algorithm step 4:} At final, the conservation of $\chi$ gives the third component of ${\bf{U}}_{(3)}$ and the remaining constraints determine the other Lagrange multipliers. A direct calculation determines the Lagrange multiplier $\bf U,V,W $ and ${\bf{U}}_{(3)}$ as
\begin{equation}
\begin{split}
{\bf{U}} &\approx {\bf{0}},\quad {\bf{V}}\approx {\bf{0}},\quad {\bf{W}}\approx {\bf{0}},\\
{\bf{U}}_{(3)} &\approx {\bf{q}}_{(1)}(\frac{{\bf{q}}_{(2)}\cdot {\bf{q}}_{(3)}}{||{\bf{q}}_{(1)}||^2})-\frac{\mu }{\zeta^3 ||{\bf{q}}_{(1)}||^2}[\frac{3\zeta^2}{2\mu m}{\bf{q}}_{(2)}\times {\bf{q}}_{(3)}+ m\zeta{\bf{q}}_{(3)}]\times [\frac{3\zeta^2}{2\mu m}{\bf{q}}_{(1)}\times {\bf{q}}_{(2)}+m\zeta{\bf{q}}_{(1)}]
\end{split}
\end{equation} 
under the assumption ${||{\bf{q}}_{(1)}}||^2\neq 0$. The following table summarizes the discussions have been done so far.
\begin{center}
\renewcommand{\arraystretch}{1.5}
\begin{tabular}{|c|c|}
\hline
&\textbf{Total Reduction}  \\ \hline
\textbf{Reduction} & $\mathbf{x}=\mathbf{q}_{(1)}$,~$\dot{\mathbf{x}}=%
\mathbf{q}_{(2)}$,~$\ddot{\mathbf{x}}=
\mathbf{q}_{(3)}$,  \\ \hline
\textbf{Coordinates} & $\mathbf{q}_{(1)},\mathbf{q}_{(2)},\mathbf{q}_{(3)},\boldsymbol{\lambda}_{1},\boldsymbol{\lambda}_{2},\mathbf{p}^{(1)},%
\mathbf{p}^{(2)},\mathbf{p}^{(3)},\mathbf{p}^{(\lambda_1)},\mathbf{p}^{(\lambda_2)}$ \\ \hline
\textbf{Primary} \textbf{Constraints}& $\begin{array}{cclccc}
\boldsymbol{\Phi}^{(1)}&=&\mathbf{p}^{(1)}-\boldsymbol{\lambda}_{1}&\boldsymbol{\Phi}^{(\lambda_1)}&=&\mathbf{p}^{(\lambda_1)}\\
\boldsymbol{\Phi}^{(2)}&=&\mathbf{p}^{(2)}-\boldsymbol{\lambda}_{2}&\boldsymbol{\Phi}^{(\lambda_2)}&=&\mathbf{p}^{(\lambda_2)}
\\
\boldsymbol{\Phi}^{(3)}&=&\mathbf{p}^{(3)}
\end{array}$ \\ \hline
\textbf{Secondary Constraints}& $\boldsymbol{\Phi}=\frac{\zeta^2}{2\mu m}\mathbf{q}_{(1)}\times\mathbf{q}_{(2)}-\boldsymbol{\lambda}_{2}$  \\ \hline
\textbf{Tertiary Constraints}&$\boldsymbol{\varphi}=m\zeta \mathbf{q}_{(2)}+\frac{\zeta^2}{\mu m}\mathbf{q}_{(1)}\times \mathbf{q}_{(3)}+\boldsymbol{\lambda}_{1}$ \\ \hline
\textbf{Quaternary Constraint}&$\boldsymbol{\chi}=\mathbf{q}_{(1)}\cdot(m\zeta\mathbf{q}_{(3)}+\frac{3\zeta^2}{2\mu m}\mathbf{q}_{(2)}\times \mathbf{q}_{(3)})$ \\  \hline
\end{tabular}\end{center}

\bigskip

\noindent \textbf{The total Hamiltonian function and the equations of motion:} Let us substitute the Lagrange multipliers ${\bf U}_1,{\bf U}_2,{\bf U}_3,{\bf U}_{\lambda_1}, {\bf U}_{\lambda_2},{\bf U },{\bf V }, {\bf W }$ determined in the constraint algorithm into the total Hamiltonian function $H_T^3$. So that we have the following explicit expression
\begin{equation} \label{TotRedHam}
\begin{split}
H_T^3&=\frac{m\zeta}{2}||{\bf{q}}_{(2)}||^2-\frac{\zeta^2}{2\mu m}{\bf{q}}_{(1)}\cdot{\bf{q}}_{(2)}\times {\bf{q}}_{(3)}+{\bf{q}}_{(2)}\cdot{\bf{p}}^{(1)}+{\bf{q}}_{(3)}\cdot{\bf{p}}^{(2)} \\&+[{\bf{q}}_{(1)}(\frac{{\bf{q}}_{(2)}\cdot {\bf{q}}_{(3)}}{||{\bf{q}}_{(1)}||^2})-\frac{\mu }{\zeta^3||{\bf{q}}_{(1)}||^2}[\frac{3\zeta^2}{2\mu m}{\bf{q}}_{(2)}\times {\bf{q}}_{(3)}+ m\zeta{\bf{q}}_{(3)}]\times [\frac{3\zeta^2}{2\mu m}{\bf{q}}_{(1)}\times {\bf{q}}_{(2)}+m\zeta{\bf{q}}_{(1)}]]\cdot {\bf{p}}^{(3)}
\\&+\frac{\zeta^2}{2\mu m}{\bf{p}}^{(\lambda_1)}\cdot
{\bf{q}}_{(2)}\times {\bf{q}}_{(3)}+[-m\zeta {\bf{q}}_{(2)}+\frac{\zeta^2}{2\mu m}{\bf{q}}_{(3)}\times {\bf{q}}_{(1)}-\boldsymbol{\lambda}_1 ]\cdot\bf{p}^{(\lambda_2)}.
\end{split}
\end{equation}
So that the Hamilton's equations of motion governed by the total Hamiltonian in (\ref{TotRedHam}) are computed to be
\begin{equation} \label{totclemHamEq}
\begin{split}
\dot{\bf{q}}_{(1)}&\approx {\bf{q}}_{(2)},\qquad \dot{\bf{q}}_{(2)}\approx{\bf{q}}_{(3)},\\
\dot{\bf{q}}_{(3)}&\approx {\bf{q}}_{(1)}(\frac{{\bf{q}}_{(2)}\cdot {\bf{q}}_{(3)}}{||{\bf{q}}_{(1)}||^2})-\frac{\mu }{\zeta^3||{\bf{q}}_{(1)}||^2}[\frac{3\zeta^2}{2\mu m}{\bf{q}}_{(2)}\times {\bf{q}}_{(3)}+ m\zeta{\bf{q}}_{(3)}]\times [\frac{3\zeta^2}{2\mu m}{\bf{q}}_{(1)}\times {\bf{q}}_{(2)}+m\zeta{\bf{q}}_{(1)}],
\\
\dot{\bf{p}}^{(1)}&=\dot{\boldsymbol{\lambda} }_1\approx\frac{\zeta^2}{2\mu m}{\bf{q}}_{(2)}\times {\bf{q}}_{(3)}, \\ \dot{\bf{p}}^{(2)}&=\dot{\boldsymbol{\lambda}}_2 \approx-m\zeta {\bf{q}}_{(2)}+\frac{\zeta^2}{2\mu m}{\bf{q}}_{(3)}\times {\bf{q}}_{(1)}-\boldsymbol{\lambda}_1
\quad \dot{\bf{p}}^{(3)}\approx \frac{\zeta^2}{2\mu m}{\bf{q}}_{(1)}\times {\bf{q}}_{(2)}-\boldsymbol{\lambda}_2,\\
\dot{\bf{p}}^{(\lambda_1)}&\approx0, \qquad \dot{\bf{p}}^{(\lambda_2)}\approx0.
\end{split}
\end{equation} 
In order to retrieve the Euler-Lagrange equations (\ref{clee}) generated by the Cl\`{e}ment Lagrangian from the Hamilton's  equations, we simply substitute the momenta into the equation of motion governing $\bf{p}^{(1)}$. The rest of the equations are then trivially satisfied. \\

\noindent \textbf{The Dirac bracket:} 

We can also derive the Hamilton's equations using Dirac bracket. All constraints for the Total reduction case are
\begin{align}
&\boldsymbol{\Phi}^{(1)}=\mathbf{p}^{(1)}-\boldsymbol{\lambda}_{1},\quad\boldsymbol{\Phi}^{(\lambda_1)}=\mathbf{p}^{(\lambda_1)},\quad\boldsymbol{\Phi}^{(2)}=\mathbf{p}^{(2)}-\boldsymbol{\lambda}_{2},\quad\boldsymbol{\Phi}^{(\lambda_2)}=\mathbf{p}^{(\lambda_2)},\quad
\boldsymbol{\Phi}^{(3)}=\mathbf{p}^{(3)}, \\&\boldsymbol{\Phi }=\frac{\zeta^2}{2\mu m}\mathbf{q}_{(1)}\times\mathbf{q}_{(2)}-\boldsymbol{\lambda}_{2},\quad \boldsymbol{\varphi}=m\zeta \mathbf{q}_{(2)}+\frac{\zeta^2}{\mu m}\mathbf{q}_{(1)}\times \mathbf{q}_{(3)}+\boldsymbol{\lambda}_{1},\quad\boldsymbol{\chi}=\mathbf{q}_{(1)}\cdot(m\zeta\mathbf{q}_{(3)}+\frac{3\zeta^2}{2\mu m}\mathbf{q}_{(2)}\times \mathbf{q}_{(3)})
\end{align}
second class. Recalling the definition in equation (\ref{diracbrac}) of the Dirac bracket, we derive some of Dirac brackets
\begin{align*}
\{{q}_{(1)}^i,{q}_{(1)}^j\}_{DB}&=\{{q}_{(1)}^i,\lambda_{(2)}^j\}_{DB}=\{{q}_{(1)}^i,p^{(2)}_j\}_{DB}=\{{q}_{(1)}^i,p^{(3)}_j\}_{DB}=\{{q}_{(1)}^i,p^{(\lambda_1)}_j\}_{DB}=\{{q}_{(1)}^i,p^{(\lambda_2)}_j\}_{DB}\\&=\{{q}_{(2)}^i,p^{(\lambda_1)}_j\}_{DB}=\{{q}_{(2)}^i,p^{(\lambda_2)}_j\}_{DB}=\{{q}_{(2)}^i,p^{(3)}_j\}_{DB}=\{\lambda_{(1)}^i,p^{(3)}_j\}_{DB}=\{\lambda_{(2)}^i,p^{(3)}_j\}_{DB}\\&=\{\lambda_{(1)}^i,p^{(\lambda_1)}_j\}_{DB}=\{\lambda_{(1)}^i,p^{(\lambda_2)}_j\}_{DB}=\{\lambda_{(2)}^i,p^{(\lambda_1)}_j\}_{DB}=\{\lambda_{(2)}^i,p^{(\lambda_2)}_j\}_{DB}=0\\
\{{q}_{(1)}^i,{q}_{(2)}^j\}_{DB}&=-\frac{1}{m\zeta||\mathbf{q}_{(1)}||^2}{q}_{(1)}^i{q}_{(1)}^j\\
\{{q}_{(1)}^i,{\lambda}_{(1)}^j\}_{DB}&=\{{q}_{(1)}^i,\delta^{jk}p^{(1)}_k\}_{DB}=\delta^{ij}+\frac{\zeta}{2\mu m^2||\mathbf{q}_{(1)}||^2}{q}_{(1)}^i\delta^{jk}\epsilon_{klr} {q}_{(1)}^l{q}_{(2)}^r\\
\{{q}_{(2)}^i,{q}_{(2)}^j\}_{DB}&=-\frac{\mu m}{\zeta^2||\mathbf{q}_{(1)}||^2}\epsilon_{ijk} {q}_{(1)}^k-\frac{1}{2m\zeta||\mathbf{q}_{(1)}||^2}\epsilon^{ijk}\epsilon_{klr}{q}_{(1)}^l{q}_{(2)}^r\\
\{{q}_{(2)}^i,{\lambda}_{(1)}^j\}_{DB}&=\frac{-{q}_{(1)}^j{q}_{(2)}^i +{ \delta^{ij}}\mathbf{q_{(1)}\cdot\mathbf{q_{(2)}}}}{2||\mathbf{q}_{(1)}||^2}+\frac{\zeta}{4\mu m^2||\mathbf{q}_{(1)}||^2}{q}_{(2)}^i\delta^{jr}\epsilon_{rkl} {q}_{(2)}^k{q}_{(1)}^l+\frac{\zeta}{\mu m^2||\mathbf{q}_{(1)}||^2}{q}_{(1)}^i\delta^{jr}\epsilon_{rkl} {q}_{(1)}^k{q}_{(3)}^l\\
\{{q}_{(2)}^i,{\lambda}_{(2)}^j\}_{DB}&=\{{q}_{(2)}^i,\delta^{jk}p^{(2)}_k\}_{DB}=\frac{-{q}_{(1)}^j{q}_{(1)}^i +{ \delta^{ij}}||\mathbf{q_{(1)}||^2}}{2||\mathbf{q}_{(1)}||^2}+\frac{\zeta}{4\mu m^2||\mathbf{q}_{(1)}||^2}{q}_{(1)}^i\delta^{jr}\epsilon_{rkl} {q}_{(1)}^k{q}_{(2)}^l\\
\{{\lambda}_{(1)}^i,{\lambda}_{(1)}^j\}_{DB}&=\{{\lambda}_{(1)}^i,\delta^{jk}p^{(1)}_k\}_{DB}=-\frac{\zeta^2}{4\mu m||\mathbf{q}_{(1)}||^2}\epsilon^{ijk}\delta_{kl}{q}_{(1)}^l(\mathbf{q_{(1)}\cdot\mathbf{q_{(2)}}})+\frac{\zeta^3}{2\mu^2 m^3||\mathbf{q}_{(1)}||^2}\epsilon^{ijk} \delta_{kl}{q}_{(1)}^l(\mathbf{q}_{(1)}\cdot \mathbf{q}_{(3)}\times \mathbf{q}_{(2)})\\
\{{\lambda}_{(1)}^i,{\lambda}_{(2)}^j\}_{DB}&=\{{\lambda}_{(1)}^i,\delta^{jk}p^{(2)}_k\}_{DB}=\{\delta^{ik}p^{(1)}_k,{\lambda}_{(2)}^j,\}_{DB}\\&=\frac{\zeta^2}{4\mu m||\mathbf{q}_{(1)}||^2}\delta^{ir}\epsilon_{rkl}{q}_{(1)}^k{q}_{(2)}^l+\frac{\zeta^2}{4\mu m} \epsilon^{ijk}\delta_{kl}{q}_{(2)}^l+ \frac{\zeta^3}{8\mu^2 m^3||\mathbf{q}_{(1)}||^2}\delta^{ir}\epsilon_{rkl}q_{(2)}^k{q}_{(1)}^l\delta^{j{r'}}\epsilon_{{r'}{k'}{l'}}{q_{(1)}^{k'}{q}_{(2)}^{l'}}\\
\{{\lambda}_{(2)}^i,{\lambda}_{(2)}^j\}_{DB}&=\{{\lambda}_{(2)}^i,\delta^{jk}p^{(2)}_k\}_{DB}=- \frac{\zeta^2}{4\mu m}\epsilon^{ijk}\delta_{kl}{q}_{(1)}^l
\end{align*} 
which are required to derive Hamilton equations. Equations of motion generated by canonical Hamiltonian function H given in (\ref{HamforTotClem}) can be evaluated from 
\begin{align}
{\dot{X}}^i=\{X^i,H\}_{DB}&=-\frac{\zeta^2}{2\mu m}\epsilon_{jkl}{q}_{(2)}^k{q}_{(3)}^l\{X^i,{q}_{(1)}^j\}_{DB}+ \frac{\zeta^2}{2\mu m}\epsilon_{jkl}{q}_{(3)}^k{q}_{(1)}^l\{X^i,{q}_{(2)}^j\}_{DB}\notag\\&+ \delta_{jk}{q}_{(2)}^k\{X^i,\lambda_{(1)}^j\}+\delta_{jk}{q}_{(3)}^k\{X^i,\lambda_{(2)}^j\}
\end{align}
using the Dirac brackets of coordinates.

\subsection{Partial reduction I} 

We start with Cl\`{e}ment
Lagrangian (\ref{LC}) once more but, in this case, we will apply the partial reduction presented in (\ref{parred1}). Here is the reduced first order Lagrangian derived from the Cl\`{e}ment Lagrangian 
\begin{align}
L_{C_1}^{C}&=-{\frac{m\zeta}{2}}||\mathbf{q}_{(2)}||^2+{\frac{\zeta^{2}}{2\mu m }%
}\mathbf{q}_{(1)}\cdot\mathbf{q}_{(2)}\times\mathbf{\dot{q}}_{(2)}+\boldsymbol{%
\lambda}_1\cdot(\mathbf{\dot{q}}_{(1)}-\mathbf{q}_{(2)}),
\label{constraint lagrangian}
\end{align}
where we have employed the coordinate transformations $\mathbf{x}=\mathbf{q}_{(1)}$, $\mathbf{\dot{x%
}}=\mathbf{q}_{(2)}$, $\mathbf{\ddot{x}}=\mathbf{\dot{q}}_{(2)}$. Here, $\boldsymbol{\lambda}_{1}$ is a set of Lagrange multipliers. The Euler-Lagrange equations generated by $L_{C_1}^{C}$ is computed to be
\begin{align}
\boldsymbol{\dot{\lambda}}
_1={\frac{\zeta^{2}}{2\mu m}}\mathbf{q}_{(2)}\times \mathbf{\dot{q}}_{(2)},\qquad \boldsymbol{\lambda}_1=-m\zeta \mathbf{q}_{(2)}-{%
\frac{\zeta^{2}}{\mu m}}\mathbf{q}_{(1)}\times\mathbf{\dot{q} }_{(2)}+{\frac{%
\zeta^{2}}{2\mu m}}\mathbf{q}_{(2)}\times\mathbf{\dot{q}}_{(1)},\qquad \dot{\mathbf{q}}_{(1)}-\mathbf{q}_{(2)}=\mathbf{0}. \label{lambda1}
\end{align}

We introduce the conjugate momenta $(\mathbf{p}^{(1)},\mathbf{p}^{(2)},\mathbf{p}^{\lambda_1})$ dual to the velocities $(\mathbf{\dot{q}}_{(1)},\mathbf{\dot{q}}_{(2)},\boldsymbol{\dot{\lambda}}
_1)$. The Legendre transformation leads to the following relationships
\begin{align}
{\mathbf{p}}^{(1)} =\boldsymbol{\lambda}_1,\quad \mathbf{p}^{(2)} =%
\frac{\zeta^{2}}{2\mu m}\mathbf{q}_{(1)}\times\mathbf{q}_{(2)} ,\quad 
\mathbf{p}^{\lambda_1} =0.   \label{canonical momentum3}
\end{align}
Since neither of the velocities $\mathbf{%
\dot{q}}_{(1)}, \mathbf{\dot{q}}_{(2)}$ and $\boldsymbol{\dot{\lambda}}_1$  can be solved from these relations,
we introduce the primary constraints
\begin{align}
\boldsymbol{\Phi}^{(1)} = \mathbf{p}^{(1)}-\boldsymbol{\lambda}%
_{1},\quad \boldsymbol{\Phi}^{(2)} =\mathbf{p}^{(2)}-\frac{
\zeta^{2}}{2\mu m} \mathbf{q}_{(1)}\times\mathbf{q}_{(2)},\quad \boldsymbol{\Phi}%
^{(\lambda_1)}=\mathbf{p}^{\lambda_1}.   \label{prim2}
\end{align}
In accordance with this, the total Hamiltonian function is defined to be
\begin{align}
H_{T} & =H_{C_1}+\mathbf{u}_{(1)}\cdot\boldsymbol{\Phi}^{(1)}+\mathbf{u}_{(2)}\cdot%
\boldsymbol{\Phi}^{(2)}+\mathbf{u}_{(\lambda_1)} \cdot\boldsymbol{\Phi}^{(\lambda_1)},
\label{consthamil}
\end{align}
where $\mathbf{u}_{(1)},\mathbf{u}_{(2)}$ and $\mathbf{u}_{(\lambda_1)}$ are Lagrange multipliers whereas $H_{C_1}$ is the canonical Hamiltonian function 
computed as
\begin{align}
H_{C_1} ={\frac{m\zeta}{2}}||\mathbf{q}_{(2)}||^2+\mathbf{p} ^{(1)}\cdot\mathbf{q}%
_{(2)}.  \label{Hamilforfirst}
\end{align}
Here are the steps of the Dirac-Bergmann constraint algorithm. 

\bigskip

\noindent \textbf{Dirac-Bergmann constraint algorithm step 1:} 
Consistency checks of each of the primary constraints exhibited in the equations (\ref{prim2}) read
\begin{equation}
 \begin{split}
\boldsymbol{\dot{\Phi}}^{(1)}&=\{\boldsymbol{\Phi}^{(1)},H_{T}\} \approx \frac{%
\zeta^{2}}{2\mu m}\mathbf{q}_{(2)}\times\mathbf{u}_{(2)}-\mathbf{u}%
_{(\lambda_1)},  \label{eqnconst0} \\
\boldsymbol{\dot{\boldsymbol{\Phi}}}^{(2)}&=\{\boldsymbol{\Phi}^{(2)},H_{T}\}
\approx m\zeta\mathbf{q}_{(2)}+\mathbf{p}^{(1)}-\frac{\zeta^{2}}{2\mu m}%
\mathbf{u}_{(1)}\times\mathbf{q}_{(2)}+\frac{\zeta^{2}}{\mu m} \mathbf{u}%
_{(2)}\times \mathbf{q}_{(1)},  \\
\boldsymbol{\dot{\Phi}}^{(\lambda_1)}&=\{\boldsymbol{\Phi}^{(\lambda_1)},H_{T}\}%
\approx \mathbf{u}_{(1)}.
\end{split}
\end{equation}
From these expressions, we determine the Lagrange multipliers $\mathbf{u}_{(1)}\approx0$ and $\mathbf{u}_{(\lambda_1)}\approx\frac{\zeta^{2}}{2\mu m}\mathbf{q}%
_{(2)}\times\mathbf{u}_{(2)}.$ From Eq.(\ref{eqnconst0}), we can solve only two components of the $\mathbf{u}_{(2)}$. So that there remains a secondary constraint \begin{equation}
{\Phi}=\big( m\zeta\mathbf{q}_{(2)}+\mathbf{p}^{(1)}\big)\cdot 
\mathbf{q}_{(1)}.  \label{secondary constraint}
\end{equation}
By adding this secondary constraint to the total Hamiltonian function $H_T$ in (\ref{consthamil}), we revise the Hamiltonian as  $H_T^1=H_T+ u\Phi$, where $u$ being a Lagrange multiplier. 

\bigskip

\noindent \textbf{Dirac-Bergmann constraint algorithm step 2:} Consistency of the secondary constraint $\Phi$ can be checked by the following calculation
\begin{equation}  \label{eqnconst2}
\dot{\Phi} =\{\Phi,H_{T}^1\} \approx (m\zeta\mathbf{q}_{(1)}+\mathbf{p}^{(1)})\cdot%
\mathbf{q}_{(2)} +\mathbf{u}_{(2)}\big( m\zeta\mathbf{q}_{(1)}+\frac{\zeta^{2}}{%
2\mu m}\mathbf{q}_{(2)}\times \mathbf{q}_{(1)}\big).
\end{equation}
Notice that, one may determine the third component of $\mathbf{u}_{(2)}$ using Eq.(\ref{eqnconst2}) whereas consistency equation of the constraint $\boldsymbol{\Phi}^{(2)}$ leads us to determine $u$. Hence, all Lagrange multipliers are computed to be
\begin{align}
\begin{split}
\mathbf{u}_{(1)} & \approx\mathbf{0}, \\
\mathbf{u}_{(2)} &\approx \frac{\mu m}{\zeta^{2}{||\mathbf{q}_{(1)}}||^{2}}\mathbf{q}%
_1\times\mathbf{D} - \frac{3}{2m\zeta \mathbf{q}_{(1)}^{2}}(\mathbf{D}\cdot 
\mathbf{q}_{(2)})\mathbf{q}_{(1)}, \\
\mathbf{u}_{(\lambda_1)} & \approx\frac{\zeta^{2}}{2\mu m} \mathbf{q}_{(2)}\times\big(%
\frac{\mu m}{\zeta^{2}{||\mathbf{q}_{(1)}}||^{2}}\mathbf{q}_{(1)}\times\mathbf{D} - 
\frac{3}{2m\zeta ||\mathbf{q}_{(1)}||^{2}}(\mathbf{D}\cdot\mathbf{q}_{(2)})\mathbf{q}%
_{(1)}\big),\\
u&\approx-\frac{\Phi}{m\zeta ||\mathbf{q}_{(1)}||^{2}}
\end{split}
\end{align}
where we have used the abbreviation $\mathbf{D}=m\zeta\mathbf{q}_{(2)}+\mathbf{p}^{(1)}
$. 

\bigskip

\noindent \textbf{The total Hamiltonian function and the equations of motion:} Substitutions
of the Lagrange multipliers  $u, \mathbf{u}_{(1)},\mathbf{u}_{(2)}$ and $\mathbf{u}_{(\lambda_1)}$  into $H_T^1$ determines the total Hamiltonian function 
\begin{align}
H_{T}^1 & =\frac{1}{2}\mathbf{p}^{(1)}\cdot \mathbf{q}_{(2)}+\frac{1}{2||
\mathbf{q}_{(1)}||^2}(\mathbf{q}_{(1)}\cdot\mathbf{q}_{(2)})(\mathbf{D}\cdot\mathbf{q}_{(1)})+%
\frac{\mu m}{\zeta^{2}{{||\mathbf{q}}_{(1)}}||^{2}}\mathbf{p} ^{(2)}\cdot
\mathbf{q}_{(1)}\times\mathbf{D}  \notag \\
&- \frac{3}{2m\zeta ||{\mathbf{q}}_{(1)}||^{2}}(\mathbf{D}\cdot\mathbf{q}_{(2)})({%
\mathbf{q}}_{(1)}\cdot\mathbf{p} ^{(2)})+\frac{1}{2||{\mathbf{q}}_{(1)}||^{2} }%
\mathbf{p}^{\lambda_1}\cdot {\mathbf{q}}_{(2)}\times(\mathbf{q}_{(1)}\times 
\mathbf{D})  \notag \\
&-\frac{3\zeta}{4\mu m^2||\mathbf{q}_{(1)}||^{2}}\mathbf{p}^{\lambda_1}\cdot{%
\mathbf{q}}_{(2)}\times{\mathbf{q}}_{(1)}(\mathbf{D}\cdot\mathbf{q}_{(2)})-\frac{(%
\mathbf{D}\cdot\mathbf{q}_{(1)})^2}{m\zeta ||\mathbf{q}_{(1)}||^2}-\frac{\Phi^2}{m\zeta ||\mathbf{q}_{(1)}||^{2}}.  \label{HamforClem1}
\end{align}
The Hamilton's equations generated by the total Hamiltonian function $(\ref%
{HamforClem1})$ are 
\begin{align}
\begin{split}
\dot{\mathbf{q}}_{(1)} &\approx\frac{1}{2}\mathbf{q}_{(2)}+\frac{1}{2||\mathbf{q}_{(1)}||^2%
}(\mathbf{q}_{(1)}\cdot\mathbf{q}_{(2)})\mathbf{q}_{(1)}+\frac{\mu m}{\zeta^2 ||\mathbf{q}%
_1||^2}\mathbf{p}^{(2)}\times\mathbf{q}_{(1)} \\
\dot{\mathbf{q}}_{(2)}&\approx \frac{\mu m}{\zeta^2 ||\mathbf{q}_{(1)}||^2}\mathbf{q}%
_{(1)}\times\mathbf{D}-\frac{3}{2m\zeta ||\mathbf{q}_{(1)}||^2}(\mathbf{D}\cdot\mathbf{q}%
_{(2)})\mathbf{q}_{(1)} \\
\dot{\boldsymbol{\lambda}}_{1} & \approx\frac{1}{2||\mathbf{q}_{(1)}||^2}\mathbf{q}%
_{(2)}\times(\mathbf{q}_{(1)}\times\mathbf{D})-\frac{3\zeta}{4\mu m^2||\mathbf{q}_{(1)}||^2}%
\mathbf{q}_{(2)}\times\mathbf{q}_{(1)}(\mathbf{D}\cdot\mathbf{q}_{(2)}) \\
\dot{\mathbf{p}}^{(1)} & \approx\frac{-1}{2||\mathbf{q}_{(1)}||^2}(\mathbf{q}%
_{(1)}\cdot\mathbf{q}_{(2)})\mathbf{D}-\frac{\mu m}{\zeta^2\mathbf{q}_{(1)}^2}\mathbf{D}%
\times\mathbf{p}^{(2)}+\frac{3}{2m\zeta ||\mathbf{q}_{(1)}||^2}\mathbf{p}^{(2)}(%
\mathbf{D}\cdot\mathbf{q}_{(2)})   \\
&+\frac{2\mu m}{\zeta^2 ||\mathbf{q}_{(1)}||^4}(\mathbf{p}^{(2)}\cdot\mathbf{q}%
_{(1)}\times\mathbf{D})\mathbf{q}_{(1)} \\
\dot{\mathbf{p}}^{(2)} & \approx-\frac{1}{2}\mathbf{p}^{(1)}-\frac{%
m\zeta}{2||\mathbf{q}_{(1)}||^2}(\mathbf{q}_{(1)}\cdot\mathbf{q}_{(2)})\mathbf{q}_{(1)}-\frac{%
\mu m^2}{\zeta ||\mathbf{q}_{(1)}||^2}\mathbf{p}^{(2)}\times\mathbf{q}_{(1)} \\
\dot{\mathbf{p}}^{\lambda_1} & \approx 0.
\end{split}
\end{align}
The equation of motion for ${\mathbf{p}}^{(1)}$ gives the
Euler-Lagrange equations $(\ref{clee})$ whereas the remaining ones are
identically satisfied after the back substitution of the momenta.

%\begin{remark}
% As it is pointed out in $\cite{Po89}$ that, the constraints $\boldsymbol{\Phi%
%}^{(\lambda_1)}=\mathbf{p}^{\lambda_1}$ effect only the equation of %motion
%for $\boldsymbol{\lambda}_1$. So, we may omit to add them to the total
%Hamiltonian function $H_{T}$. But consistency condition of $%\boldsymbol{\Phi%
%}^{(\lambda_1)}=\mathbf{p}^{\lambda_1}$ leads to $\mathbf{u}_{(1)}=0$, %this means that, the constraint $%
%\boldsymbol{\Phi}^{(1)}$ does not need to be included in total %Hamiltonian $H_T$. So that, the total Hamiltonian function reduces to 
%\begin{equation}
%H_{T}=H_C+\mathbf{u}_{(2)}\cdot\boldsymbol{\Phi}^{(2)}
%\end{equation}
%with the constraints 
%\begin{align}
%\boldsymbol{\Phi}^{(2)} =\mathbf{p}^{(2)}-\frac{\zeta^2}{2\mu m}\mathbf{q
%}_{(1)}\times\mathbf{q}_{(2)}, \qquad \Phi=(m\zeta\mathbf{q}_{(2)}+\mathbf{p}^{(1)})\cdot\mathbf{q}_{(1)}.
%\end{align}
%\end{remark}

\bigskip

\noindent \textbf{The Dirac bracket:} 
We shall derive the Hamilton's equations using the Dirac bracket. To do this, we record here the set of second class constraints 
\begin{align}
\boldsymbol{\Phi}^{(1)} & \equiv\mathbf{p}^{(1)}-\boldsymbol{\lambda}_{1}, \qquad 
\boldsymbol{\Phi}^{(2)} \equiv\mathbf{p}^{(2)}-\frac{\zeta^{2}}{2\mu m}%
\mathbf{q}_{(1)}\times\mathbf{q}_{(2)} ,\qquad  \boldsymbol{\Phi}^{(\lambda_1)} \equiv%
\mathbf{p}^{\lambda_1},\qquad  \Phi\equiv(m\zeta\mathbf{q}_{(2)}+\mathbf{p}%
^{(1)})\cdot\mathbf{q}_{(1)}.  \label{prim3}
\end{align}
In the present case, the Dirac bracket (\ref{diracbrac}) turns out to be 
\begin{align}
\{F,G\}_{DB} &
=\{F,G\}-\{F,\Phi_{n}^{(1)}\}C_{1\lambda}^{nn^{\prime}}\{\delta_{n^{\prime}l}%
\Phi_{(\lambda_1)}^{l},G\}-\{F,\Phi_{n}^{(2)}\}C_{11}^{nn^{\prime}}\{\Phi_{n^{%
\prime}}^{(2)},G\}  \notag \\
&-\{F,\Phi_{n}^{(2)}\}C_{1\lambda}^{nn^{\prime}}\{\delta_{n^{\prime}l}\Phi_{%
\lambda}^{l},G\}-\{F,\Phi_{n}^{(2)}\}C_{1\Phi}^{n}\{\Phi,G\}  \notag \\
&
-\{F,\delta_{nl}\Phi_{\lambda}^{l}\}C_{\lambda0}^{nn^{\prime}}\{\Phi_{n^{%
\prime}}^{(0)},G\}-\{F,\delta_{nl}\Phi_{\lambda}^{l}\}C_{\lambda_1
1}^{nn^{\prime}}\{\Phi_{n^{\prime}}^{(1)},G\}  \notag \\
&
-\{F,\delta_{nl}\Phi_{\lambda}^{l}\}C_{\lambda\lambda}^{nn^{\prime}}\{%
\delta_{n^{\prime}r}\Phi_{n}^{r},G\}-\{F,\delta_{nl}\phi_{\lambda}^{l}\}C_{%
\lambda\Phi}^{n}\{\Phi,G\}  \notag \\
& -\{F,\Phi\}C_{\Phi1}^{n^{\prime}}\{\Phi^{(1)}_{n^{\prime}},G\}-\{F,\Phi
\}M_{\Phi1}^{n^{\prime}}\{\delta_{n^{\prime}l}\Phi_{\lambda}^{l},G\}. 
\label{diracbracforconst}
\end{align}
Here, the matrix $C$ is computed to be
\begin{equation*}
C = \left( 
\begin{array}{cccc}
0 & {\zeta^{2}}\epsilon_{n^{\prime}nk^{\prime}}q_{(2)}^{k^{%
\prime}}/ {2\mu m} & -\delta_{n}^{n^{\prime}} & -D_{n} \\ 
-{\zeta^{2}}\epsilon_{nn^{\prime}k^{\prime}}q_{(2)}^{k^{\prime}}/{2\mu m}
& {\zeta^{2}}\epsilon_{nn^{\prime}k}q_{(1)}^{k}/{\mu m} & 0 & -E_n \\ 
\delta_{n^{\prime}}^{n} & 0 & 0 & 0 \\ 
D_{n^{\prime}} & E_{n^{\prime }} & 0 & 0%
\end{array}
\right) 
\end{equation*}
with determinant ${\zeta^{6}||{q}_{(1)}||^{2}}/{\mu^{2}}$. The inverse of $%
C$ is 
\begin{equation}
\begin{split}
C^{-1}&=\left( 
\begin{array}{cccc}
C_{00}^{nn^{\prime}} & C_{01}^{nn^{\prime}} & C_{0\lambda}^{nn^{\prime}} & 
C_{0\Phi}^{n} \\ 
C_{10}^{nn^{\prime}} & C_{11}^{nn^{\prime}} & C_{1\lambda}^{nn^{\prime}} & 
C_{1\Phi}^{n} \\ 
C_{\lambda0}^{nn^{\prime}} & C_{\lambda1}^{nn^{\prime}} & C_{\lambda\lambda
}^{nn^{\prime}} & C_{\lambda\Phi}^{n} \\ 
C_{\Phi0}^{n^{\prime}} & C_{\Phi1}^{n^{\prime}} & C_{\Phi\lambda}^{n^{%
\prime}} & C_{\Phi\Phi}%
\end{array}
\right)  \\&=\left( 
\begin{array}{cccc}
0 & 0 & \delta^{nn^{\prime}} & 0 \\ 
0 & {-\mu\epsilon^{nn^{\prime}k}E_k}/{\zeta^{3}||\mathbf{q}_{(1)}||^{2}} & 
A^{nn^{\prime}} & {q_{(1)}^{n}}/{m\zeta||\mathbf{q}_{(1)}||^{2}} \\ 
-\delta^{nn^{\prime}} & B^{nn^{\prime}} & C^{nn^{\prime}} & -\frac {%
\zeta\epsilon_{krs}\delta^{nk}q_{(1)}^{r}q_{(2)}^{s}}{2m^{2}\mu ||\mathbf{q}_{(1)}||^{2}} \\ 
0 & -{q_{(1)}^{n^{\prime}}}/{m \zeta||\mathbf{q}_{(1)}||^{2}} & {%
\zeta\epsilon_{krs}\delta^{n^{\prime}k}q_{(1)}^{r}q_{(1)}^{s}}/{2m^{2}\mu ||\mathbf{q}%
_{(1)}^{2}||} & 0%
\end{array}
\right)
\end{split}
\end{equation}
where $$A^{n}_{n^{\prime }} =\frac{1}{2m\zeta ||\mathbf{q}_{(1)}||^2}\epsilon^{nrk}\epsilon_{n^{%
\prime }rl}q_{(2)}^lE_k+\frac{1}{m\zeta ||\mathbf{q}_{(1)}||^2}q_1^nD_{n^{\prime }}, \qquad
B^{nn^{\prime}} =-\frac{\epsilon_{rkp}\delta^{nk}q_{(2)}^{p}\epsilon
^{rn^{\prime}s}E_s}{2m\zeta ||\mathbf{q}_{(1)}||^{2}}+\frac{D^nq_{(1)}^{n^{\prime}}}{||\mathbf{q}%
_{(1)}||^{2}}. 
$$
The Dirac brackets of the coordinates are 
\begin{align}
\begin{split}
\{q_{(1)}^{i},q_{(2)}^{j}\}_{DB}&=-\frac{q_{(1)}^{i}q_{(1)}^{j}}{m\zeta|| \mathbf{q}_{(1)}||^{2}},
\\
\{q_{(1)}^{i},\lambda_{1}^{j}\}_{DB}& =\delta_{ij}-\frac{q_{(1)}^{i}q_{(1)}^{j}}{m\zeta||\mathbf{q}_{(1)}||^{2}%
},
\\
 \{q_{(1)}^{i},{p}^{(1)}_{j}\}_{DB}&=\delta_{j}^{i}-\frac{\zeta\epsilon _{jk^{\prime}n^{\prime}}q_{(1)}^{n^{\prime}}q_{(2)}^{k^{\prime}}}{2m^{2}\mu
||\mathbf{q}_{(1)}||^{2}} 
\\
\{q_{(2)}^{i},q_{(2)}^{j}\}_{DB} &=-\frac{\mu\epsilon^{ijk}E_k}{\zeta^{3} ||\mathbf{q}_{(1)}||^{2}}%
,\\
\{q_{(2)}^{i},\lambda_1^{j}\}_{DB} &=A^{ij},
\\ \{q_{(2)}^{i},{p}^{(1)}_{j}\}_{DB} &=-\frac{1}{2m\zeta ||\mathbf{q}_{(1)}||^{2}}\epsilon^{irk}
\epsilon_{jrl}q_{(2)}^{l}E_k-\frac{q_{(1)}^{i}D_j}{m\zeta ||\mathbf{q}_{(1)}||^{2}},~~ \\
\{q_{(2)}^{i},{p}^{(2)}_{j}\}_{DB}  &=\delta_{j}^{i}-\frac{1}{2m\zeta ||\mathbf{q}_{(1)}||^{2}}%
\epsilon^{in^{\prime
}k}\epsilon_{jn^{\prime}k^{\prime}}q_{(1)}^{k^{\prime}}E_k-\frac{q_{(1)}^{i}\delta_{jl} q_{(1)}^{l}}{
||\mathbf{q}_{(1)}||^{2}}, \\
\{\lambda_{1}^{i},\lambda_{1}^{j}\}_{DB} &=C^{ij} \\
\{\lambda_{1}^{i},{p}^{(1)}_{j}\}_{DB} & = \frac{\zeta^2}{2\mu m}%
B^{ir}\epsilon_{jkr q_{(2)}^k}+\frac{\zeta}{2m^{2}\mu ||\mathbf{q}_{(1)}||^{2} }%
\delta^{ik}\epsilon_{krs}q_{(1)}^{r} q_{(2)}^{s}D_j \\
\{\lambda_1^{i},{p}^{(2)}_{j}\}_{DB} &=\frac{\zeta^2}{2\mu m}%
B^{ir}\epsilon_{jrk}q_{(1)}^k+\frac{\zeta^2}{2m\mu ||\mathbf{q}_{(1)}||^{2} }%
\epsilon_{irs}q_{(1)}^{r} q_{(2)}^{s}(\delta _{jl}q_{(1)}^{l}) \\
\{{p}^{(1)}_{i},{p}^{(1)}_{j}\}_{DB} & =\frac{\zeta\epsilon_{ik^{%
\prime}j}q_{(2)}^{k^{\prime}}E_kq_{(2)}^{k}}{4\mu m^{2}||\mathbf{q}_{(1)}||^{2}}-\frac{%
\zeta\epsilon_{ikn}q_{(2)}^{k}q_{(1)}^{n}D_j}{2\mu m^{2}||\mathbf{{q}}_{(1)}||^{2}}+\frac{%
\zeta\epsilon_{jk^{\prime}n^{\prime}}q_{(2)}^{k^{\prime}}q_{(1)}^{n^{\prime}} D_i%
}{2\mu m^{2}||\mathbf{q}_{(1)}||^2} \\
\{{p}^{(1)}_{i},{p}^{(2)}_{j}\}_{DB} & =-\frac{\zeta\epsilon_{ik^{%
\prime}j}q_{(2)}^{k^{\prime}}D_iq_{(1)}^{k}}{4\mu m^{2}||\mathbf{q}_{(1)}||^{2}}+\frac{%
\zeta\epsilon_{ik^{\prime}n}q_{(2)}^{k^{\prime}}q^{n}_{(1)}E_j}{4\mu m^{2}||\mathbf{q}_{(1)}||^{2}%
},\\
\{{p}^{(2)}_{i},{p}^{(2)}_{j}\}_{DB} &=-\frac{\zeta}{4\mu m^3||\mathbf{q}_{(1)}||^2}%
\epsilon_{ink^{\prime}}q_{(1)}^{k^{\prime}}\epsilon^{nrk}E_k
\epsilon_{jrl}q_{(1)}^{l}
\end{split}
\end{align}
and all the others are zero. Here we used abbreviations 
\begin{equation}
D_i=m\zeta%
\delta_{ij}q_{(2)}^j+{p}^{(1)}_i, \qquad  E_i=m\zeta\delta_{ij}q_{(1)}^j+{p}^{(2)}_i,\qquad 
\delta_{ij}q_{(1)}^iq_{(1)}^j=||\mathbf{q}_{(1)}||^2
.
\end{equation}
Calculation of the Hamilton's equations of motion using the Dirac bracket of
coordinates is a matter of a direct calculation.
\subsubsection{Unconstrained Variational Formalism}

Now, we shall present the case of unconstrained Lagrangian in this framework. We point out that, Dirac analysis of this realization is much more simple. Let us start with unconstraint Lagrangian $(\ref{LU1})$ corresponding to the first order Cl\`{e}%
ment Lagrangian $(\ref{constraint lagrangian})$ which is 
\begin{align}
L_{U_1}^{C} ={\frac{m\zeta}{2}}\mathbf{q}_{(2)}\cdot\mathbf{q}_{(2)}-m\zeta\mathbf{q}%
_{(2)}\cdot\mathbf{\dot{q}} _{(1)}-{\frac{\zeta^2}{2\mu m}}\mathbf{q}_{(1)}\cdot%
\mathbf{q}_{(2)}\times\mathbf{\dot{q}} _{(2)}+{\frac{\zeta^2}{\mu m}}\mathbf{q}%
_{(1)}\cdot\mathbf{\dot{q}}_{(1)}\times\mathbf{\dot{q}}_{(2)}
\label{unconstraint lagrangian}
\end{align}
by substituting $\boldsymbol{\lambda}_1$ in $(\ref{lambda1})$ into $%
L^{C}_{C_1}$ in $(\ref{constraint lagrangian})$. The conjugate momenta turn out to be
\begin{align}
{\mathbf{p}}^{(1)} =-m\zeta\mathbf{q}_{(2)}-\frac{\zeta ^{2}}{\mu m}%
\mathbf{q}_{(1)}\times\mathbf{\dot{q}}_{(2)}, \qquad  {\mathbf{p}}^{(2)} =-\frac{%
\zeta^{2}}{2\mu m}\mathbf{q}_{(1)}\times\mathbf{q}_{(2)}+\frac{\zeta^{2}}{\mu m}%
\mathbf{q}_{(1)}\times\mathbf{\dot{q }}_{(1)}.   \label{pi^{(1)}}
\end{align}
whereas the canonical Hamiltonian function for the first order unconstraint
Lagrangian $(\ref{unconstraint lagrangian})$ is 
\begin{equation}\label{hamforunconst}
\begin{split}
H_{U_1} &=-\frac{m\zeta}{2}||\mathbf{q}_{(2)}||^2+\frac{\mu m }{\zeta^2 q_{(1)}^3}%
(p^{(2)}_2D_1-p^{(2)}_1D_2)+\frac{1}{2q_{(1)}^3}(-D_2(q_{(1)}^2q_{(2)}^3-q_{(1)}^3q_{(2)}^2)  
\\
&+D_1(q_{(1)}^3q_{(2)}^{1}-q_{(1)}^{1}q_{(2)}^3)). 
\end{split}
\end{equation}
Here, we are still using the abbreviation  ${D}_i={p}^{(1)}_i+m\zeta\delta_{ij}{q}_{(1)}^j$. Equations $(\ref{pi^{(1)}})$ lead to solve two components of $\mathbf{\dot{q}}_{(1)}$ and two components of $ \mathbf{\dot{q}}_{(2)}$. The rest determines a set of primary constraints 
\begin{align}
\Phi=\mathbf{p}^{(2)}\cdot \mathbf{q}_{(1)},\quad \phi=(\mathbf{p}^{(1)}+
m\zeta \mathbf{q}_{(2)})\cdot\mathbf{q}_{(1)}.
\end{align}
Then the total Hamiltonian
function is defined to be
\begin{align}
H_{T}=H_{U_1} +U\Phi+V\phi  \label{H_{T_1}}
\end{align}
with Lagrange multipliers $U$ and $V$. Here are the steps of the Dirac-Bergmann constraint algorithm and the computation of the total Hamiltonian function.

\bigskip

\noindent \textbf{Dirac-Bergmann constraint algorithm step 1:} 
Consistency checks of the primary constraints 
$\Phi$ and $\phi$ give no more constraint, instead we can determine $U$ and $V$ as
\begin{align}
U\approx\frac{\frac{\mu m^2}{\zeta q_{(1)}^3}(q_{(1)}^2D_1+q_{(1)}^{1}D_2)-\frac{3}{2}D_i{%
q}_{(2)}^i}{m\zeta ||{\bf{q}_{(1)}}||^2}, \qquad
V\approx\frac{\delta_{ij}q_{(1)}^iq_{(2)}^j-\frac{2\mu m}{\zeta^2q_{(1)}^3}%
(p^{(1)}_2q_{(1)}^{1}+p^{(1)}_1q_{(1)}^2)}{||{\bf{q}_{(1)}}||^2}.
\end{align}
Substitutions of $U$ and $V$ into \eqref{H_{T_1}} result with the determination of the total Hamiltonian function $H_{T}$ for the first order unconstraint
Lagrangian $(\ref{unconstraint lagrangian})$.

\bigskip

\noindent \textbf{The equations of motion:}  In this case, the Hamilton's equations of
motion are 
\begin{align}
\begin{split}
\dot{q}_{(1)}^i&\approx\frac{\mu m}{\zeta^2 q_{(1)}^3}(p^{(2)}_2\delta^i_1-p^{(2)}_1%
\delta^i_2)+\frac{1}{2q_{(1)}^3}\big(-\delta_2^i(q_{(1)}^2q_{(2)}^3-q_{(1)}^3q_{(2)}^2)+%
\delta_1^i(q_{(1)}^3q_{(2)}^{1}-q_{(1)}^{1}q_{(2)}^3))   \\
&+(\frac{\delta_{kj}q_{(1)}^kq_{(2)}^j-\frac{2\mu m}{\zeta^2q_{(1)}^3}%
(p^{(1)}_2q_{(1)}^{1}+p^{(1)}_1q_{(1)}^2)}{||{\bf{q}_{(1)}}||^2})q_{(1)}^i, \\
\dot{q}_{(2)}^i&\approx\frac{\mu m}{\zeta^2 q_{(1)}^3}(\delta^i_2D_1-\delta_1^iD_2)+%
\big( \frac{\frac{\mu m^2}{\zeta q_0^3}(q_{(1)}^2D_1+q_{(1)}^{1}D_2)-\frac{3}{2}D_i{q}%
_{(2)}^i}{m\zeta ||{\bf{q}_{(1)}}||^2} \big)q_{(1)}^i ,\\
\dot{{p}}^{(1)}_i&\approx\frac{-1 }{2 q_{(1)}^3}(-D_2(\delta^2_iq_{(2)}^3-%
\delta^3_iq_{(2)}^2)+D_1(\delta^3_iq_{(2)}^{1}-\delta^{1}_iq_{(2)}^3))+\frac{\mu m}{%
\zeta^2(q_{(1)}^3)^2}\delta^3_i(p^{(1)}_2D_1-p^{(1)}_1D_2)   \\
&-(\frac{\frac{\mu m^2}{\zeta q_{(1)}^3}(q_{(1)}^2D_1+q_{(1)}^{1}D_2)-\frac{3}{2}D_i{q}_{(2)}^i%
}{m\zeta ||{\bf{q}_{(1)}}||^2})p^{(1)}_i-(\frac{\delta_{ij}q_{(1)}^iq_{(2)}^j-\frac{2\mu m}{%
\zeta^2q_{(1)}^3}(p^{(1)}_2q_{(1)}^{1}+p^{(1)}_1q_{(1)}^2)}{||{\bf{q}_{(1)}}||^2})D_i  \\
&+\frac{1}{2(q_{(1)}^3)^2}%
\delta^3_i(-D_2(q_{(1)}^2q_{(2)}^3-q_{(1)}^3q_{(2)}^2)+D_1(q_{(1)}^3q_{(2)}^{1}-q_{(1)}^{1}q_{(2)}^3)), \\
\dot{{p}}^{(2)}_i&\approx\frac{\mu m^2}{\zeta q_{(1)}^3}({p}^{(1)}_2\delta^{(1)}_i+%
p^{(1)}_1\delta^2_i)+m\zeta\delta_{ij}q_{(2)}^j-\frac{m\zeta}{2q_{(1)}^3}%
(\delta^2_i(q_{(1)}^2q_{(2)}^3-q_{(1)}^3q_{(2)}^2)+\delta^{1}_i(q_{(1)}^3q_{(2)}^{1}-q_{(1)}^{1}q_{(2)}^3))  
\\
&-\frac{1}{2q_{(1)}^3}(-D_2(\delta^3_iq_{(1)}^2-\delta^2_iq_{(1)}^3)+D_1(%
\delta^{1}_iq_{(1)}^3-\delta^3_iq_{(1)}^{1}))  \\
&-(\frac{\delta_{ij}q_{(1)}^iq_{(2)}^j-\frac{2\mu m}{\zeta^2q_{(1)}^3}%
(p^{(1)}_2q_{(1)}^{1}+p^{(1)}_1q_{(1)}^2)}{||{\bf{q}_{(1)}}||^2})m\zeta q_{(1)}^i.
\end{split}
\end{align}
Here, the equations governing the momenta ${\mathbf{p}}^{(1)}$ give the
Euler-Lagrange equations $(\ref{clee})$, and the remaining equations are
identically satisfied.

\bigskip

\noindent \textbf{The Dirac bracket:}  Let us find the Hamilton's equations for the unconstraint Lagrangian $(\ref{unconstraint lagrangian})$ using the Dirac algebra. All constraints are second class since their
Poisson bracket 
\begin{align}
\{\Phi,\phi\}=-m\zeta||\mathbf{q}_{(1)}||^2
\end{align}
is nonzero. In this case the matrix $M$ in the definition of the Dirac bracket given in $(\ref{diracbrac})$ has a relatively simple form given by
\begin{align}
M & =\left( 
\begin{array}{cc}
\{\Phi,\Phi\} & \{\Phi,\phi\} \\ 
\{\phi,\Phi\} & \{\phi,\phi\}%
\end{array}
\right) =m\zeta ||\mathbf{q}_{(1)}||^2\left( 
\begin{array}{cc}
0 & -1 \\ 
1 & 0%
\end{array}
\right).
\end{align}
In accordance with this, we compute the Dirac bracket of two functions as follows
\begin{align}
\{F,G\}_{DB} & =\{F,G\}-\frac{1}{m\zeta ||\mathbf{q}_{(1)}||^2}\big(\{F,\Phi\}\{\phi,G\}-\{F,%
\phi \}\{\Phi,G\}\big).  \label{DIRR}
\end{align}
More explicitly, the Dirac brackets of the coordinates are 
\begin{align}
\begin{split}
\{q_{(1)}^i,q_{(2)}^j\}_{DB}&=-\frac{1}{m \zeta ||\mathbf{q}_{(1)}||^2}q_{(1)}^iq_{(2)}^j \\
\{q_{(1)}^i,{p}^{(1)}_j\}_{DB}&=\frac{1}{m \zeta ||\mathbf{q}_{(1)}||^2}q_{(1)}^i{p}^{(2)}_j+\delta^i_j \\
\{q_{(2)}^i,{p}^{(1)}_j\}_{DB}&=-\frac{1}{m \zeta ||\mathbf{q}_{(1)}||^2}q_{(1)}^i({p}^{(1)}_j+m \zeta
\delta_{jk}q_{(2)}^k) \\
\{q_{(2)}^i,{p}^{(2)}_j\}_{DB}&=\delta^i_j-\frac{1}{ ||\mathbf{q}_{(1)}||^2}q_{(1)}^iq_{(1)}^j \\
\{{p}^{(1)}_i,{p}^{(1)}_j\}_{DB}&=\frac{1}{m \zeta ||\mathbf{q}_{(1)}||^2}[{p}^{(2)}_i({p}%
^{(1)}_j+m\zeta\delta_{jk}q_{(2)}^k)-({p}^{(1)}_i+ m\zeta\delta_{ik}q_{(2)}^k){p}^{(2)}_j] \\
\{{p}^{(1)}_i,{p}^{(2)}_j\}_{DB}&=\frac{1}{ ||\mathbf{q}_{(1)}||^2}{p}^{(2)}_i\delta_{jk}q_{(2)}^k
\end{split}
\end{align}
and all the rest are zero. It is now straight forward to compute the Hamilton's equations using the Dirac
brackets. 

\subsection{Partial reduction II}

Starting with  Cl\`{e}ment
Lagrangian $(\ref{LC})$, and following the definition in $(\ref{LC2})$, we introduce the following reduced Lagrangian 
\begin{align}
L_{C_2}^{C}&=-{\frac{m\zeta}{2}}||\mathbf{\dot{q}}_{(1)}||^2+{\frac{\zeta^{2}}{%
2\mu m }}\mathbf{q}_{(1)}\cdot\mathbf{\dot{q}}_{(1)}\times\mathbf{\dot{q}}_{(2)}+%
\boldsymbol{\lambda}_2\cdot(\mathbf{\dot{q}}_{(1)}-\mathbf{q}_{(2)})
\label{constraint lagrangian0}
\end{align}
using coordinate transformations $\mathbf{x}=\mathbf{q}_{(1)}$, $\mathbf{\dot{x%
}}=\dot{\mathbf{q}}_{(1)}=\mathbf{q}_{(2)},\mathbf{\ddot{x}}=\mathbf{\dot{q}}_{(2)}$ and a set $\boldsymbol{\lambda}_{2}$ of Lagrange
multipliers. Euler-Lagrange equations are 
\begin{align}
\label{lambda}~~
\boldsymbol{\dot{\lambda}}_2=m\zeta\mathbf{\ddot{q}}_1+\frac{\zeta^{2}}{%
\mu m}\mathbf{\dot{q}}_{(1)}\times \mathbf{\dot{q}}_{(2)}-\frac{\zeta^{2}}{2\mu m%
}\mathbf{\ddot{q}}_{(2)}\times \mathbf{q}_{(1)},\quad \boldsymbol{{\lambda}}%
_2=-{\frac{\zeta^{2}}{2\mu m}}\mathbf{q}_{(1)}\times \mathbf{\ddot{q}}_{(1)},~~\dot{\mathbf{q}}_{(1)}-\mathbf{q}_{(2)}=\mathbf{0}.
\end{align}
Before to pass Hamiltonian formalism for $L^C_{C_2}$ given in $(\ref{constraint lagrangian0})$, let us rewrite the Lagrangian  in $(\ref{constraint lagrangian0})$ as
\begin{eqnarray}
L^C_{C_2}=-{\frac{m\zeta}{2}}||\mathbf{\dot{q}}_{(1)}||^2+{\frac{\zeta^{2}}{%
2\mu m }}\mathbf{q}_{(1)}\cdot\mathbf{\dot{q}}_{(1)}\times\mathbf{\dot{q}}_{(2)}+[
\boldsymbol{\lambda}_1+m\zeta\dot{\mathbf{q}}_1-\frac{\zeta^2}{2\mu m}\dot{\mathbf{q}}_{(2)}\times\mathbf{q}_{(1)}]\cdot(\mathbf{\dot{q}}_{(1)}-\mathbf{q}_{(2)})\label{lagrew}
\end{eqnarray}  
using the relation between the Lagrange multipliers $$\boldsymbol{{\lambda}}%
_2=\boldsymbol{{\lambda}}%
_1+m\zeta\dot{\mathbf{q}}_{(1)}-{\frac{\zeta^{2}}{2\mu m}} \mathbf{\dot{q}}_{(2)}\times \mathbf{q}_{(1)}.$$ After some cancellation, the Lagrangian function $(\ref{lagrew})$ turns out to be
\begin{eqnarray}
L^C_{C_2}={\frac{m\zeta}{2}}||\mathbf{\dot{q}}_{(1)}||^2+{\frac{\zeta^{2}}{%
2\mu m }}\mathbf{q}_{(1)}\cdot\mathbf{{q}}_{(2)}\times\mathbf{\dot{q}}_{(2)}-m\zeta \mathbf{\dot{q}}_{(1)}\cdot \mathbf{{q}}_{(2)} +
\boldsymbol{\lambda}_1\cdot(\mathbf{\dot{q}}_{(1)}-\mathbf{q}_{(2)}).\label{ContLag1}
\end{eqnarray}  
In order to pass Hamiltonian formalism for $L_{C_2}^{C}$ in $(
\ref{ContLag1})$, the conjugate momenta are defined by 
\begin{align}
{\mathbf{P}}^{(1)} =m\zeta\dot{\mathbf{q}}_{(1)}+\boldsymbol{\lambda}_1-m\zeta\mathbf{q}_{(2)},~~\mathbf{P}^{(2)} =%
\frac{\zeta^{2}}{2\mu m}\mathbf{q}_{(1)}\times\mathbf{q}_{(2)} ,~~ 
\mathbf{P}^{\lambda_2} =0.   \label{canonical momentum3}
\end{align}
From these momenta  it is possible to solve $\dot{\mathbf{q}}_{(1)}$. The remaining momenta lead to primary constraints
\begin{eqnarray}
 \boldsymbol{\Phi}=\mathbf{P}^{(2)}-\frac{\zeta^{2}}{2\mu m}\mathbf{q}_{(1)}\times\mathbf{q}_{(2)},\quad\boldsymbol{\Phi}^{\lambda_2}=\mathbf{P}^{\lambda_2}.
\end{eqnarray}
For the Lagrangian $L^{C}_{C_{2}}$, the canonical Hamiltonian function $(\ref{CanHam1})$ is
\begin{eqnarray}
H=\frac{1}{2m\zeta}||\mathbf{P}^{(1)}-\boldsymbol{\lambda}_1||^2+\mathbf{q}_{(2)}\cdot\mathbf{P}^{(1)}+ \frac{m\zeta}{2} ||\mathbf{q}_{(2)}||^2
\end{eqnarray}
thus total Hamiltonian function is 
\begin{equation}
H_T=H+\mathbf{U}\cdot\boldsymbol{\Phi}+\mathbf{U}_\lambda\boldsymbol{\Phi}^{\lambda_2}
\end{equation}
with the sets $\mathbf{U}$ and $\mathbf{V}$ of Lagrange multipliers.
Here are the steps of the Dirac-Bergmann Constraint algorithm and the computation of the total Hamiltonian function.

\bigskip

\noindent \textbf{Dirac-Bergmann constraint algorithm step 1:} 
Consistency of $\boldsymbol{\Phi}$ and $\boldsymbol{\Phi}^{\lambda_2}$ lead to 
\begin{equation}
\begin{split}
\dot{\boldsymbol{\Phi}}&=\{\boldsymbol{\Phi},H_T\}=-\mathbf{P}^1-m\zeta \mathbf{q}_{(2)}+\frac{\zeta}{2\mu m^2}\mathbf{q}_{(2)}\times(\boldsymbol{\mathbf{P}^1-\lambda}_1)+\frac{\zeta^2}{\mu m}\mathbf{U}\times\mathbf{q}_{(1)}\\
\dot{\boldsymbol{\Phi}}^{\lambda_2}&=\{\boldsymbol{\Phi}^{\lambda_2},H_T\}=\frac{1}{ m\zeta} (\mathbf{P}^1-\boldsymbol{\lambda}_1).
\end{split}
\end{equation}
So that we arrive at the secondary constraints $\phi=(\mathbf{P}^1+m\zeta \mathbf{q}_{(2)})\cdot \mathbf{q}_{(1)}$ and $\beta=\mathbf{P}^1-\boldsymbol{\lambda}_1$.

\bigskip

\noindent \textbf{Dirac-Bergmann constraint algorithm step 2:} From the conservations of these secondary constraints no more constraint arise and the Lagrange multipliers $\mathbf{U}$ and $\mathbf{U}_\lambda$ are determined. All constraints derived for $L_{C_2}^{C}$ are completely equivalent to $L_{C_1}^{C}$. Also Hamiltonian function for both Lagrangians are identical. That means Hamiltonian formalism for two Lagrangian system $L^{C}_{C_{1}}$ and $L^{C}_{C_{2}}$ are equivalent.

\bigskip 

\begin{remark}
 All constraints derived for the Lagrangian $L_{C_2}^{C}$ in Eq.(\ref{constraint lagrangian0}) 
are completely identical with the constraints  in (\ref{prim3}) which are derived for the Lagrangian $L_{C_1}^{C}$ in Eq. (\ref{constraint lagrangian}). Notice the following table comparing the constraints derived for the partial reductions I and II:

\begin{center}
\renewcommand{\arraystretch}{1.5}
\begin{tabular}{|c|c|c|}
\hline
 & \textbf{Partial Reduction I} &\textbf{Partial Reduction II}  \\ \hline
\textbf{Reduction} & $\mathbf{x}=\mathbf{q}_{(1)}$,~$\dot{\mathbf{x}}=%
\mathbf{q}_{(2)}$,~~$\ddot{\mathbf{x}}=%
\dot{\mathbf{q}}_{(2)}$&$\mathbf{x}=\mathbf{q}_{(1)}$,~$\dot{\mathbf{x}}=%
\dot{\mathbf{q}}_{(1)}$,~~$\ddot{\mathbf{x}}=%
\dot{\mathbf{q}}_{(2)}$ \\ \hline
\textbf{Coordinates}  & $\mathbf{q}_{(1)},\mathbf{q}_{(2)},\boldsymbol{\lambda}_{1},\mathbf{p}^{(1)},%
\mathbf{p}^{(2)},\mathbf{p}^{(\lambda_1)}$&$\mathbf{q}_{(1)},\dot{\mathbf{q}}_{(1)},\boldsymbol{\lambda}_{2},\mathbf{P}^{(1)},%
\mathbf{P}^{(2)},\mathbf{p}^{(\lambda_2)}$ \\ \hline 
\textbf{Primary} \textbf{Constraints} & $\begin{array}{ccl}
\boldsymbol{\Phi}^{(1)}&=&\mathbf{p}^{(1)}-\boldsymbol{\lambda}_{1}\\
\boldsymbol{\Phi}^{(2)}&=&\mathbf{p}^{(2)}-\frac{\zeta^2}{2\mu m}\mathbf{q}_{(1)}\times \mathbf{q}_{(2)}\\\boldsymbol{\Phi}^{(\lambda_1)}&=&\mathbf{p}^{\lambda_1}\end{array}$ & $\begin{array}{ccl}
\boldsymbol{\Phi}^{(2)}&=&\mathbf{P}^{(2)}-\frac{\zeta^2}{2\mu m}\mathbf{q}_{(1)}\times \mathbf{q}_{(2)}\\
\boldsymbol{\Phi}^{(\lambda_2)}&=&\mathbf{P}^{\lambda_2}\end{array}$\\ 
\hline
\textbf{Secondary} \textbf{Constraints} &$\boldsymbol{\Phi}=(\mathbf{p}^{(1)}+m\zeta\mathbf{q}_{(2)})\cdot\mathbf{q}_{(1)}$&$\begin{array}{ccl}
\boldsymbol{\Phi}&=&(\mathbf{P}^{(1)}+m\zeta\mathbf{q}_{(2)})\cdot\mathbf{q}_{(1)}\\
\boldsymbol{\Phi}^{(1)}&=&\mathbf{P}^{(1)}-\boldsymbol{\lambda}_{1}\end{array}$ \\  \hline
\end{tabular}
\end{center}
So the constraints are almost the same for two methods. This gives that the Dirac bracket for the method of Partial Reduction II will be the same with the Dirac bracket for the method of Partial Reduction I.
\end{remark}
  
\begin{remark} Unconstraint variational formalism is useful if the Lagrange multipliers does not contain second order derivatives. Notice that, the Lagrange multipliers presented in (\ref{lambda}) contain  second order derivatives. So that, it is not feasible to study the unconstraint formalism in this present case.
  \end{remark}
\subsection{Deriglazov's trick / Schmidt's method}

This is the last method for the reduction of the Cl\`{e}ment Lagrangian (\ref{LC}) into the first order formalism. In this subsection, we shall employ Deriglazov's trick / Schmidt's method presented in subsection (\ref{alexei}). By referring to the reduced Lagrangian function in Eq.(\ref{AlexeiLag}), we introduce the following first order Lagrangian function
\begin{align}
L^C_A=-\frac{m\zeta}{2}||\dot{\mathbf{x}}||^2+\frac{\zeta^2}{2\mu m}\mathbf{x}%
\cdot\dot{\mathbf{x}}\times{\mathbf{s}}-\mathbf{\dot{\boldsymbol{\gamma}}}\cdot\mathbf{\dot{{x}%
}}-\mathbf{\boldsymbol{\gamma}}\cdot\mathbf{s}  \label{CleAlex}
\end{align}
using the coordinate transformations $$\mathbf{\ddot{x}}=\mathbf{s}, \qquad 
\mathbf{\boldsymbol{\gamma}}=\frac{\partial L^C_A}{\partial \mathbf{s}}=\frac{\zeta^2}{2\mu m}\mathbf{x}
\times\dot{\mathbf{x}}.$$ Here, the Lagrangian $L^C_A$ depends on the velocity components $(\mathbf{\dot{x},\dot{s},\dot{\gamma}})$. This Lagrangian is degenerate. In the dual picture, the conjugate momenta are $(\mathbf{p}^{(x)},\mathbf{p}^{(s)},\mathbf{p}^{(\gamma)})$. The Legendre transformation reads the following definitions for the momenta
\begin{align}
\mathbf{p}^{(x)}=-m\zeta \dot{\mathbf{x}}+\frac{\zeta^2}{2\mu m}{\mathbf{s}}%
\times\mathbf{x}-\mathbf{\dot{\boldsymbol{\gamma}}},\quad \mathbf{p}^{(\gamma)}=-\mathbf{\dot{x}},\quad 
\mathbf{p}^{(s)}=0.
\end{align}
From these relationships, it is possible to solve velocities $\dot{\mathbf{x}}$ and $\dot{\boldsymbol{\gamma}}$ in terms of the momenta as follows 
\begin{align}
\dot{\mathbf{x}}=-\mathbf{p}^{(\gamma)},\quad\dot{\boldsymbol{\gamma}}=m\zeta\mathbf{p}^{(\gamma)}+%
\frac{\zeta^2}{2\mu m}\mathbf{s}\times\mathbf{x}-\mathbf{p}^{(x)}.
\end{align}
We cannot solve the velocity $\dot{\mathbf{s}}$ in terms of the momenta, this induces a primary constraint $
\boldsymbol{\varphi}^s=\mathbf{p}^{(s)}=0$ in the cotangent bundle. 

Let us write the canonical Hamiltonian function for the first order Lagrangian (\ref%
{CleAlex}) as 
\begin{align}
H&=\mathbf{p}^{(x)}\cdot\mathbf{\dot{x}}+\mathbf{p}^{(\gamma)}\cdot\mathbf{\dot{\gamma}}+%
\mathbf{p}^{(s)}\cdot\mathbf{\dot{s}}-L^C_A  \notag \\
&=-\mathbf{p}^{(x)}\cdot\mathbf{p}^{(\gamma)}+\frac{m\zeta}{2}||\mathbf{p}^{(\gamma)}||^2+\frac{%
\zeta^2}{2\mu m}\mathbf{p}^{(\gamma)}\cdot\mathbf{s}\times\mathbf{x}+\boldsymbol{\gamma}\cdot%
\mathbf{s}.
\end{align}
We introduce the total Hamiltonian function by adding the primary constraints into the canonical Hamiltonian function. So that we have
\begin{align}
H_T=H+\mathbf{v}_s\cdot\boldsymbol{\varphi}^s,  \label{TolHamAlx}
\end{align}
Here,  $\mathbf{v}_s$ denoted a set of Lagrange multipliers. Here are the steps of the Dirac-Bergmann constraint algorithm and the explicit computation of the total Hamiltonian function.

\bigskip

\noindent \textbf{Dirac-Bergmann constraint algorithm step 1:}
Consistencies of the primary constraints $\boldsymbol{\varphi}^s$ 
\begin{align}
\dot{\boldsymbol{\varphi}}^s=\{\boldsymbol{\varphi}^s,H_T\}\approx-\frac{\zeta^2}{%
2\mu m}\mathbf{x}\times\mathbf{p}^{(\gamma)}-\boldsymbol{\gamma}
\end{align}
lead to a set of secondary constraints $\boldsymbol{\varphi}=\frac{\zeta^2}{2\mu m}%
\mathbf{x}\times\mathbf{p}^{(\gamma)}+\mathbf{\gamma}$. The total Hamiltonian should be revised as $H_T^1=H_T+ \bf{v}\cdot\boldsymbol{\varphi}$.

\bigskip

\noindent \textbf{Dirac-Bergmann constraint algorithm step 2:} Consistency of these secondary constraints
\begin{align}
\dot{\boldsymbol{\varphi}}=\{\boldsymbol{\varphi},H_{T}^1\}\approx\frac{\zeta^2}{%
2\mu m}\mathbf{s}\times\mathbf{x}-\mathbf{p}^{(x)}+m\zeta\mathbf{p}^{(\gamma)}+\frac{\zeta^2}{\mu m}\bf{v \times x}
 \label{dotvarp}
\end{align}
 result with the determination of the Lagrange multiplier $\bf{v}$ but one more constraint  $\psi=\bf{x}\cdot(\mathbf{p}^{(x)}-m\zeta\mathbf{p}^{(\gamma)})$ arise by taking the dot product of the Eq. (\ref{dotvarp}) by $\bf{x}$. From the conservation of $\psi$, the Lagrange multipliers $\bf v$ are determined.  Thus the Eq.(\ref{dotvarp}) gives a set of tertiary constraints $$\boldsymbol{\Phi}=\frac{\zeta^2}{%
2\mu m}\mathbf{s}\times\mathbf{x}-\mathbf{p}^{(x)}+m\zeta\mathbf{p}^{(\gamma)}.$$ Revision of the total Hamiltonian function result with $H_T^2=H_T^1+ \bf{w}\cdot \boldsymbol{\Phi}$
\bigskip

\noindent \textbf{Dirac-Bergmann constraint algorithm step 3:}
For the tertiary constraints, we compute
\begin{align}
\dot{\boldsymbol{\Phi}}&=\{\boldsymbol{\Phi},H_{T}^2\}\approx\frac{3\zeta^2}{2\mu m}%
\mathbf{p}^{(\gamma)}\times\mathbf{s}-m\zeta\mathbf{s}+\frac{\zeta^2}{2\mu m}\mathbf{v
}_s\times\mathbf{x}+ \frac{2\zeta^2}{\mu m}\bf{w}\times \bf{s}
\label{xx}
\end{align}
and for the secondary and primary constraints, we get
\begin{align}
\dot{\boldsymbol{\varphi}}&=\{\boldsymbol{\varphi},H_{T}^2\}\approx \boldsymbol{\varphi}+m\zeta \bf{w}-\frac{\zeta^2}{\mu m}\bf{w\times \mathbf{p}^{(\gamma) }},\quad
\dot{\boldsymbol{\varphi}^s}=\{\boldsymbol{\varphi}^s,H_{T}^2\}\approx \boldsymbol{\varphi}+\frac{\zeta^2}{\mu m}\bf{w\times \mathbf{x}}.
\end{align}
These equations determine $\bf{w}\approx 0$ and substitution of $\bf w$ into $(\ref{xx})$ leads to the determination of two components of $\mathbf{{v}}_s$, and a new constraint $$\chi=(\frac{3\zeta^2}{2\mu m}%
\mathbf{p}^{(\gamma)}\times\mathbf{s}-m\zeta\mathbf{s})\cdot\mathbf{x}$$ which can also be computed by
taking dot product of $(\ref{xx})$ with $\mathbf{x}$. Now, total Hamiltonian function becomes $H_T^3=H_T^2+ z \chi$.

\bigskip

\noindent \textbf{Dirac-Bergmann constraint algorithm step 4:} Finally from the conservation of $\chi$ we
can determine third component of $\mathbf{v}_s$ and no more
constraint arise. Thus $\mathbf{v}_s$ is determined as 
\begin{align}
\mathbf{v}_s\approx\frac{\frac{\mu m^2}{\zeta}(\mathbf{s}\cdot\mathbf{p}^{(\gamma)})\mathbf{%
x}+\frac{\mu m}{\zeta^2}(\frac{3\zeta^2}{2\mu m}\mathbf{x}\times\mathbf{p}^{(\gamma)}-m\zeta \mathbf{x})\times(\frac{3\zeta^2}{2\mu m}\mathbf{p}^{(\gamma)}\times\mathbf{%
s}-m\zeta\mathbf{s})}{m\zeta ||\mathbf{x}||^2}.
\end{align}
We summarize the results of computations with acceleration bundle in the following table.

\begin{center}
\renewcommand{\arraystretch}{1.5}
\begin{tabular}{|c|c|}
\hline
&\textbf{Deriglazov/Schmidt's Method}    \\ \hline
\textbf{Reduction} & $\ddot{\mathbf{x}}=%
\mathbf{s}$,~$\boldsymbol{\gamma}=\frac{\zeta^2}{2\mu m}\mathbf{x}\times\dot{\mathbf{x}}$,  \\ \hline
\textbf{Coordinates} & $\mathbf{x},\boldsymbol{\gamma},\boldsymbol{s},\mathbf{p}^{(x)},%
\mathbf{p}^{(\gamma)},\mathbf{p}^{(s)}$ \\ \hline
\textbf{Primary} \textbf{Constraints}& $\boldsymbol{\varphi}^s=\mathbf{p}^{(s)}$\\ \hline
\textbf{Secondary Constraints}&$\boldsymbol{\varphi}=\frac{\zeta^2}{2\mu m}\mathbf{x}\times\mathbf{p}^{(\gamma)}+\boldsymbol{\gamma}$  \\
\hline
\textbf{Tertiary Constraints}&$\boldsymbol{\Phi}=m\zeta \mathbf{p}^{(\gamma)}+\frac{\zeta^2}{2\mu m}\mathbf{s}\times \mathbf{x}-\boldsymbol{p}^{(x)}$ \\
\hline
\textbf{Quaternary Constraints}
&$\boldsymbol{\chi}=\mathbf{x}\cdot(m\zeta\mathbf{s}+\frac{3\zeta^2}{2\mu m}\mathbf{s}\times \mathbf{p}^{(\gamma)})$ \\  \hline
\end{tabular}
\end{center}

\bigskip

\noindent \textbf{The total Hamiltonian function and the equations of motion:} 
Substitution of $\mathbf{v}_s$ into (\ref%
{TolHamAlx}) results with the explicit expression of the total Hamiltonian function  
\begin{align}
H_T&=-\mathbf{p}^{(x)}\cdot\mathbf{p}^{(\gamma)}+\frac{m\zeta}{2}(\mathbf{p}^{(\gamma)})^2+\frac{%
\zeta^2}{2\mu m}\mathbf{p}^{(\gamma)}\cdot\mathbf{s}\times\mathbf{x}+\mathbf{{\gamma}}\cdot%
\mathbf{x}  \notag \\
&+\left[\frac{\frac{\mu m^2}{\zeta}(\mathbf{s}\cdot\mathbf{p}^{(\gamma)})\mathbf{x}+%
\frac{\mu m}{\zeta^2}(\frac{3\zeta^2}{2\mu m}\mathbf{x}\times\mathbf{p}%
^{(\gamma)}-m\zeta \mathbf{x})\times(\frac{3\zeta^2}{2\mu m}\mathbf{p}^{(\gamma)}\times\mathbf{%
s}-m\zeta\mathbf{s})}{m\zeta ||\mathbf{x}||^2}\right]\cdot\boldsymbol{p}^s.
\end{align}
In accordance with this, we compute the Hamilton's equations of motion as follows
\begin{align}
\begin{split}
\dot{\mathbf{x}}&\approx-\mathbf{p}^{(\gamma)}, \\ \dot{\mathbf{\gamma}}&\approx-\mathbf{p%
}^{(x)}+m\zeta \mathbf{p}^{(\gamma)}+\frac{\zeta^2}{2\mu m}\mathbf{s}\times\mathbf{x} \\
\dot{\boldsymbol{s}}&\approx\frac{\frac{\mu m^2}{\zeta}(\mathbf{s}\cdot\mathbf{p}%
^{(\gamma)})\mathbf{x}+\frac{\mu m}{\zeta^2}(\frac{3\zeta^2}{2\mu m}\mathbf{x}\times%
\mathbf{p}^{(\gamma)}-m\zeta \mathbf{x})\times(\frac{3\zeta^2}{2\mu m}\mathbf{p}^
{(\gamma)}\times\mathbf{s}-m\zeta\mathbf{s})}{m\zeta ||\mathbf{x}||^2} \\
\dot{\mathbf{p}}^{(x)}&\approx-\frac{\zeta^2}{2\mu m}\mathbf{p}^{(\gamma)}\times\mathbf{s}%
, \\ \dot{\mathbf{p}}^{(\gamma)}&\approx-\boldsymbol{s}, \\ \dot{\mathbf{p}}%
^{(s)}&\approx-\frac{\zeta^2}{2\mu m}\mathbf{x}\times\mathbf{p}^{(\gamma)}-\mathbf{\gamma}.
\end{split}
\end{align}
The Hamilton's equations governing the motion of $\mathbf{p}^{(x)}$ gives
Euler-Lagrange equations in $(\ref{clee})$, and the remaining equations identically satisfied after a back substitution of the momenta in terms of the velocities.

\begin{remark} An alternative reduction of the the second order Cl\`{e}ment Lagrangian (\ref{LC}) can be achieved in the following way. Define the first order Lagrangian
\begin{align}
L^C_T=-\frac{m\zeta}{2}||\dot{\mathbf{q}}_{(1)}||^2-\frac{\zeta^2}{2\mu m}\dot{%
\mathbf{q}}_{(1)}\cdot\dot{\mathbf{q}}_{(2)}+\boldsymbol{\tau}\cdot(\mathbf{q}%
_{(1)}\times\dot{\mathbf{q}}_{(1)}-\mathbf{q}_{(2)})  \label{firstClemAct}
\end{align}
using action angle coordinates $\mathbf{x}=\mathbf{q}_{(1)}$ and $%
\mathbf{x}\times\mathbf{\dot{x}}=\mathbf{q}_{(2)}$. Here, $\boldsymbol{\tau}$ is a set of Lagrange multipliers. See that the Lagrangian (\ref{firstClemAct}) is the same with the one in (\ref{CleAlex}) after employing the identifications  ${\zeta^2}\mathbf{s}/{2\mu m}=\boldsymbol{\tau}$ and ${\zeta^2}\mathbf{q}_{(2)}/{2\mu m}=\boldsymbol{\gamma}$.

\end{remark}
\section{Sar\i o\u{g}lu-Tekin Lagrangian} \label{STLAG}

We start with a $6$-dimensinal manifold $Q$ with local coordinates $(\mathbf{%
x,y})$ consisting of two $3$-dimensional vectors. The higher order tangent
bundles are equipped with the following induced sets of coordinates 
\begin{equation}
\begin{split}
(\mathbf{x,y,\dot{x},\dot{y}}) &\in TQ \\
(\mathbf{x,y,\dot{x},\dot{y},\ddot{x},\ddot{y}}) &\in T^{2}Q \\
(\mathbf{x,y,\dot{x},\dot{y},\ddot{x},\ddot{y},\dddot{x},\dddot{y}}) &\in
T^{3}Q.
\end{split}
\end{equation}%
In \cite{st06}, Sar\i o\u{g}lu and Tekin proposed a degenerate second order
Lagrangian on $T^{2}Q$ given by%
\begin{equation}
L^{ST}[\mathbf{x},\mathbf{y}]=\frac{1}{2}\left[ a(\mathbf{\dot{x}}^{2}+%
\mathbf{\dot{y}}^{2})+\frac{2}{\mu }\dot{\mathbf{y}}\cdot \ddot{\mathbf{x}}%
-m^{2}(\mathbf{y}^{2}+\mathbf{x}^{2})\right] .  \label{LasST}
\end{equation}%
In this case, the Euler-Lagrange equations (\ref{EL2}) take the
particular form 
\begin{equation}
m^{2}\mathbf{x}+a\mathbf{\ddot{x}}=\frac{1}{\mu }\mathbf{y}^{(3)},\text{\ \
\ }m^{2}\mathbf{y}+a\mathbf{\ddot{y}}=-\frac{1}{\mu }\mathbf{x}^{(3)}.
\label{ste}
\end{equation}

\subsection{Total Reduction} 
At first, we are applying the total reduction method exhibited in the subsection (\ref{totred}) to the Sar\i o\u{g}lu and Tekin Lagrangian (\ref{LasST}) in order to arrive at a first order formalism. Accordingly, writing the Lagrangian function defined in (\ref{L-1}) for the the Sar\i o\u{g}lu and Tekin Lagrangian, we compute the following particular form
\begin{equation}
L^{ST}_C=\frac{1}{2}\left[ a(\mathbf{{q}}_{(1)}^{2}+%
\mathbf{{q}}_{(3)}^{2})+\frac{2}{\mu }{\mathbf{q}}_{(3)}\cdot{\mathbf{q}}_{(2)}%
-m^{2}(\mathbf{y}^{2}+{\mathbf{x}}^{2})\right]+\boldsymbol{\lambda}_1\cdot(\dot{\bf x}-\mathbf{q}_{(1)})+\boldsymbol{\lambda}_2\cdot(\dot{\bf q}_{(1)}-\mathbf{q}_{(2)})+\boldsymbol{\lambda}_3\cdot(\dot{\bf y}-\mathbf{q}_{(3)})\label{totSARTEKLag}
\end{equation}
using coordinate transformations $\dot{\bf x}=\mathbf{q}_{(1)}$, $\dot{\bf q}_{(1)}=\mathbf{q}_{(2)}$, $\dot{\bf y}=\mathbf{q}_{(3)}$ and the Lagrange multipliers $\boldsymbol{\lambda}_1,\boldsymbol{\lambda}_2, \boldsymbol{\lambda}_3$.
The conjugate momenta for (\ref{totSARTEKLag}) are computed as
\begin{eqnarray}
{\bf{p}}^{(1)}=\boldsymbol{\lambda}_2,\quad {\bf{p}}^{(2)}=0,~~{\bf{p}}^{(3)}=0,~~{\bf{p}}^{(\lambda_1)}=0,~~\bf{p}^{(\lambda_2)}=0,~~{\bf{p}}^{(\lambda_3)}=0,~~{\bf{p}}^{(x)}=\boldsymbol{\lambda}_1,~~{\bf{p}}^{(y)}=\boldsymbol{\lambda}_3.
\end{eqnarray}  
It is not possible to solve the velocities in terms of the momenta instead we have the following set of primary constraints
\begin{equation}
\begin{split}
{\bf\Phi}^{(1)}={\bf{p}}^{(1)}-\boldsymbol{\lambda}_2, \quad {\bf\Phi}^{(2)}&={\bf{p}}^{(2)}, \quad{\bf\Phi}^{(3)}={\bf{p}}^{(3)}, \quad{\bf\Phi}^{(x)}={\bf{p}}^{(x)}-\boldsymbol{\lambda}_1, \quad{\bf\Phi}^{(y)}={\bf{p}}^{(y)}-\boldsymbol{\lambda}_3 \\{\bf\Phi}^{(\lambda_1)}&={\bf{p}}^{(\lambda_1)}, \quad {\bf\Phi}^{(\lambda_2)}={\bf{p}}^{(\lambda_2)}, \quad {\bf\Phi}^{(\lambda_3)}={\bf{p}}^{(\lambda_3)}.
\end{split}
\end{equation}
The canonical Hamiltonian function is 
\begin{equation}
H=-\frac{1}{2}\left[ a(\mathbf{{q}}_{(1)}^{2}+%
\mathbf{{q}}_{(3)}^{2})+\frac{2}{\mu }{\mathbf{q}}_{(3)}\cdot{\mathbf{q}}_{(2)}%
-m^{2}(\mathbf{y}^{2}+{\mathbf{x}}^{2})\right]+\boldsymbol{\lambda}_1\cdot\mathbf{q}_{1}+\boldsymbol{\lambda}_2\cdot\mathbf{q}_{2}+\boldsymbol{\lambda}_3\cdot\mathbf{q}_{3}\label{canHamSARTEK}
\end{equation} 
and the total Hamiltonian function is 
\begin{equation}
H_T=H+{\bf U}_1\cdot\Phi^{(1)}+{\bf U}_2\cdot\Phi^{(2)}+{\bf U}_3\cdot{\bf\Phi}^{(3)}+{\bf U}_{\lambda_1}{\bf\Phi}^{(\lambda_1)}+{\bf U}_{\lambda_2}\cdot{\bf\Phi}^{(\lambda_2)}+{\bf U}_{\lambda_3}\cdot{\bf\Phi}^{(\lambda_3)}+{\bf U}_{x}\cdot{\bf\Phi}^{(x)}+{\bf U}_{y}\cdot{\bf\Phi}^{(y)}.\label{totHamSARTEK}
\end{equation}
Here are the steps of the Dirac-Bergmann constraint algorithm and the computation of the total Hamiltonian function.

\bigskip

\noindent \textbf{Dirac-Bergmann constraint algorithm step 1:}
Conservations of primary constraints
\begin{equation}
\begin{split}
\dot{{\bf\Phi}}^{(1)}&=\{\boldsymbol{\Phi}^{(1)},H_{T}\}\approx a{\bf{q}}_{(1)}-\boldsymbol{\lambda}_1-{\bf U}_{\lambda_2}, \quad\dot{{\bf\Phi}}^{(2)}=\{\boldsymbol{\Phi}^{(2)},H_{T}\}\approx \frac{1}{\mu} {\bf{q}}_{(3)}-\boldsymbol{\lambda}_2, \\ \dot{{\bf\Phi}}^{(3)}&=\{\boldsymbol{\Phi}^{(3)},H_{T}\}\approx a {\bf{q}}_{(3)}+\frac{1}{\mu}{\bf{q}}_{(2)}-\boldsymbol{\lambda}_3,\quad \dot{{\bf\Phi}}^{(\lambda_1)}=\{\boldsymbol{\Phi}^{(\lambda_1)},H_{T}\}\approx -{\bf{q}}_{(1)}+{\bf U}_x, \\ \dot{{\bf\Phi}}^{(\lambda_2)}&=\{\boldsymbol{\Phi}^{(\lambda_2)},H_{T}\}\approx {\bf U}_1-{\bf{q}}_{(2)}, \quad \dot{{\bf\Phi}}^{(\lambda_3)}=\{\boldsymbol{\Phi}^{(\lambda_3)},H_{T}\}\approx {\bf U}_y-{\bf{q}}_{(3)}
\\ \dot{{\bf\Phi}}^{(x)}&=\{\boldsymbol{\Phi}^{(x)},H_{T}\}\approx -m^2 {\bf{x}}-{\bf U}_{\lambda_1}, \quad \dot{{\bf\Phi}}^{(y)}=\{\boldsymbol{\Phi}^{(y)},H_{T}\}\approx -m^2 {\bf{y}}-{\bf U}_{\lambda_3}
\end{split}
\end{equation}
lead to the determinations of ${\bf U}_1,{\bf U}_x,{\bf U}_y,{\bf U}_{\lambda_1},{\bf U}_{\lambda_2},{\bf U}_{\lambda_3}$ and there arise the following secondary constraints 
\begin{equation}
\boldsymbol{\Phi}=\frac{1}{\mu} {\bf{q}}_{(3)}-\boldsymbol{\lambda}_2,~~~~\boldsymbol{\phi}=a {\bf{q}}_{(3)}+\frac{1}{\mu} {\bf{q}}_{(2)}-\boldsymbol{\lambda}_3.
\end{equation}
In accordance with this, by substitutions of the Lagrange multipliers and addition of secondary constraints we revise the total Hamiltonian function as
 \begin{equation} \label{totHamSARTEK-1}
 \begin{split}
H_T^1&=H_T+ \bf{U}\cdot\boldsymbol{\Phi}+\bf{V}\cdot \boldsymbol{\phi} \\
&=-\frac{1}{2}\left[ a(\mathbf{{q}}_{(1)}^{2}+%
\mathbf{{q}}_{(3)}^{2})+\frac{2}{\mu }{\mathbf{q}}_{(3)}\cdot{\mathbf{q}}_{(2)}%
-m^{2}(\mathbf{y}^{2}+{\mathbf{x}}^{2})\right]+\boldsymbol{p}^1\cdot\mathbf{q}_{2}-m^2(\bf{p}^{(\lambda_1)}\cdot \bf{x}+\bf{p}^{(\lambda_3)}\cdot \bf{y}) \\
&+{\bf{p}}^{(\lambda_2)}\cdot [a{\bf{q}}_{(1)}-\boldsymbol{\lambda}_1]+{\bf{p}}^y\cdot{\bf{q}}_{(3)}+{\bf U}_{2}\cdot{\bf\Phi}^{2}+{\bf U}_{3}\cdot{\bf\Phi}^{3}+{\bf U}\cdot{\bf\Phi}+{\bf V}\cdot{\boldsymbol{\phi}}.
\end{split}
\end{equation}

\bigskip

\noindent
\textbf{Dirac-Bergmann constraint algorithm step 2:}
Consistency checks of the secondary constraints ${\boldsymbol{\Phi}}$ and ${\boldsymbol{\phi}}$ are
\begin{equation}
\dot{{\boldsymbol \Phi}}=\{\boldsymbol{\Phi},H_{T}^1\}\approx\frac{1}{\mu} {\bf{U}}_{3}-a{\bf{q}}_{(1)}+\boldsymbol{\lambda}_1,~~\dot{{\boldsymbol \phi}}=\{{\boldsymbol{\phi}},H_{T}^1\}\approx a{\bf{U}}_3+ \frac{1}{\mu} {\bf{U}}_{2}+m^2 \bf{y}.
\end{equation} 
These calculations determine the Lagrange multipliers ${\bf{U}}_{2}$ and ${\bf{U}}_{3}$. Conservation of ${\bf\Phi}^{(\lambda_2)}$ and ${\bf\Phi}^{(\lambda_3)}$ give the Lagrange multipliers $\bf{U}\approx \mu ({\boldsymbol \phi}-{\boldsymbol \Phi}^y)\approx 0$ and $\bf{V}\approx \mu ({\boldsymbol \Phi}^1-{\boldsymbol \Phi})\approx 0$. Here is the table of constraints:

\begin{center}
\renewcommand{\arraystretch}{1.5}
\begin{tabular}{|c|c|}
\hline
&\textbf{Total Reduction}    \\ \hline
\textbf{Reduction} & $\dot{\mathbf{x}}=\mathbf{q}_{(1)}$,~$\ddot{\mathbf{x}}=%
\mathbf{q}_{(2)}$,~$\dot{\mathbf{y}}=
\mathbf{q}_{(3)}$,  \\ \hline
\textbf{Coordinates} & $\mathbf{x},\mathbf{y},\mathbf{q}_{(1)},\mathbf{q}_{(2)},\mathbf{q}_{(3)},\boldsymbol{\lambda}_{1},\boldsymbol{\lambda}_{2},\boldsymbol{\lambda}_{3},\mathbf{p}^x,\mathbf{p}^y,\mathbf{p}^{(1)},%
\mathbf{p}^{(2)},\mathbf{p}^{(3)},\mathbf{p}^{(\lambda_1)},\mathbf{p}^{(\lambda_2)},\mathbf{p}^{(\lambda_3)}$ \\ \hline
\textbf{Primary}~\textbf{Constraints} &$\begin{array}{cclcccccc}
\boldsymbol{\Phi}^{(1)}&=&\mathbf{p}^{(1)}-\boldsymbol{\lambda}_{2} &\boldsymbol{\Phi}^{(x)}&=&\mathbf{p}^{(x)}-\boldsymbol{\lambda}_1&\boldsymbol{\Phi}^{(\lambda_1)}&=&\mathbf{p}^{(\lambda_1)}\\
\boldsymbol{\Phi}^{(2)}&=&\mathbf{p}^{(2)}&\boldsymbol{\Phi}^{(y)}&=&\mathbf{p}^{(y)}-\boldsymbol{\lambda}_3&\boldsymbol{\Phi}^{(\lambda_2)}&=&\mathbf{p}^{(\lambda_2)}\\
\boldsymbol{\Phi}^{(3)}&=&\mathbf{p}^{(3)}&&&&\boldsymbol{\Phi}^{(\lambda_3)}&=&\mathbf{p}^{(\lambda_3)}
\end{array}$  \\ \hline
\textbf{Secondary constraints}&$\boldsymbol{\Phi}=\frac{1}{\mu}\mathbf{q}_{(3)}-\boldsymbol{\lambda}_{2}$,~~$ \boldsymbol{\phi}=a\bf{q}_{(3)}+\frac{1}{\mu}\mathbf{q}_{(2)}-\boldsymbol{\lambda}_{3}$ \\   \hline
\end{tabular}
\end{center}

\bigskip

\noindent \textbf{The total Hamiltonian function and the equations of motion:} Substitutions of the Lagrange multipliers into (\ref{totHamSARTEK-1}) give the total Hamiltonian function
\begin{equation}
\begin{split}
H_T^1 = & -\frac{1}{2}\left[ a(\mathbf{{q}}_{(1)}^{2}+%
\mathbf{{q}}_{(3)}^{2})+\frac{2}{\mu }{\mathbf{q}}_{(3)}\cdot{\mathbf{q}}_{(2)}
-m^{2}(\mathbf{y}^{2}+{\mathbf{x}}^{2})\right]+{\bf{p}}^{(1)}\cdot\mathbf{q}_{2} \\ & - \mu m^2 {\bf{p}}^{(2)}\cdot\mathbf{y}(a\mu-1)+ \mu {\bf{p}}^{(3)}\cdot(-\boldsymbol{\lambda}_1+a \mathbf{q}_{(1)})  -m^2{\bf{p}}^{(\lambda_1)}\cdot \mathbf{x}+ {\bf{p}}^{(\lambda_2)}\cdot(a \mathbf{q}_{(1)}- \boldsymbol{\lambda}_1) \\& -m^2{\bf{p}}^{(\lambda_3)}\cdot \mathbf{y}+\mathbf{p}^x\cdot\mathbf{q}_{(1)}+\mathbf{p}^y\cdot\mathbf{q}_{(3)}+\mu \boldsymbol \Phi\cdot({\boldsymbol \phi}-{\boldsymbol \Phi}^y) +\mu \boldsymbol{\phi}\cdot({\boldsymbol \Phi}^1-{\boldsymbol \Phi}).
\end{split}
\end{equation} 
Accordingly, the Hamilton's equations are computed to be
\begin{equation}
\begin{split}
\dot{\bf{q}}_{(1)}&\approx {\bf{q}}_{(2)}, \qquad\dot{\bf{q}}_{(2)}\approx-a\mu^2 m^2{\bf{y}}+\mu m^2{\bf{y}}, \qquad\dot{\bf{q}}_{(3)}\approx-\mu(\boldsymbol{\lambda}_1-a\mathbf{q}_{(1)}),\quad \dot{\bf{x}}\approx {\bf{q}}_{(1)}, \qquad \dot{\bf{y}}\approx{\bf{q}}_{(3)},
\\
\dot{\boldsymbol{\lambda}_1 }& \approx-m^2{\bf{x}}, \qquad 
\dot{\boldsymbol{\lambda}_2 }\approx a {\bf{q}}_{(1)}-\boldsymbol{\lambda}_1,
\qquad 
\dot{\boldsymbol{\lambda}_3 }\approx-m^2{\bf{y}},\\ \dot{\bf{p}}^{(1)}& \approx a{\bf{q}}_{(1)}- {\bf{p}}^{x}-a\bf{p}^{\lambda_2},\qquad \dot{\bf{p}}^{(2)}\approx\frac{1}{\mu}{\bf{q}}_{(3)}-{\bf{p}}^{(1)},\qquad 
\dot{\bf{p}}^{(3)} \approx a{\bf{q}}_{(3)}+\frac{1}{\mu} {\bf{q}}_{(2)}-\mathbf{p}^y, 
\\
\dot{\bf{p}}^{(\lambda_1)}&\approx  0, \qquad \dot{\bf{p}}^{(\lambda_2)}\approx 0,~~\dot{\bf{p}}^{(\lambda_3)}\approx 0,\quad
\dot{\bf{p}}^{x} \approx -m^2 \mathbf{x}, \qquad \dot{\bf{p}}^{y}\approx -m^2 \mathbf{y},
\end{split}
\end{equation}
with the constraints ${\bf{p}}^{(2)}={\bf{p}}^{(3)}={\bf{p}}^{(\lambda_2)}={\bf{p}}^{(\lambda_3)}=0$.

\bigskip

\noindent \textbf{The Dirac bracket:} All constraints for the first order Lagrangian (\ref{totSARTEKLag})
\begin{align*}
\boldsymbol{\Phi}^{(1)}=\mathbf{p}^{(1)}-\boldsymbol{\lambda}_{2} ,\quad\boldsymbol{\Phi}^{(x)}=\mathbf{p}^{(x)}-\boldsymbol{\lambda}_1,\quad\boldsymbol{\Phi}^{(\lambda_1)}=\mathbf{p}^{(\lambda_1)},\quad
\boldsymbol{\Phi}^{(2)}=\mathbf{p}^{(2)},\quad\boldsymbol{\Phi}^{(y)}=\mathbf{p}^{(y)}-\boldsymbol{\lambda}_3,\\ \boldsymbol{\Phi}^{(\lambda_2)}=\mathbf{p}^{(\lambda_2)},\quad
\boldsymbol{\Phi}^{(3)}=\mathbf{p}^{(3)},\quad\boldsymbol{\Phi}^{(\lambda_3)}=\mathbf{p}^{(\lambda_3)}
,\quad \boldsymbol{\Phi}=\frac{1}{\mu}\mathbf{q}_{(3)}-\boldsymbol{\lambda}_{2},~~\boldsymbol{\phi}=a\bf{q}_{(3)}+\frac{1}{\mu}\mathbf{q}_{(2)}-\boldsymbol{\lambda}_{3}
\end{align*}
are second class since their bracket is nonzero. Poisson brackets of constraint defined in (\ref{cmatrix}) becomes
\begin{displaymath}
\mathbf{C} =
\left( \begin{array}{cccccccccc}
\mathbf{0} & \mathbf{0} & \mathbf{0}&\mathbf{0}&-\delta_{ij}&\mathbf{0}&\mathbf{0}&\mathbf{0}&\mathbf{0}&\mathbf{0} \\
\mathbf{0} & \mathbf{0} & \mathbf{0}&\mathbf{0}&\mathbf{0}&\mathbf{0}&\mathbf{0}&\mathbf{0}&\mathbf{0}&-\frac{1}{\mu}\delta_{ij}  \\
\mathbf{0} & \mathbf{0} & \mathbf{0}&\mathbf{0}&\mathbf{0}&\mathbf{0}&\mathbf{0}&\mathbf{0}&-\frac{1}{\mu}\delta_{ij}&-a\delta_{ij} \\
\mathbf{0} & \mathbf{0} & \mathbf{0}&\mathbf{0}&\mathbf{0}&\mathbf{0}&\delta_{ij}&\mathbf{0}&\mathbf{0}&\mathbf{0} \\
\delta_{ij} & \mathbf{0} & \mathbf{0}&\mathbf{0}&\mathbf{0}&\mathbf{0}&\mathbf{0}&\mathbf{0}&\delta_{ij}&\mathbf{0} \\
\mathbf{0} & \mathbf{0} & \mathbf{0}&\mathbf{0}&\mathbf{0}&\mathbf{0}&\mathbf{0}&\delta_{ij}&\mathbf{0}&\delta_{ij}\\
\mathbf{0}&\mathbf{0}&\mathbf{0}&-\delta_{ij}&\mathbf{0}&\mathbf{0}&\mathbf{0}&\mathbf{0}&\mathbf{0}&\mathbf{0}
 \\
\mathbf{0}&\mathbf{0}&\mathbf{0}&\mathbf{0}&\mathbf{0}&-\delta_{ij}&\mathbf{0}&\mathbf{0}&\mathbf{0}&\mathbf{0}
\\
\mathbf{0}&\mathbf{0}&\frac{1}{\mu}\delta_{ij}&\mathbf{0}&-\delta_{ij}&\mathbf{0}&\mathbf{0}&\mathbf{0}&\mathbf{0}&\mathbf{0}\\
\mathbf{0}&\frac{1}{\mu}\delta_{ij}&a\delta_{ij}&\mathbf{0}&\mathbf{0}&-\delta_{ij}&\mathbf{0}&\mathbf{0}&\mathbf{0}&\mathbf{0}
\end{array} \right)
\end{displaymath}
and substitution of the inverse of C into Dirac bracket (\ref{diracbrac}) leads to
\begin{align}
\{F,G\}_{DB}&=\{F,G\}+a\mu^2\{F,\Phi^{(1)}_k\}\delta^{kl}\{\Phi^{(2)}_l,G\}-\mu\{F,\Phi^{(1)}_k\}\delta^{kl}\{\Phi^{(3)}_l,G\}-\{F,\Phi^{(1)}_k\}\delta^{kl}\{\Phi^{(\lambda_2)}_l,G\}\notag\\
&-a\mu^2\{F,\Phi^{(2)}_k,G\}\delta^{kl}\{\Phi^{(1)}_l,G\}+\mu\{F,\Phi^{(2)}_k\}\delta^{kl}\{\Phi^{(y)}_l,G\}+a\mu^2\{F,\Phi_k\}\delta^{kl}\{\Phi^{(y)}_l,G\}-\{F,\Phi^{(2)}_k\}\delta^{kl}\{\phi_l,G\}\notag\\
&+\mu\{F,\Phi^{(3)}_k\}\delta^{kl}\{\Phi^{(1)}_l,G\}-\mu\{F,\Phi^{(3)}_k\}\delta^{kl}\{\Phi_l,G\}+\{F,\Phi^{(\lambda_1)}_k\}\delta^{kl}\{\Phi^{(x)}_l,G\}+\{F,\Phi^{(\lambda_2)}_k\}\delta^{kl}\{\Phi^{(1)}_l,G\}\notag\\
&+\{F,\Phi^{(\lambda_3)}_k\}\delta^{kl}\{\Phi^{(y)}_l,G\}-\{F,\Phi^{(x)}_k\}\delta^{kl}\{\Phi^{(\lambda_1}_l,G\}-\mu\{F,\Phi^{(y)}_k\}\delta^{kl}\{\Phi^{(2)}_l,G\}-\{F,\Phi^{(y)}_k\}\delta^{kl}\{\Phi^{(\lambda_3)}_l,G\}\notag\\
&-a\mu^2\{F,\Phi_k\}\delta^{kl}\{\Phi^{(2)}_l,G\}+\mu\{F,\Phi_k\}\delta^{kl}\{\Phi^{(3)}_l,G\}+\mu\{F,\phi_k\}\delta^{kl}\{\Phi^{(2)}_l,G\}.
\end{align}
Dirac brackets of coordinates are
\begin{align*}
\{q_{(1)}^i,q_{(2)}^j\}_{DB}&=-a\mu^2\delta^{ij},\quad \{q_{(1)}^i,q_{(3)}^j\}_{DB}=\mu\delta^{ij},\quad \{q_{(1)}^i,\lambda_2^j\}_{DB}=\delta^{ij},\quad \{q_{(1)}^i,p^{(1)}_j\}_{DB}=\delta^i_{j}\\
\{q_{(2)}^i,y^j\}_{DB}&=-\mu\delta^{ij}\quad\{x^i,\lambda^j_1\}_{DB}=\delta^{ij}, \quad \{y^i,\lambda^j_3\}_{DB}=\delta^{ij},\quad\{x^i,p^{(x)}_j\}_{DB}=\delta^i_{j},\quad\{y^i,p^{(y)}_j\}_{DB}=\delta^i_{j}
\end{align*}
and all the remaining Dirac brackets of coordinates are zero. Using these Dirac brackets of coordinates and Hamilton function (\ref{canHamSARTEK}) one can easily determine Hamilton equations of motion.
\subsection{Partial reduction I}  
\label{First LC0ST} 
We now apply the partial reduction I (c.f. \ref{parred1}) to the Sar\i o\u{g}lu and Tekin Lagrangian (\ref{LasST}). So that we rewrite the first order Lagrangian function given in $(\ref{LC1})$ for the case of the Sar\i o\u{g}lu and Tekin Lagrangian and arrive at 
\begin{align}
L_{C_{1}}^{ST}&=\frac{1}{2}\left[ a(\mathbf{\dot{q}}_{(1)}^2+\mathbf{\dot{y}}%
^2)+\frac{2}{\mu}\mathbf{\dot{y}}\cdot\mathbf{\dot{q}}_{(2)}-m^{2}(\mathbf{y}%
^2+\mathbf{q}_{(1)}^2)\right]+\boldsymbol{\lambda}_{1}\cdot (\mathbf{\dot{q}}%
_{(1)}-\mathbf{q}_{(2)})  \label{LC0ST} 
\end{align}
using coordinate transformations $\mathbf{x} =\mathbf{q}_{(1)}, 
\mathbf{\dot{x}} =\mathbf{\dot{q}}_{(1)}=\mathbf{q}_{(2)}, \mathbf{\ddot{x}}=%
\mathbf{\dot{q}}_{(2)}$
and Lagrange multipliers $\boldsymbol{\lambda}_{1}$. The dual coordinates are 
$(\mathbf{q}_{(1)},\mathbf{q}_{(2)},\boldsymbol{\lambda}_1,\mathbf{y})$. Using this first order Lagrangian, we compute the conjugate momenta as
\begin{align}
\mathbf{p}^{(1)} & =a\mathbf{\dot{q}}_{(1)}+\boldsymbol{\lambda}_{1},\quad 
\mathbf{p}^{(2)}=\frac{1}{\mu}\mathbf{\dot{y}},\quad \mathbf{p}%
^{\lambda_{1}} =\mathbf{0},\quad \mathbf{p}^{(y)}=a\mathbf{\dot{y}}+%
\frac{1}{\mu}\mathbf{\dot{q}}_{(2)}.   \label{mompiy}
\end{align}
From these equations we compute the velocities $\mathbf{\dot{y}}$, $\mathbf{\dot{q}}_{(1)}$ and $\mathbf{\dot{q}}_{(2)}$ in terms of the momenta as follows
\begin{align}
\mathbf{\dot{y}} =\mu\mathbf{p}^{(2)},\quad \mathbf{\dot{q}}_{(1)} =\frac{1%
}{a}( \mathbf{p}^{(1)}-\boldsymbol{\lambda}_{1}),\quad 
\mathbf{\dot{q}}_{(2)} =\mu( \mathbf{p}^{(y)}-a\mu\mathbf{p}%
^{(1)})
\end{align}
whereas from the equation involving the momenta $\mathbf{p}^{\lambda_1}$, we have a primary
constraint $\boldsymbol{\Phi}^{(\lambda_1)}=\mathbf{p}^{\lambda_1}=0.$
In this case, the canonical Hamiltonian function is  
\begin{align}
H^{ST}_{C_{1}}=\mu\mathbf{p}^{(2)}\cdot\mathbf{p}^{(y)}-\frac{%
a\mu^{2}}{2}({\mathbf{p}}^{(2)})^2+\frac{m^{2}}{2}(\mathbf{y}^2+\mathbf{q
}^2_{(1)} )+\boldsymbol{\lambda}_1\cdot\mathbf{q}_{(2)} +\frac{1}{2a}(%
\mathbf{p}^{(1)}-\boldsymbol{\lambda}_{1})^2
\label{hamforfirstorderSarandTekin0}
\end{align}
whereas the total Hamiltonian $(\ref{TotHam1})$ defined to be
\begin{align}
H_{T_1}^{ST} & =H^{ST}_{C_{1}}+\mathbf{u}_{(\lambda_1)}\cdot\boldsymbol{\Phi}^{(\lambda_1)} 
\label{tothamforsar0}
\end{align}
by adding the primary constraint with a Lagrange multiplier $\mathbf{u}_{(\lambda_1)}$.
Here are the steps of the Dirac-Bergmann constraint algorithm and the computation of the total Hamiltonian function.

\bigskip

\noindent \textbf{Dirac-Bergmann constraint algorithm step 1:}
Consistency of the primary constraint $\boldsymbol{\Phi}^{(\lambda_1)}$ 
\begin{align}
\boldsymbol{\dot{\Phi}}^{(\lambda_1)}&=\frac{1}{a}(%
\mathbf{p}^{(1)}-\boldsymbol{\lambda}_{1})-\mathbf{q}_{(2)} 
\label{conservation of psi}
\end{align}
leads us to a secondary constraint $\boldsymbol{\Phi}=\frac{1}{a}(%
\mathbf{p}^{(1)}-\boldsymbol{\lambda}_1)-\mathbf{q}_{(2)}. $ Note that Eq. $(%
\ref{conservation of psi})$ will vanish weakly when we use $\mathbf{\dot{q}}%
_{(1)}=\mathbf{q}_{(2)}$. In this
case, it is not possible to determine the Lagrange multiplier $\mathbf{u}_{(\lambda_1)}$, thus
equation of motion for $\boldsymbol{\lambda}_1$ remains arbitrary. To solve
this, we consider $\boldsymbol{\Phi}$ as a secondary constraint. Revised the
total Hamiltonian as 
\begin{align}
H_{T_2}^{ST} & =H^{ST}_{C_{1}}+\mathbf{u}_{(\lambda_1)}\cdot\boldsymbol{\Phi}^{(\lambda_1)}+%
\mathbf{u}\cdot\boldsymbol{\Phi}   \label{tothamforsar1}
\end{align}
by adding secondary constraint with Lagrange multiplier $\mathbf{u}$. 

\bigskip

\noindent \textbf{Dirac-Bergmann constraint algorithm step 2:} The
consistency of the secondary constraint $\boldsymbol{\Phi}$ can be checked
through 
\begin{align}
\boldsymbol{\dot{\Phi}}& =\{\boldsymbol{\Phi},H^{ST}_{T_2}\}\approx-\frac{%
m^{2}}{a}\mathbf{q}_{(1)}+a\mu^{2}\mathbf{p}^{(2)}-\mu \mathbf{p}%
^{(y)}-\frac{1}{a}\mathbf{u}_{(\lambda_1)}
\end{align}
which leads us to determine the Lagrange multiplier $\mathbf{u}_{(\lambda_1)}$ as $\mathbf{%
u}_{(\lambda_1)}\approx-{m^{2}}\mathbf{q}_{(1)}+a^{2}\mu^{2}\mathbf{p}^{(2)}-a\mu%
\mathbf{p}^{(y)}$. On the other hand, the consistency of $\boldsymbol{%
\Phi}^{(\lambda_1)}$ 
\begin{align}
\boldsymbol{\dot{\Phi}}^{(\lambda_1)}& =\{\boldsymbol{\Phi}^{(\lambda_1)},H^{ST}_{T_2}\}%
\approx \boldsymbol{\Phi}+\frac{1}{a}\mathbf{u}
\end{align}
leads to us to determine $\mathbf{u}\approx-a\boldsymbol{\Phi%
}.$

\bigskip

\noindent \textbf{The total Hamiltonian function and the equations of motion:} Substitutions of the Lagrange multipliers $\mathbf{u}_{(\lambda_1)}$ and $\mathbf{u}$ into (\ref{tothamforsar1}) give the total Hamiltonian function 
\begin{align}
H^{ST}_{T_2} & =\mu\mathbf{p}^{(2)}\cdot\big(\mathbf{p}^{(y)}-\frac{%
a\mu} {2}{\mathbf{p}}^{(2)}\big)+\frac{m^{2}}{2}(\mathbf{y}^2+\mathbf{q}%
^2_{(1)})+\boldsymbol{\lambda}_{1}\cdot\mathbf{q}_{(2)}+\frac{1}{2a}(\boldsymbol{p}^{(1)}-\boldsymbol{\lambda}_{1})^2  \notag \\
& -\big( {m^{2}}\mathbf{q}_{(1)}-a^{2}\mu^{2}{\mathbf{p}}^{(2)}+a\mu 
\mathbf{p}^{(y)}\big)\cdot\mathbf{p}^{\lambda_1}-\frac{1}{a}\big(%
\mathbf{p}^{(1)}-\boldsymbol{\lambda}_{1}-\mathbf{q}_{(2)}\big)^2.
\label{tothamforsarson}
\end{align}
The Hamilton's equations of motion using are 
\begin{align}
\mathbf{\dot{{q}}}_{(1)} &\approx\frac{1}{a}(\mathbf{p}^{(1)}-\boldsymbol{\lambda%
}_{1}),\quad \dot{\mathbf{q}}_{(2)} \approx\mu(\mathbf{p}^{(y)}-a\mu^{2}%
\mathbf{p}^{(2)}),\quad \mathbf{\dot{y}} \approx\mu{\mathbf{p}}^{(2)}
\notag \\
\dot{\boldsymbol{\lambda}}_{1} & \approx-m^{2}\mathbf{q}_{(1)}+ a^{2}\mu^{2}{%
\mathbf{p}}^{(2)}-a\mu{\mathbf{p} }^{(y)},\quad \dot{{\boldsymbol{p
}}}^{(1)}\approx-m^{2} \mathbf{q}_{(1)}  \label{HameqforALx} \\
\quad\dot{{\mathbf{p}}}^{(2)} & \approx-\boldsymbol{\lambda}_{1},\quad \dot{{%
\mathbf{p}}}^{\lambda_1}\approx\frac{1}{a}({\mathbf{p}}^{(1)}-\boldsymbol{
\lambda}_{1})-\mathbf{q}_{(2)},\quad \dot{{\mathbf{p}}}^{(y)}\approx-m^{2}%
\mathbf{y}.  \notag
\end{align}

\bigskip

\noindent \textbf{The Dirac bracket:} 
The constraints 
\begin{align}  \label{conss}
\boldsymbol{\Phi} =\frac{1}{a}(\mathbf{p}^{(1)}-\boldsymbol{\lambda}%
_1)-\mathbf{q}_{(2)},\quad \boldsymbol{\Phi}^{(\lambda_1)} =\mathbf{p}^{\lambda_1}
\end{align}
for the first order Lagrangian $L_{C_{(1)}}^{ST}$ are of the second class since their  Poisson bracket $\{\Phi^{i},\Phi_{j}^{(\lambda_1)}\}=\frac{1}{a}\delta_j^{i}$
is nonzero. Recall the definition of the Dirac bracket presented in $(\ref%
{diracbrac})$. In particular, for the constraints (\ref{conss}), we arrive at
\begin{align}
\{F,G\}_{DB} &
=\{F,G\}+a\{F,\Phi^{k}\}\delta_{k}^n\{\Phi_{n}^{(\lambda_1)},G\}-a\{F,\Phi
_{k}^{(\lambda_1)}\}\delta^{k}_n\{\Phi^{n},G\}   \label{dirbrac}
\end{align}
after the substitution of the inverse matrix of 
\begin{align}
C = \left[ 
\begin{array}{cc}
\{\Phi^{k},\Phi^{n}\} & \{\Phi^{k}, \Phi_{n}^{(\lambda_1)}\} \\ 
\{\Phi_{n}^{(\lambda_1)},\Phi^{k}\} & \{\Phi_{n}^{(\lambda_1)},\Phi_{k}^{(\lambda_1)}\}%
\end{array}
\right] =\frac{1}{a}\left[ 
\begin{array}{cc}
0_{3\times3} & \delta^{k}_n \\ 
-\delta^{k}_n & 0_{3\times3}%
\end{array}
\right].
\end{align}
We compute the Dirac brackets of the coordinates as follows
\begin{equation}
\begin{split}
\{q_{(1)}^{i},\lambda_{1}^{j}\}_{DB} & =\delta^{ij},\quad\{q_{(1)}^{i},{p}%
^{(1)}_{j}\}_{DB} =\delta_{j}^{i},\quad \{q_{(2)}^{i},{p}^{(2)}_{j}\}_{DB}
=\delta^{i}_{j} \\
\{\lambda^{i}_{1},{p}^{(2)}_{j}\}_{DB} &=-a\delta^i_{j},\quad \{y^{i},{p}%
^{(y)}_{j}\}_{DB} =\delta^{i}_{j}
\end{split}
\end{equation}
and all rest is zero. Using these Dirac brackets of coordinates and
Hamiltonian function $(\ref{hamforfirstorderSarandTekin0})$ we can recover
the equations of motion (\ref{HameqforALx}) after a direct computation.

\subsection{Partial reduction II}

In this subsection, we are applying the method of partial reduction II, which is presented in the subsection (\ref{parred2}), to the Sar{\i}o\u{g}lu-Tekin Lagrangian (\ref{LasST}). Accordingly, we compute 
 \begin{align}
L_{C_{2}}^{ST}&=\frac{1}{2}\left[ a(\mathbf{q}_{(2)}^2+\mathbf{\dot{y}}^2)+%
\frac{2}{\mu}\mathbf{\dot{y}}\cdot\mathbf{\dot{q}}_{(2)}-m^{2}(\mathbf{y}^2+%
\mathbf{q}_{(1)}^2)\right]+\boldsymbol{\lambda}_{2}\cdot (\mathbf{\dot{q}}_{(1)}-%
\mathbf{q}_{(2)})  \label{LC0ST1}
\end{align}
using the coordinate transformations $\mathbf{x} =\mathbf{q}_{(1)}$, $\mathbf{\dot{x}} =\mathbf{\dot{q}}_{(1)}=\mathbf{q}_{(2)}$, and $\mathbf{\ddot{x}}=%
\mathbf{\dot{q}}_{(2)}$. Here, $\boldsymbol{\lambda}_{2}$ stands for a set of Lagrange multipliers. 
The fiber derivatives of $L^{ST}_{C_{2}}$ establish the relationship between
the velocities and the momenta as follows 
\begin{align}
\mathbf{p}^{(1)} =\boldsymbol{\lambda}_{2},\quad \mathbf{p}^{(2)} =%
\frac{1}{\mu}\mathbf{\dot{y}},\quad \mathbf{p}^{\lambda_2} =0,\quad 
\mathbf{p}^{(y)}=a\mathbf{\dot{y}}+\frac{1}{\mu}\mathbf{\dot{q}}_{(2)}.
\label{Pi^{(1)}}
\end{align}
From the conjugate momenta $(\ref{Pi^{(1)}})$, it is possible to
solve $\dot{\mathbf{y}}$ and $\dot{\mathbf{q}}_{(2)}$ as functions of
coordinates and momenta
\begin{align}
\mathbf{\dot{y}}=\mu\mathbf{p}^{(2)},\quad \mathbf{\dot{q}}_{(2)}
=\mu\left( \mathbf{p}^{(y)}-a\mu\mathbf{p}^{(2)}\right),
\end{align}
but, the others lead to primary constraints 
\begin{align}
\boldsymbol{\Phi}^{(1)} ={\mathbf{p}}^{(1)}-\boldsymbol{\lambda}_{2},\quad 
\boldsymbol{\Phi}^{(\lambda_2)}={\mathbf{p}}^{\lambda_2} .  \label{consss1}
\end{align}
In this case, the canonical Hamiltonian function becomes 
\begin{align}
H^{ST}_{C_{2}}& =(\mu{\mathbf{p}}^{(2)}\cdot{\mathbf{p}}^{(y)}-\frac{%
a\mu^{2}}{2}(\mathbf{p}^{(2)})^2)-\frac{a}{2}\mathbf{q}_{(2)}+{\boldsymbol{
p} }^{(1)}\cdot\mathbf{q}_{(2)}+\frac{m^{2}}{2}(\mathbf{y}^2+\mathbf{q}^2_{(1)})
\label{hamforfirstorderSarandTekin}
\end{align}
whereas the total Hamiltonian function is defined to be 
\begin{align}
H^{ST}_{T} & =H^{ST}_{C_{2}}+\mathbf{u}_{(1)}\cdot\boldsymbol{\Phi}^{(1)}+%
\mathbf{u}_{(\lambda_2)}\cdot \boldsymbol{\Phi}^{(\lambda_2)}.\label{HST1}
\end{align}
Here, $\mathbf{u}_{(1)}$ and $\mathbf{u}_{(\lambda_2)}$ Lagrange multipliers. 
Here are the steps of the Dirac-Bergmann constraint algorithm.

\bigskip

\noindent \textbf{Dirac-Bergmann constraint algorithm step 1:} The
consistency checks of the primary constraints 
\begin{align}
\boldsymbol{\dot{\Phi}}^{(1)}=\{\boldsymbol{\Phi}^{(1)},H_{T}^{ST}\} \approx-m^{2}%
\mathbf{q}_{(1)}-\mathbf{u}_{(\lambda_2)},\quad \boldsymbol{\dot{\Phi}}%
^{(\lambda_2)}=\{\boldsymbol{\Phi}^{(\lambda_2)},H_{T}^{ST}\} \approx \mathbf{u}_{(1)}
\end{align}
allow us to determine the multipliers as $\mathbf{u}_{(1)}\approx0$ and $\mathbf{u}_{(\lambda_2)}\approx -m^{2}%
\mathbf{q}_{(0)}$, respectively. 

\bigskip

\noindent \textbf{The total Hamiltonian function and the equations of motion:} Substitutions of the Lagrange multipliers $\mathbf{u}_{(1)}$ and $\mathbf{u}_{(\lambda_2)}$ into (\ref{HST1}) determine
the total Hamiltonian $H^{ST}_{T}$ in the following explicit form 
\begin{align}
H_{T}^{ST} =\mu\mathbf{p}^{(2)}\cdot(\mathbf{p}^{(y)}-\frac{a\mu}{2}{%
\mathbf{p}}^{(2)})-\frac{a}{2}\mathbf{q}_{(2)}^2+{\mathbf{p} }%
^{(1)}\cdot\mathbf{q}_{(2)}+\frac{m^{2}}{2}(\mathbf{y}^2+\mathbf{q}^2_{(1)})-m^{2}%
\mathbf{q}_{(1)}\cdot\mathbf{p}^{\lambda_2}.  \label{hamforLSTC1}
\end{align}
The Hamilton's equations are 
\begin{align}
\mathbf{\dot{q}}_{(1)}&\approx\mathbf{q}_{(2)},\quad \mathbf{\dot{q}}_{(2)}\approx\mu{\mathbf{p} }^{(y)}-a\mu^{2}{\mathbf{p}}^{(2)},\quad \mathbf{\dot{y}}%
\approx\mu{\mathbf{p}}^{(2)},\quad \boldsymbol{\dot{\lambda}}_2 \approx-m^{2}%
\mathbf{q}_{(1)}  \notag \\
\dot{{\mathbf{p}}}^{(2)} & \approx a\mathbf{q}_{(2)}-{\mathbf{p}}%
^{(1)},\quad \dot{{\mathbf{p}}}^{\lambda_2} \approx 0,\quad \dot{{\mathbf{p}%
}}^{(1)} \approx-m^{2} \mathbf{q}_{(1)},\quad \dot{{\mathbf{p}}}^{(y)}
\approx-m^{2}\mathbf{y} .  \label{hameqALEx1}
\end{align}
%The equations governing $\mathbf{q}_{(1)},\mathbf{q}_{(1)},\mathbf{y},{%
%\mathbf{p}}^{(2)}$ and ${\mathbf{p}}^{\lambda_2}$ are identically satisfied
%using the definitions of momenta. The equations governing ${%
%\mathbf{p}}^{(1)}$ and $\boldsymbol{\lambda}_2$ give one half of the
%Euler-Lagrange equations $(\ref{ste})$. The equations governing the other
%momenta ${\mathbf{p}}^{(y)}$ give the rest half of the Euler-Lagrange
%equations.\newline

\noindent \textbf{The Dirac bracket:} Now we are going to arrive at the Hamilton's equations
by defining the Dirac bracket 
\begin{align}
\{F,G\}_{DB} &
=\{F,G\}-\{F,\Phi^{(1)}_{k}\}\delta^{kn}\{\Phi^{(\lambda_2)}_{n},G\}+\{F,\Phi^{
(\lambda_2)}_{k}\}\delta^{kn}\{\Phi^{(1)}_{n},G\},  \label{DBforsar-tek}
\end{align}
where we substitute the inverse of matrix
\begin{align}
C & =\left( 
\begin{array}{cc}
\{\Phi^{(1)}_{k},\Phi^{(1)}_{n}\} & \{\Phi^{(1)}_{k},\Phi^{(\lambda_2)}_{n}\} \\ 
\{\Phi^{(\lambda_2)}_{k},\Phi^{(1)}_{n}\} & \{\Phi^{(\lambda_2)}_{k},\Phi_{%
\lambda}^{n}\}%
\end{array}
\right) =\left( 
\begin{array}{cc}
{0}_{3\times3} & -\delta_{kn} \\ 
\delta_{kn} & {0}_{3\times3}%
\end{array}
\right).
\end{align}
For the constraint space defined by $(\ref{consss1})$, the Dirac brackets of the
coordinates are 
\begin{align}
\{q_{(1)}^{i},\lambda_2^{j}\}_{DB}=\delta^{ij},~~ \{q_{(1)}^{i},{p}%
^{(1)}_{j}\}_{DB}=\delta_{j}^{i},~~ \{q_{(2)}^{i},{p}^{(2)}_{j}\}_{DB}=%
\delta_{j}^{i},~~ \{y^{i},{p}^{(y)}_{j}\}_{DB}=\delta_{j}^{i}
\label{DBL_{C1}}
\end{align}
and the rest is zero.

\begin{remark}[Constraints for partial reductions I and II] In the following table we list the constraints derived from the Lagrangian functions $L_{C_{(1)}}^{ST}$ in Eq.(\ref{LC0ST}) and $L_{C_{(2)}}^{ST}$ in Eq.(\ref{LC0ST1}). Observe that, to arrive at an identification between these two constraint systems, it is enough to take  
$\boldsymbol{\lambda}_{1}=\boldsymbol{\lambda}_{2}-a\mathbf{q}_{2}$. 

\begin{center}
\renewcommand{\arraystretch}{1.5}
\begin{tabular}{|c|c|c|}
\hline
 & \textbf{Partial Reduction I} &\textbf{Partial Reduction II}  \\ \hline
\textbf{Reduction} & $\mathbf{x}=\mathbf{q}_{(1)}$,~$\dot{\mathbf{x}}=%
\mathbf{q}_{(2)}$,~~$\ddot{\mathbf{x}}=%
\dot{\mathbf{q}}_{(2)}$&$\mathbf{x}=\mathbf{q}_{(1)}$,~$\dot{\mathbf{x}}=%
\dot{\mathbf{q}}_{(1)}$,~~$\ddot{\mathbf{x}}=%
\dot{\mathbf{q}}_{(2)}$ \\ \hline
\textbf{Coordinates}  & $\mathbf{q}_{(1)},\mathbf{q}_{(2)},\mathbf{y},\boldsymbol{\lambda}_{1},\mathbf{p}^{(1)},%
\mathbf{p}^{(2)},\mathbf{p}^{(y)},\mathbf{p}^{(\lambda_1)}$&$\mathbf{q}_{(1)},\mathbf{q}_{(2)},\mathbf{y},\boldsymbol{\lambda}_{2},\mathbf{p}^{(1)},%
\mathbf{p}^{(2)},\mathbf{p}^{(y)},\mathbf{p}^{(\lambda_2)}$ \\ \hline
\textbf{Primary} \textbf{Constraints}   & $\boldsymbol{\Phi}^{(\lambda_1)}=\mathbf{p}^{(\lambda_1)}$ &$\begin{array}{ccl}
\boldsymbol{\Phi}^{(1)}&=&\mathbf{p}^{(1)}-\boldsymbol{\lambda}_2\\
\boldsymbol{\Phi}^{(\lambda_2)}&=&\mathbf{p}^{\lambda_2}
\end{array}$ \\ 
\hline
\textbf{Secondary} \textbf{Constraints} 
 &$\boldsymbol{\Phi}=\frac{1}{a}(\mathbf{p}^{(1)}-\boldsymbol{\lambda}_1)-\mathbf{q}_{(2)}$& \\  \hline
\end{tabular}
\end{center}

\end{remark}

\subsection{Deriglazov's trick / Schmidt's method}

Finally, we are now applying the method in the subsection (\ref{alexei}) to the Sar\i o\u{g}lu-Tekin Lagrangian \eqref{LasST}. Accordingly, following the definition in (\ref{AlexeiLag}), we introduce the first order Lagrangian 
\begin{align}
L^{ST}_A=\frac{1}{2}\left[a(\mathbf{\dot{x}}^2+\mathbf{\dot{y}}^2)+\frac{2}{%
\mu}\mathbf{\dot{y}}\cdot\mathbf{q}_{(3)}-m^2(\mathbf{x}^2+\mathbf{y}^2)\right]-%
\boldsymbol{\dot{\gamma}}\cdot\mathbf{\dot{x}}-\boldsymbol{\gamma}\cdot\mathbf{q}_{(3)}
\end{align}
depending on the base variables $(\mathbf{x,y,q}_{(3)},\boldsymbol{\gamma})$ and the velocity variables $(\mathbf{\dot{x},\dot{y},\dot{q}}_{(3)},\boldsymbol{\dot{\gamma}})$. Introduce the 
dual coordinates as $(\mathbf{p}^{(x)},\mathbf{p}^{(y)},\mathbf{p}^{(3)},\mathbf{p}^{(\gamma)})$. The Legendre transformation reads the following equations 
\begin{align}
\mathbf{p}^{(x)}=a\mathbf{\dot{x}}-\boldsymbol{\dot{\gamma}},~\mathbf{p}^{(y)}=a\mathbf{\dot{y%
}}+\frac{1}{\mu}\mathbf{q}_{(3)},~\mathbf{p}^{(3)}=\mathbf{0},~\mathbf{p}^{(\gamma)}=-\mathbf{%
\dot{x}}.
\end{align}
We can solve the velocities $\mathbf{\dot{x}},\mathbf{\dot{y}},%
\boldsymbol{\dot{\gamma}}$ in terms of the momenta and there remains a set of primary constraint $\boldsymbol{\varphi}=\mathbf{p}^{(3)}\approx 0$. The total Hamiltonian function is defined to be
\begin{align}
H_T=-\mathbf{p}^{(x)}\cdot\mathbf{p}^{(\gamma)}+\frac{1}{2}\left[-a(\mathbf{p}^{(\gamma)})^2+\frac{1%
}{a}(\mathbf{p}^{(y)}-\frac{1}{\mu}\mathbf{q}_{(3)})^2\right]+\frac{m^2}{2}(\mathbf{x}%
^2+\mathbf{y}^2)+\boldsymbol{\gamma}\cdot\mathbf{q}_{(3)}+\boldsymbol{w}\cdot \mathbf{p}^{(3)} \label{AlexeiHAM}
\end{align}
where $\boldsymbol{w}$ is a set of Lagrange multipliers.
Here are the steps of the Dirac-Bergmann constraint algorithm and the computation of the total Hamiltonian function.

\bigskip

\noindent \textbf{Dirac-Bergmann constraint algorithm step 1:} 
Consistency condition of primary constraint $\boldsymbol{\varphi}$ 
\begin{align}
\dot{\boldsymbol{\varphi}}=\{\boldsymbol{\varphi},H_T\}\approx\frac{1}{a}(\mathbf{p}^{(y)}-\frac{1}{\mu}\mathbf{q}_{(3)})-\boldsymbol{\gamma}
\end{align}
gives secondary constraint $\boldsymbol{\Phi}=\frac{1}{a}(\mathbf{p}^{(y)}-\frac{%
1}{\mu}\mathbf{q}_{(3)})-\boldsymbol{\gamma}$. We are revising the total Hamiltonian as $%
H_{T1}=H_T+\bar{\boldsymbol{w}}\cdot\boldsymbol{\Phi}$.

\bigskip

\noindent \textbf{Dirac-Bergmann constraint algorithm step 2:} Conservation of the secondary
constraint $\boldsymbol{\Phi}$ 
\begin{align}
\dot{\boldsymbol{\Phi}}=\{\boldsymbol{\Phi},H_{T1}\}\approx-\frac{1}{a}m^2\mathbf{y%
}-\frac{1}{a\mu}\boldsymbol{w}+\mathbf{p}^{(x)}-a\mathbf{p}^{(\gamma)}
\end{align}
determine $\boldsymbol{w}$ and conservation of $\boldsymbol{\varphi}$ 
\begin{align}
\dot{\boldsymbol{\varphi}}=\{\boldsymbol{\varphi},H_{T1}\}\approx\boldsymbol{\Phi}+%
\frac{1}{a\mu} \bar{\boldsymbol{w}}
\end{align}
gives $ \bar{\boldsymbol{w}}\approx 0$. 

\bigskip

\noindent \textbf{The total Hamiltonian function and the equations of motion:}  Thus substitutions of $ {\boldsymbol{w}}$ and $ \bar{\boldsymbol{w}}$
into total Hamilton $H_{T1}$ give
\begin{align}
H_{T1}&=-\mathbf{p}^{(x)}\cdot\mathbf{p}^{(\gamma)}+\frac{1}{2}\left(-a(\mathbf{p}^{(\gamma)})^2+%
\frac{1}{a}(\mathbf{p}^{(y)}-\frac{1}{\mu}\mathbf{q}_{(3)})^2\right)+\frac{m^2}{2}(%
\mathbf{x}^2+\mathbf{y}^2)+\boldsymbol{\gamma}\cdot\mathbf{q}_{(3)}  \notag \\
&+\mathbf{p}^{(3)}\cdot(-{m^2}\mu\mathbf{y}+a\mu^2(\mathbf{p}^{(x)}+a\mathbf{p}
^{(\gamma)}))-a\mu(\frac{1}{a}(\mathbf{p}^{(y)}-\frac{1}{\mu}\mathbf{q}_{(3)})-\boldsymbol{\gamma})^2.
\end{align}
The Hamilton's equations of motion generated by this total Hamiltonian function are 
\begin{align}
\mathbf{\dot{x}}&\approx-\mathbf{p}^{(\gamma)},~\mathbf{\dot{y}}\approx\frac{1}{a}(%
\mathbf{p}^{(y)}-\frac{1}{\mu}\mathbf{q}_{(3)}), ~\boldsymbol{\dot{\gamma}}\approx-\mathbf{p}
^x-a\mathbf{p}^{(\gamma)}  \notag \\
\mathbf{\dot{q}}_{(3)}&\approx-{m^2}\mu\mathbf{y}+a\mu^2(\mathbf{p}^{(x)}+a\mathbf{p}
^{(\gamma)}),~\mathbf{\dot{p}}^{(\gamma)}\approx-\mathbf{q}_{(3)},~\mathbf{\dot{p}}^{(3)}\approx0,
\label{AlexeiHamEQ} \\
\mathbf{\dot{p}}^{(y)}&\approx-m^2\mathbf{y},\mathbf{\dot{p}}^{(x)}\approx-m^2%
\mathbf{x}  \notag
\end{align}
From these equations, first two lines of equations satisfied identically and
the last one give Euler-Lagrange equations $(\ref{ste})$ using the definitions
of momenta. Here is the list of constraints.
\begin{center}
\renewcommand{\arraystretch}{1.5}
\begin{tabular}{|c|c|}
\hline
&\textbf{Deriglazov's trick / Schmidt's method}    \\ \hline
\textbf{Reduction} & $\ddot{\mathbf{x}}=%
\mathbf{q}_{(3)}$,~$\boldsymbol{\gamma}=\frac{\partial L}{\partial \mathbf{q}_{(3)}}=\frac{1}{\mu}\dot{\mathbf{y}}-\boldsymbol{\gamma}$,  \\ \hline
\textbf{Coordinates} & $\mathbf{x},\mathbf{y},\mathbf{q}_{(3)},\boldsymbol{\gamma},\mathbf{p}^{(x)},\mathbf{p}^{(y)},\mathbf{p}^{(3)},%
\mathbf{p}^{(\gamma)}$ \\ \hline
\textbf{Primary} \textbf{Constraints}& $\boldsymbol{\varphi}=\mathbf{p}^{(3)}$\\ \hline
\textbf{Secondary constraints} &$\boldsymbol{\Phi}=\frac{1}{a}(\mathbf{p}^{(y)}-\frac{1}{\mu}\mathbf{q}_{(3)})-\boldsymbol{\gamma}$  \\ \hline
\end{tabular} 
\end{center}

\bigskip

\noindent \textbf{The Dirac bracket:} 
All the constraints 
\begin{eqnarray}
\boldsymbol{\varphi}=\mathbf{p}^{(3)},\quad \boldsymbol{\Phi}=\frac{1}{a\mu}(%
\mathbf{p}^{(y)}-\frac{1}{\mu}\mathbf{ q}_{(3)})-\boldsymbol{\gamma}
\end{eqnarray}
are second class since the Poisson brackets $\{\boldsymbol{\varphi}_i,%
\boldsymbol{\Phi}_j\}=\frac{1}{a\mu^2}\delta_{ij}$ are nonzero. In this case Dirac bracket $(%
\ref{diracbrac})$ becomes
\begin{equation}
\{{F},{G}\}_{DB}=\{{F},{G}\}+\{{F},%
\boldsymbol{\varphi}\}a\mu^2\{\boldsymbol{\Phi},{G}\}-\{{F}%
,\boldsymbol{\Phi}\}a\mu^2\{\boldsymbol{\varphi},{G}\}.  \label{DbcAlexei}
\end{equation}
The Dirac brackets of coordinates are
\begin{align}
\{x^i,p^{(x)}_j\}_{DB}&=\{y^i,p^{(y)}_j\}_{DB}=\{\gamma^i,p^{(\gamma)}_j\}_{DB}=\delta^i_j  \notag
\\
\{y^i,q_{(3)}^j\}_{DB}&=\mu\delta^{ij},\quad\{q_{(3)}^i,p^{(\gamma)}_j\}_{DB}=-a\mu^2\delta^i_j
\end{align}
and the rest are identically zero. From these Dirac brackets of coordinates and
Hamiltonian function $(\ref{AlexeiHAM})$ one obtains the Hamilton's equations (\ref{AlexeiHamEQ}) after a straight forward calculation.

\section{Conclusions}
We have presented several different ways enabling to reduce a second order Lagrangian function to a first order one in Section \ref{red}. The geometric construction presented in sections 2 and 3 clarifies  the relationships between these methods. We have particularly interested in two second order degenerate Lagrangian theories available in the theory of topologically massive gravity, namely the Cl\`{e}ment Lagrangian (\ref{clemlag}), and the Sar\i o\u{g}lu-Tekin Lagrangian (\ref{stlag}). For each of them, using the methods  presented in Section $3$ we  have introduced different first order Lagrangian functions so that different Hamiltonian realizations in Sections (\ref{Clement-Lag}) and (\ref{STLAG}), respectively. As a result, we have completed the Hamiltonian analysis of the Cl\`{e}ment, and the Sar\i o\u{g}lu-Tekin Lagrangians.

Both the Cl\'{e}ment and the Sar\i o\u{g}lu-Tekin Lagrangians have rotational symmetry. In \cite{GaHoRa11}, the higher dimensional version of Lagrangian reduction theory \cite{CeMaRa01,MaRa98} has been presented. Motivated by this, we are planning to exhibit formal reductions of these theories under rotational symmetry in a future study. 

\section{Acknowledgment}

We are grateful Prof. Alexei Deriglazov for his comments attracting our attention to \cite{de2017}.

\bigskip

\bigskip

\bigskip

\bigskip

\end{document}